\documentclass[journal,11pt,doublecolumn]{IEEEtran}
\usepackage{bbm,dsfont,mathrsfs,fixmath}
\usepackage{stmaryrd}
\usepackage{amsmath,amstext,amsfonts,amssymb}
\usepackage{epsfig,float}
\usepackage{graphics,dblfloatfix}
\usepackage{cite}
\usepackage{psfrag}
\usepackage{subfigure}
\usepackage{url}
\usepackage{shadow,color,pifont,times,rotate}

\newcommand{\mb}{\mathbf}

\newcommand{\mc}{\mathcal}

\def\ex{\times }

\def\esp{{\mathbb E}}  
\newcommand{\mkv}{-\!\!\!\!\minuso\!\!\!\!-}
 \DeclareMathAlphabet{\mathpzc}{OT1}{pzc}{m}{it}
\newtheorem{theorem}{Theorem}
\newtheorem{lemma}{Lemma}

\newtheorem{corollary}{Corollary}
\newtheorem{remark}{Remark}
\newtheorem{definition}{Definition}
\newenvironment{proof}[1][Proof]{\begin{trivlist}
\item[\hskip \labelsep {\bfseries #1}]}{\end{trivlist}}
\usepackage{enumerate}    
\ifCLASSINFOpdf
		\usepackage{epstopdf}
\else
\fi

\begin{document}
\title{Cooperative Strategies for Simultaneous \\ and Broadcast Relay Channels}

 \author{Arash~Behboodi,~\IEEEmembership{Student Member,~IEEE,}
 and Pablo~Piantanida,~\IEEEmembership{Member,~IEEE,} 

\thanks{This research was partially supported by the \emph{Institute Carnot C3S}.
The material in this paper was presented in part at the 2010 and 2011 IEEE Intern. Symp. on Information Theory; 2010 48th Annual Allerton Conference on  Communication, Control, and Computing and 2010 Proc. of the 8th Intern. Symp. on Modeling and Optimization in Mobile, Ad Hoc and Wireless Networks.}
\thanks{Arash Behboodi is with the Department of Telecommunications, SUPELEC, 91192 Gif-sur-Yvette, France, Email: arash.behboodi@supelec.fr. 

Pablo  Piantanida is with the Department of Telecommunications, SUPELEC, 91192 Gif-sur-Yvette, France, Email: pablo.piantanida@supelec.fr.}
}

\maketitle

%%*********************************************************************************************************
%%Abstract
%%*********************************************************************************************************
%\vspace{-10mm}
\begin{abstract}
Consider the \emph{simultaneous relay channel} (SRC) which consists of a set of relay channels where the source wishes to transmit common and private information to each of the destinations. This problem is recognized as being equivalent to that of sending common and private information to several destinations in presence of helper relays where each channel outcome becomes a branch of the \emph{broadcast relay channel} (BRC). Cooperative schemes and capacity region for a set with two memoryless relay channels are investigated. The proposed coding schemes, based on \emph{Decode-and-Forward} (DF) and \emph{Compress-and-Forward} (CF) must be capable of transmitting information simultaneously to all destinations in such set. 

Depending on the quality of source-to-relay and relay-to-destination channels, inner bounds on the capacity of the general BRC are derived. Three cases of particular interest are considered: cooperation is based on DF strategy for both users --referred to as DF-DF region--, cooperation is based on CF strategy for both users --referred to as CF-CF region--, and cooperation is based on DF strategy for one destination and CF for the other --referred to as DF-CF region--. These results can be seen as a generalization and hence unification of previous works. An outer-bound on the capacity of the general BRC is also derived. Capacity results are obtained for the specific cases of semi-degraded and degraded Gaussian  simultaneous relay channels. Rates are evaluated for Gaussian models where the source must guarantee a minimum amount of information to both users while additional information is sent to each of them. %Application of these results arises in the context of cooperative cellular  networks. 
\end{abstract}

\begin{IEEEkeywords}
Capacity, cooperative strategies, simultaneous relay channels, broadcast relay channel, decode-and-forward, compress-and-forward, broadcasting.
\end{IEEEkeywords}
%\IEEEpeerreviewmaketitle
%%*********************************************************************************************************
%%I. Introduction
%%*********************************************************************************************************
\newpage
\section{Introduction}
\begin{figure*}[t]
\centering
\subfigure[Simultaneous relay channel (SRC)]{
\includegraphics [width=.33 \textwidth] {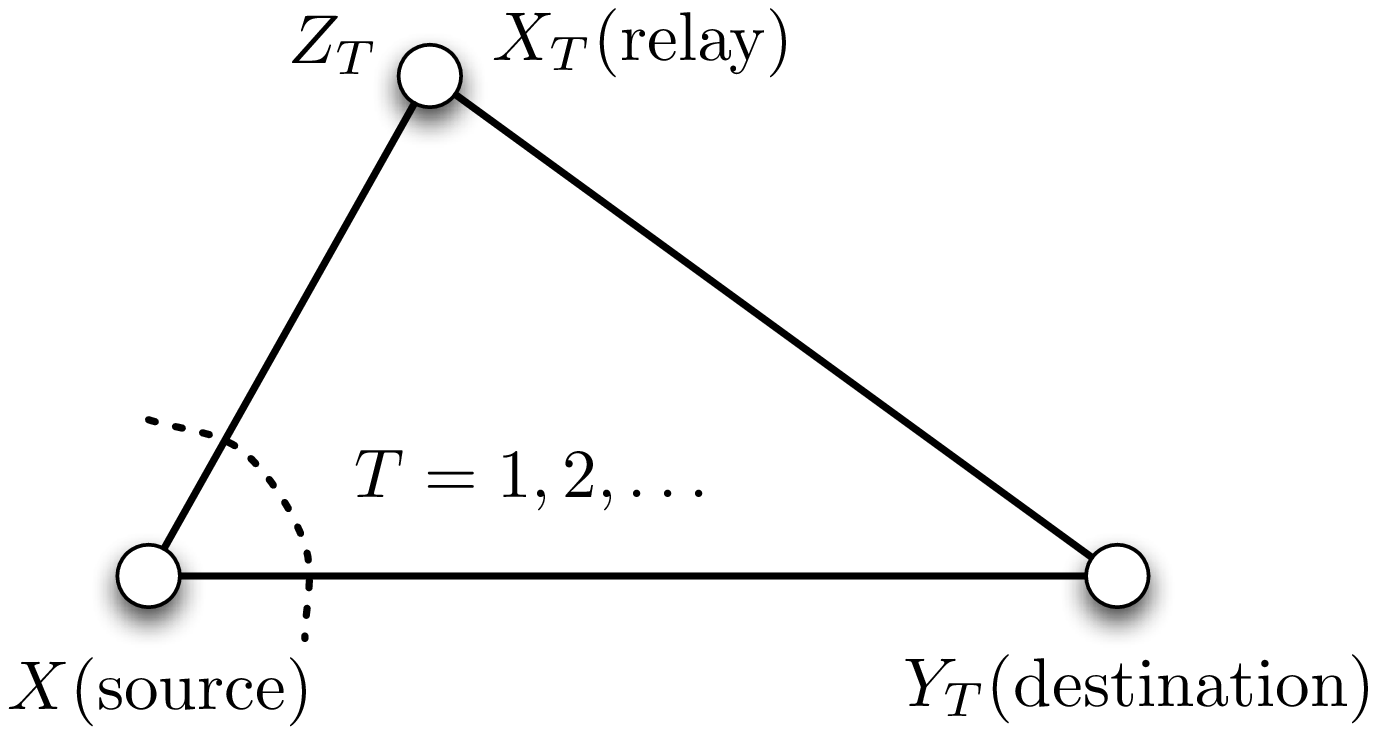}
\label{figCH1:I-1}
}
\subfigure[BRC with two relays]{
	\includegraphics [width=.25 \textwidth] {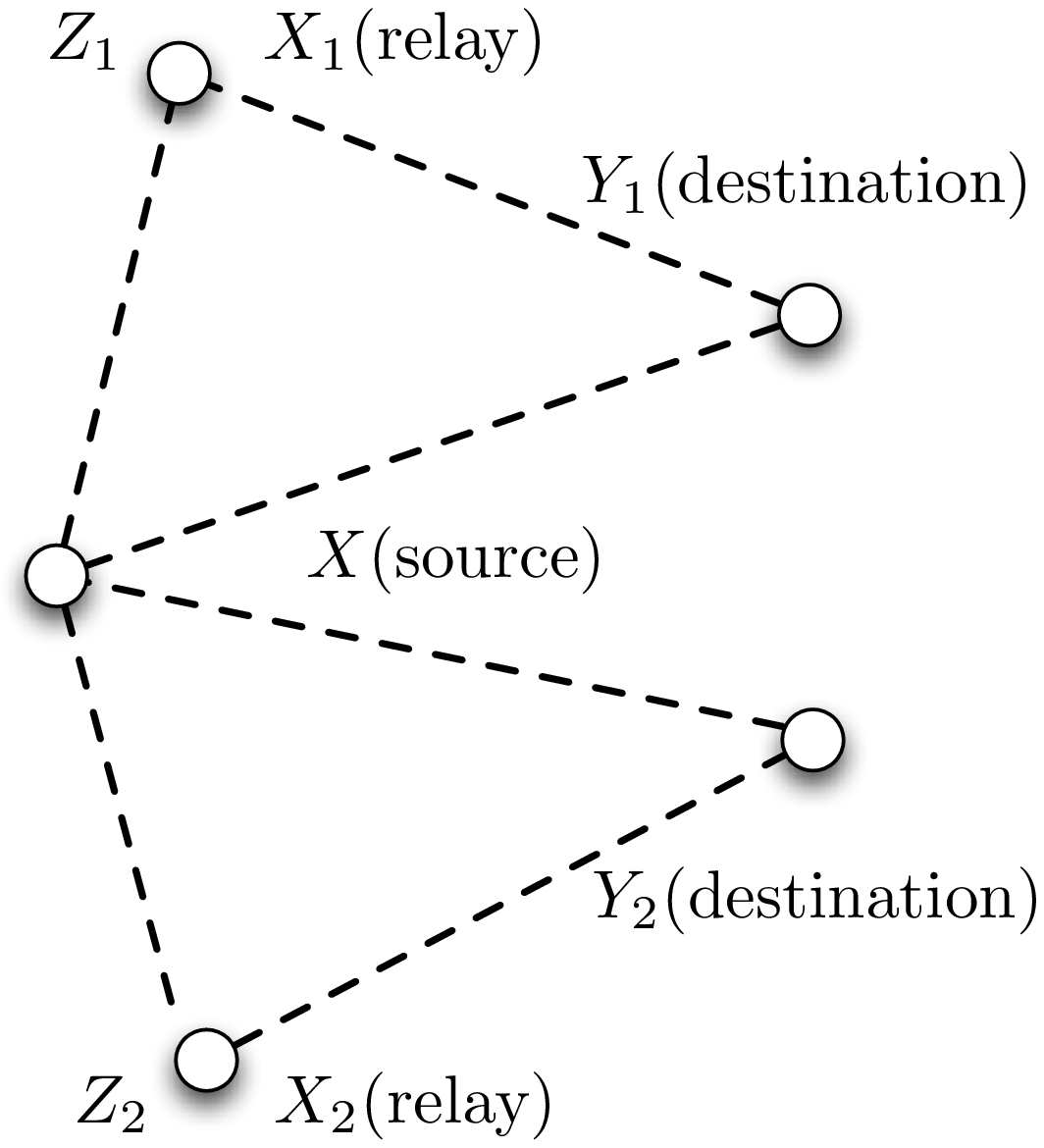}
\label{figCH1:I-2}
}
\subfigure[BRC with common relay]{
	\includegraphics [width=.22 \textwidth] {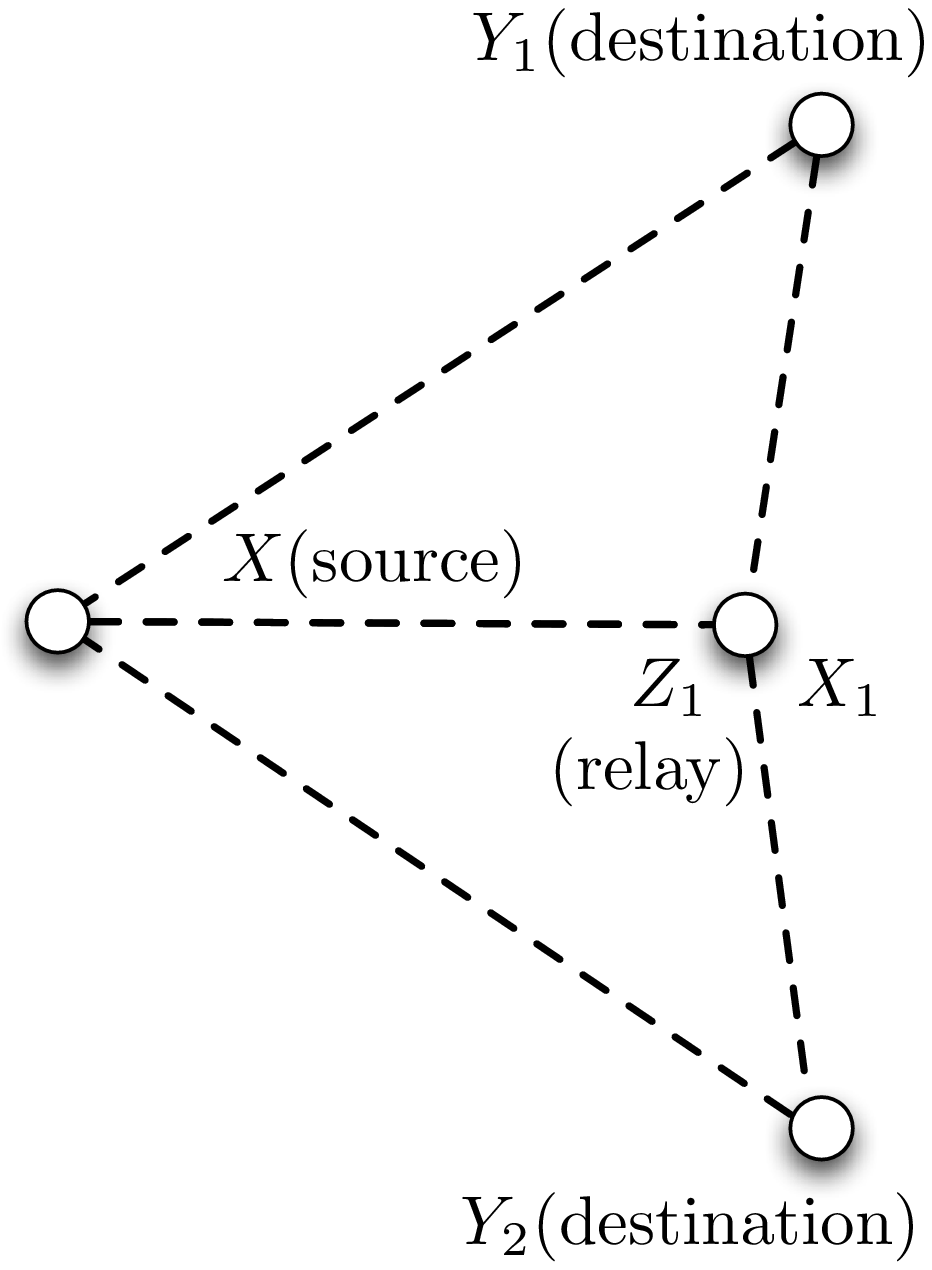}
\label{figCH1:I-3}
}
\caption{Simultaneous and broadcast relay channels.}
\end{figure*}

\IEEEPARstart{T}{he} simultaneous relay channel (SRC) is defined by a set of relay channels where the source wishes to communicate common and private information to each of the destinations in the set. In order to send common information regardless of the intended channel, the source must simultaneously consider the presence of all channels as described in Fig. \ref{figCH1:I-1}. This scenario offers a perspective of practical applications e.g., downlink communication on cellular networks where the base station --source-- may be aided by relays and opportunistic cooperation on ad-hoc networks where the source may not be aware of the presence of a nearby relay. 

Cooperative networks have been of huge interest during recent years between researchers as a possible candidate for future wireless networks \cite{Politis2004,Pabst2004}. Using the multiplicity of information in nodes, provided by the appropriate coding strategy, these networks can increase capacity and reliability, and diversity as addressed in \cite{Laneman2004,Sendonaris2003A,Sendonaris2003B} where multiple relays were introduced as an antenna array using distributed space-time coding. The simplest of cooperative networks is the relay channel. First introduced in \cite{Meulen1971}, it consists of a sender-receiver pair whose communication is aided by a relay node. In other words, it consists of a channel input $X$, a relay input $X_1$, a channel output $Y_1$ and a relay output $Z_1$, where the relay input depends only on the past observations. A significant contribution was made by Cover and El Gamal \cite{Cover1979}, where the main strategies of Decode-and-Forward (DF) and Compress-and-Forward (CF), and 
a max-flow min-cut upper bound were developed for this channel. Moreover the capacity of the degraded and the reversely degraded relay channel were established by the authors. A general theorem that combines DF and CF in a single coding scheme was also presented. The capacity of semi-deterministic relay channels and the capacity of cascaded relay channels were found in \cite{Aref1982,Aref1980}. A converse for the relay channel has been developed in \cite{Zhang1988}.  The capacity  of orthogonal relay channels was found in \cite{Elgamal2005} while the relay channel with private messages was discussed in \cite{Tannious2007}. The capacity of a class of modulo-Sum relay channels was also found in \cite{Aleksic2009}. More recently,  Compute-and-Forward strategy based on (linear) structured coding was proposed in \cite{Nazer2011}. It has been shown that the use of lattice codes outperforms DF strategy in some settings.

In general, the performance of DF and CF schemes are directly related to the noise condition between the relay and the destination. More precisely, it is well-known that DF scheme performs much better than CF when the source-to-relay channel is quite strong. Whereas CF scheme is more suitable when the relay-to-destination channel  is strong. Indeed,  inner bounds based on DF and CF strategies can be obtained using different coding and decoding techniques. Coding techniques can be classified \cite{Kramer2005} into \emph{regular and irregular coding}. Irregular coding exploits the codebooks of different sizes that are involved between relay and source while regular coding requires the same size. Decoding techniques also can roughly be classified into \emph{successive} and \emph{simultaneous decoding}. Successive decoding method decodes the transmitted codebooks in a consecutive manner. In each block, the decoder starts with a group of codebooks (e.g. relay codewords) and then afterward it moves to the next 
group (e.g. source codewords). 
However, simultaneous decoding decodes jointly all codebooks in a given block.  Generally speaking, the latter provides the better results than the former. Cover and El Gamal \cite{Cover1979} have proposed irregular coding with successive decoding. In fact, regular coding with simultaneous decoding  was first developed in \cite{King1978}. It can be exploited  for decoding with the channel outputs of a single or multiple blocks. For instance, the author in \cite{Carleial1982} by relying on this property introduces the notion of \emph{sliding window decoding} to perform decoding based on the outputs of two consecutive blocks. The notion of \emph{backward decoding} was  proposed in \cite{Willems1982} and it consists of a decoder who waits until the last block to start decoding from the last to the first message. Backward coding is shown to provide better performances than other schemes based on simultaneous decoding \cite{Zeng1989,Laneman2004B} such as sliding window. Backward decoding can use a single block as 
 in \cite{Willems1982} or multiple blocks as in \cite{Katz2009} to perform decoding. The best known  lower bound on the capacity of the relay channel was derived in \cite{Chong2007}, by using a generalized backward decoding strategy.

Extension to multiple relay networks have been studied in \cite{Xie2005} and practical scenarios were also considered, like the Gaussian relay channel\cite{Elgamal2006,Gastpar2005,Liang2005}, and the Gaussian parallel relay network \cite{Schein2000,Rezaei2009,Maric2004}. The combination of the relay channel with other networks  has been studied. The multiple access relay channel (MARC) was analyzed in \cite{Kramer2000,Sankar2009,Xie2007}. The relay-broadcast channel (RBC) where a user which can be either the receiver or a distinct node, serves as a relay for transmitting the information to the receivers, was also studied. An achievable rate region for the dedicated RBC was obtained in \cite{Kramer2005}. Preliminary works on the RBC were done in \cite{Liang2004,Dabora2005,Reznik2005} and the capacity region of physically degraded RBC was found in \cite{Dabora2006}. Inner and outer rate regions for the RBC were developed further in \cite{Liang2007A,Liang2007B,Bross2009}. The capacity of Gaussian dedicated RBC  with degraded relay channel was reported in \cite{Bhaskaran2008}.

Compound channels were introduced and further investigated in \cite{Dobrusin1959,Wolfowitz1960,Blackwell1960}. Extensive research has been undertaken for years (see  \cite{Lapidoth1998A} and references therein). This class of channels model communications over a set of possible channels where the encoder aims to maximize the worst-case capacity. Actually, the compound relay channel has a similar definition to the SRC. The SRC guarantees common and private rates for every channel in the set while the compound relay channel only guarantees a common rate. However, both terms are kept throughout this paper to indicate the difference in the code definition utilized with each model. An interesting relation between compound and broadcast channels was first mentioned in \cite{coverbroadcast-1972}, where it was suggested that the compound channel problem can be investigated via the broadcast channel. Indeed, this concept of broadcasting has been used as a method to mitigate the effect of channel uncertainty in 
numerous contributions \cite{Steiner2006,Katz2009,Katz2006,shamai1997,shamai2003}. Moreover, the SRC was also investigated through broadcast channels in \cite{Behboodi2009,Simeone2009B,Behboodi2010}. This strategy facilitates rate adaptation to the current channel in operation without requiring feedback information from the destination to the transmitter.

The broadcast channel (BC) was introduced in \cite{coverbroadcast-1972} along with the capacity of binary symmetric, product, push-to-talk and orthogonal BCs. The capacity of the degraded BC was established in \cite{Bergmans1973,Bergmans1974,Gallager1974,Ahlswede1975}. It was shown that feedback does not increase capacity of physically degraded BCs \cite{Gamal1981,Gamal1978}, but it does for Gaussian BCs \cite{Ozarow1984}. The capacity of the BC with degraded message sets was found in \cite{Korner1977} while that of more capable and less-noisy were established in \cite{ElGamal1979}. The best known inner bound for general BCs is due to Marton \cite{Marton1979} and an alternative proof was given in \cite{elgamal-vandermeulen-1981} (see \cite{Cover1998} and reference therein). This inner bound was shown to be tight for channels with one deterministic component \cite{Gelfand1980} and deterministic channels \cite{Pinsker1978,Han1981}. An outer-bound for the general BC was established in \cite{Marton1979} and  improved later in \cite{Nair2006,Nair2007}.

In this paper, we study different coding strategies and capacity region for the general memoryless broadcast relay channel (BRC) with two relays and destinations, as depicted in Fig. \ref{figCH1:I-2}. This model is equivalent to the SRC with two simultaneous memoryless relay channels. It should be emphasized that, by adding adequate Markov chains such that relays only affect a single destination, the BRC can be considered as being equivalent to the SRC. Nevertheless, for sake of generality we will not explicitly constrain the results trough this paper to the SRC. The rest of the paper is organized as follows. Section II introduces the main definitions and the problem statement. Inner bounds on the capacity region are derived for three cases of particular interest: 

\begin{itemize}
\item
Source-to-relay channels are stronger\footnote{The notion of \emph{stronger channel}  means that if channel A is stronger than channel B then the messages intended to decoder B  can fully be decoded  at decoder A. However, we shall not provide any formal definition to this since it is not needed for the proofs.} than the others and hence cooperation is based on DF strategy for both users (referred to as DF-DF region), corresponding to the SRC with DF relays. 

\item
Relay-to-destination channels are stronger than the others and hence cooperation is based on CF strategy for both users (referred to as as CF-CF region), corresponding to the SRC with CF relays.

\item
The source-to-relay channel of one destination is stronger than its corresponding  relay-to-destination channel. Whereas for the other destination the relay-to-destination channel is stronger than its source-to-relay channel. Hence cooperation is based on DF strategy for one destination and CF for the other one (referred to as DF-CF region). This case corresponds to the SRC where a different coding strategy is employed at each relay.
\end{itemize}

Section III examines general outer-bounds and capacity results for several classes of BRCs. In particular, the case of the broadcast relay channel with common relay (BRC-CR) is investigated, as shown in Fig. \ref{figCH1:I-3}. We show that the DF-DF region improves existent results  \cite{Kramer2005} on BRC-CR. Capacity results are obtained for the specific cases of semi-degraded and degraded Gaussian simultaneous relay channels. In Section IV, rates are computed for the case of distant based additive white Gaussian noise (AWGN) relay channels. Achievability and converse proofs are relegated to the appendices while summary and discussion are presented in Section V.

%------------------------------------------------------------------------------
\subsection*{Notation}

For any sequence~$(x_i)_{i\in \mathbb{N}_+}$, notation $\underline{x}$ stands for the collection $x_1^n=(x_1,x_{2},\dots, x_n)$.
Entropy is denoted by $H(\cdot)$, and mutual information by $I(\cdot;\cdot)$. The differential entropy  function is denoted by $h(\cdot) $. We denote $\epsilon$-typical  and conditional $\epsilon$-typical sets by $\textsl{A}^n_\epsilon(X)$ and $\textsl{A}^n_\epsilon(Y|X)$, respectively  (see \cite{cover-book} for details). Let $X$, $Y$ and $Z$ be three random variables on some alphabets with probability distribution~$p$. If $p(x|yz)=p(x|y)$ for each $x,y,z$, then they form a Markov chain, denoted by $X\mkv Y \mkv  Z$. 
Logarithms are taken in base $2$ and denoted by $\log(\cdot)$. The capacity function is defined as $\mathcal{C}(x)=\frac{1}{2}\log (1+x)$.

%%%%%%%%%%%%%%%%%%%%%%%%%%%%%%%%%%%%%%%%%%%%%%%%%%%%%%%%%
%%%%%%%%%%%%%%%%%%%%%%%%%%%%%%%%%%%%%%%%%%%%%%%%%%%%%%%%%%%%%%%%%%%%%%%
\section{Main Definitions and Achievable Regions}\label{achievable-regions}
In this section, we first formalize the problem of the simultaneous relay channel and then  present achievable rate regions for the cases of DF-DF strategy (DF-DF region), CF-CF strategy (CF-CF region) and DF-CF strategy (DF-CF region). 

%%*********************************************************************************************************
%%II.1 Main Defintions
%%*********************************************************************************************************
\begin{figure*}[t]
\centering
	\includegraphics [width=.7 \textwidth] {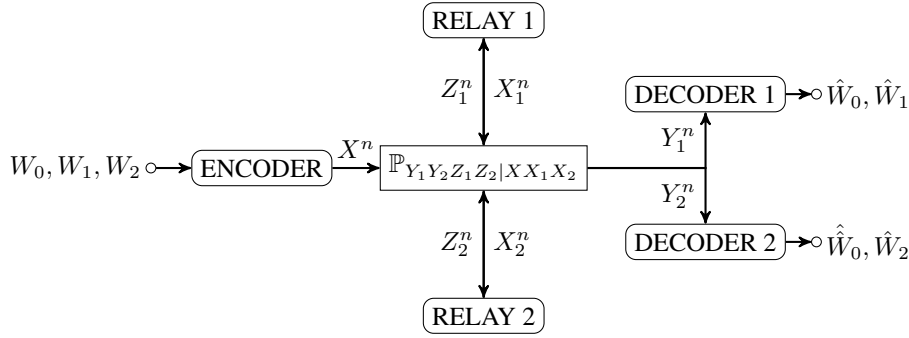}
\caption{Broadcast relay channel (BRC).}  %  \vspace{-4mm}
\label{figCH1:I-4}
\end{figure*}

\subsection{Problem statement}

The simultaneous relay channel \cite{Behboodi2009} with discrete source and relay inputs $x\in \mathcal{X}$, $x_T\in \mathcal{X}_T$, discrete channel and relay outputs $y_T\in \mathcal{Y}_T$, $z_T\in \mathcal{Z}_T$, is characterized by a set of relay channels, each of them defined by a conditional probability distribution (PD)  
$$
\mathcal{P}_{\textrm{SRC}}=\left\{ P_{Y_TZ_T |XX_T}:\mathcal{X} \ex \mathcal{X}_T  \longmapsto \mathcal{Y}_T  \ex \mathcal{Z}_T\right\},
$$ 
where $T$ denotes the channel index. The SRC models the situation in which only one single channel is present  at once, but it does not change during the communication. However, the transmitter is not cognizant of the realization of $T$ governing the communication. In this setting, $T$ is assumed to be known at the destination and the relay ends. The transition PD of the $n$-memoryless extension  with inputs $(\mb{x},\mb{x}_T)$ and  outputs  $(\mb{y}_T,\mb{z}_T)$ is given by  
$$
P^n_{Y_TZ_T |XX_T}(\mb{y}_T,\mb{z}_T|\mb{x},\mb{x}_T)=\prod\limits_{i=1}^n  P_T (y_{T,i},z_{T,i}| x_i, x_{T,i}).
$$
The focus is on the case where ${T=\{1,2\}}$, in other words there are two relay channels in the set.\\

\begin{definition}[code] \label{def-codeCH1}
A code for the SRC consists of: 
\begin{itemize}
\item An encoder mapping $\{ \varphi:\mc{W}_{0} \ex\mc{W}_{1} \ex\mc{W}_{2}  \longmapsto \mathcal{X}^n  \}$,  
\item Two decoder mappings $\{ \psi_T:\mathcal{Y}_T^n \longmapsto \mc{W}_{0} \ex\mc{W}_{T} \}$, 
\item A set of relay functions $\{f_{T,i}\}_{i=1}^n$ such that $$\{ f_{T,i} :\mathcal{Z}_T^{i-1}  \longmapsto \mathcal{X}_T^n  \}_{i=1}^n,$$ 
\end{itemize}
for $T=\{1,2\}$ and some finite sets of integers  $\mc{W}_{b}=\big\{ 1,\dots, M_{b} \big\}_{b=\{0,1,2\}}$. The rates of such code are $n^{-1} \log M_{b}$ and the corresponding maximum error probabilities for $T=\{1,2\}$ are defined as 
\begin{align*}
P_{e,T}^{(n)}\big( \varphi,\psi,& \{f_{T,i}\}_{i=1}^n\big) = \\
&\max_{(w_0,w_T)\in\mc{W}_{0}\times\mc{W}_{T}} \Pr \left\{  \psi (\mb{Y} _T ) \neq (w_0,w_T)  \right\}.
%\label{errorprob_defCH1}   
\end{align*}     %\vspace{1mm}
\end{definition}

\begin{definition}[achievability and capacity] 
For any positive numbers $0 < \epsilon, \gamma< 1$, a triple of non-negative numbers $(R_0,R_1,R_2)$ is said achievable for the SRC if for every sufficiently large $n$, there exists a $n$-length block code whose error probability satisfies 
$$
P_{e,T}^{(n)}\big( \varphi,\psi, \{f_{T,i}\}_{i=1}^n\big) \leq \epsilon
$$  
for $T=\{1,2\}$ and the rates $\frac{1}{n} \log M_{b} \geq R_b-\gamma$ for $b=\{0,1,2\}$. The set of all achievable rates $\mathcal{C}_{\textrm{SRC}}$ is called the capacity region of the SRC. We emphasize that no prior distribution on $T$ is assumed and thus the encoder must exhibit a code that yields small error probability for every $T=\{1,2\}$. A similar definition can be offered for the common-message SRC with a single message set $\mc{W}_{0}$, $n^{-1} \log M_{0}$ and rate $R_0$. The common-message SRC is equivalent to the compound relay channel and so its achievable rate is similarly defined. 
\end{definition}

\begin{figure*}[t]
\centering
\subfigure[Diagram of auxiliary random variables]{
	\includegraphics  [width=.35 \textwidth] {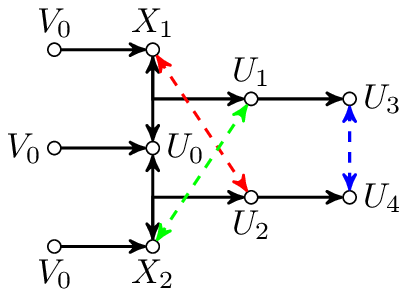}
\label{figCH1:II-3}
}
\subfigure[Message reconfiguration]{
	\includegraphics  [width=.25 \textwidth] {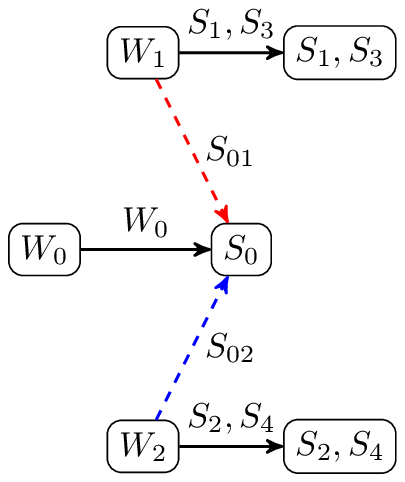}
\label{figCH1:II-3-1}
}
\caption{Description of encoding techniques for DF strategy.}
\vspace{-4mm}
\end{figure*}
 \begin{figure*} [!b]
\hrule
\begin{align}
%\begin{array} {l}
I_i= &\min\big\{I(U_0,U_i;Z_i\vert V_0,X_i)+ I(U_{i+2};Y_i\vert U_0,V_0,X_i,U_i),I(U_0,V_0,U_i,U_{i+2},X_i;Y_i)\big\},\nonumber\\
J_i= &\min\big\{I(U_i;Z_i\vert U_0,V_0,X_i) +I(U_{i+2};Y_i\vert U_0,V_0,X_i,U_i),I(U_{i+2},U_i,X_i;Y_i\vert U_0,V_0)\big\},\nonumber \\
I_M= &I(U_3;U_4\vert U_1,U_2,X_1,X_2,U_0,V_0),
\label{thm:II2-1-defs}
\end{align}
\end{figure*}
\begin{figure*} [!b]
\hrule
\begin{equation}
\begin{array} {c}
\\
\mathcal{Q}=\big \{  P_{U_0V_0U_1U_2U_3U_4X_1X_2X}= P_{U_3U_4X|U_1U_2}P_{U_1U_2|U_0X_1X_2}\,P_{U_0|X_1X_2V_0}P_{X_2|V_0}P_{X_1|V_0} P_{V_0}  \vspace{1mm} \\ 
\textrm{ satisfying }\\  \vspace{1mm}
(U_0,V_0,U_1,U_2,U_3,U_4)\mkv (X_1,X_2,X) \mkv (Y_1,Z_1,Y_2,Z_2) \big \}.  % \vspace{1mm}
\end{array}     
\label{thm:II2-1-probset}
\end{equation}
\end{figure*}\vspace{2mm}

\begin{remark}
We emphasize that both relay and destination are assumed to be cognizant of the realization of $T$ and hence the problem of coding for the SRC can be turned into that of the broadcast relay channel  (BRC) \cite{Behboodi2009}. Because the source is uncertain about the actual channel, it has to count for each of them and therefore assume the simultaneous presence of both. This leads to an equivalent broadcast model consisting of two sub-channels (or branches) for $T=\{1,2\}$, where each one corresponds to a single-relay channel, as illustrated in Fig.  \ref{figCH1:I-2} and Fig.  \ref{figCH1:I-4}. The encoder sends common and private messages  $(W_0,W_T)$ to destination $T$ at rates $(R_0,R_T)$. The general BRC is defined by the PD 
\begin{align*}
\mathcal{P}_{\textrm{BRC}}=\Big\{&\mathbb{P}_{Y_1Z_1Y_2Z_2 |XX_1X_2}:\\	
&\mathcal{X} \ex \mathcal{X}_1  \ex \mathcal{X}_2  \longmapsto \mathcal{Y}_1  \ex \mathcal{Z}_1 \ex \mathcal{Y}_2  \ex \mathcal{Z}_2 \Big\},
\end{align*}
with channel and relay inputs $(X,X_1,X_2)$ and channel and relay outputs  $(Y_1,Z_1,Y_2,Z_2)$. Notions of achievability for rates $(R_0,R_1,R_2)$ and capacity remain the same as for conventional BCs (see \cite{coverbroadcast-1972}, \cite{Kramer2005} and \cite{Liang2007A}). Similar to the case of conventional BCs, the capacity region of the BRC depends only on the marginal PDs: $P_{Y_1|XX_1X_2Z_1Z_2}$, $P_{Y_2|XX_1X_2Z_1Z_2}$ and $P_{Z_1Z_2 |XX_1X_2}$. 
\end{remark}\vspace{2mm}

\begin{remark}
The definition of the BRC does not dismiss the possibility of dependence of destination $Y_1$ (respect to destination $Y_2$) on the relay input $X_2$ (respect to the relay input $X_1$). Therefore, it appears to be more general than the SRC. In other words, the current definition of BRC corresponds to that of the SRC with the additional constraints that $(Y_1,Z_1) \mkv  (X,X_1) \mkv (Y_2,Z_2,X_2)$ and $(Y_2,Z_2) \mkv  (X,X_2) \mkv (Y_1,Z_1,X_1)$. These Markov chains guarantee that $(Y_T,Z_T)$ only depend on inputs $(X,X_T)$, for $T=\{1,2\}$. Despite the fact that this condition is not necessary until converse proofs, the achievable region developed below are more adapted to the SRC. Nevertheless, these achievable rate regions do not require any additional assumption and thus are valid for the general BRC as well.
\end{remark}

The next subsections provide achievable rate regions for three different coding strategies. 
%%*********************************************************************************************************
%%II.2 DF DF strategies
%%*********************************************************************************************************
\subsection{Achievable region based on DF-DF strategy}
Consider the situation where the source-to-relay channels are stronger than the others. In this case, the best known coding strategy for both relays turns out to be Decode-and-Forward (DF). The source should broadcast the information to the destinations based on a broadcast code combined with DF scheme. Both relays help the common information  using a common description, namely $V_0$. The private information for each destination is sent partly by the help of the corresponding relay and partly by direct transmission. The next theorem presents the achievable rate region \cite{Behboodi2010}. \vspace{1mm}

\begin{theorem}(DF-DF region) An inner bound on the capacity region $\mathcal{R}_{\textrm{DF-DF}}\subseteq \mathcal{C}_{\textrm{BRC}}$ of the broadcast relay channel is given by      
\begin{align*}
\mathcal{R}_{\textrm{DF-DF}} = \displaystyle{co\bigcup\limits_{P\in\mathcal{Q}}} & \Big\{(R_0\geq 0,R_1\geq 0,R_2 \geq 0): \\ 
R_0+R_1 \leq &I_1-I(U_0,U_1;X_2\vert X_1,V_0), \\
R_0+R_2 \leq &I_2-I(U_0,U_2;X_1\vert X_2,V_0), \\
R_0+R_1+R_2 \leq &I_1+J_2-I(U_0,U_1;X_2\vert X_1,V_0)\\
&-I(U_1,X_1;U_2\vert X_2,U_0,V_0)-I_M \\
R_0+R_1+R_2 \leq &J_1+I_2-I(U_0,U_2;X_1\vert X_2,V_0)\\
&-I(U_1;U_2,X_2\vert X_1,U_0,V_0)-I_M \\
2R_0+R_1+R_2 \leq &I_1+I_2-I(U_0,U_1;X_2\vert X_1,V_0)\\
&-I(U_0,U_2;X_1\vert X_2,V_0)\\
&-I(U_1;U_2\vert X_1,X_2,U_0,V_0)-I_M    \Big\},
\label{eqCH1:II2-1}     
\end{align*}
where $(I_i,J_i,I_M)$ with $i=\{1,2\}$ are as in \eqref{thm:II2-1-defs}. $co\{\cdot\}$ denotes the convex hull and the union is over all joint PDs $P_{U_0V_0U_1U_2U_3U_4X_1X_2X}\in \mathcal{Q}$, with $\mathcal{Q}$ given by \eqref{thm:II2-1-probset}.
\label{thm:II2-1}   
\end{theorem}
\begin{proof}
The proof of this theorem is relegated to Appendix \ref{proof-1CH1}. Instead, here we provide an overview of it. First, the original messages are reorganized via rate-splitting  into new messages, as shown in Fig. \ref{figCH1:II-3-1}, where we add part of the private messages together with the common message into new messages, which is similarly to \cite{Kramer2005}. The general coding idea of the proof is depicted in Fig. \ref{figCH1:II-3}. 

The description $V_0$ represents the common part of $(X_1,X_2)$ (the information sent by the relays), which is intended to help the common information encoded in $U_0$. Private information is sent in two steps, first using the relay help through $(U_1,U_2)$ and based on DF strategy. Then, the direct links  between source and destinations are used to decode $(U_3,U_4)$. Marton coding is used to allow correlation between the descriptions according to the arrows in Fig. \ref{figCH1:II-3}. To make a random variable simultaneously correlated with multiple random variables (RVs), we used multi-level Marton coding. 

\begin{table*}[t] 
	\caption{DF strategy with $b=\{1,2\}$}
	\label{tab:II2-1}
	 \centering
\begin{tabular}{| l | l | } 
\hline
  $\underline{v}_0(t_{0(i-1)})$ & $\underline{v}_0(t_{0(i)})$  \\
  \hline
  $\underline{u}_0(t_{0(i-1)},t_{0i})$ & $\underline{u}_0(t_{0i},t_{0(i+1)})$  \\
  \hline
  $\underline{x}_b(t_{0(i-1)},t_{b(i-1)})$ & $\underline{x}_b(t_{0i},t_{bi})$  \\
  \hline
  $\underline{u}_b(t_{0(i-1)},t_{0i},t_{b(i-1)},t_{bi})$ & $\underline{u}_b(t_{0i},t_{0(i+1)},t_{bi},t_{b(i+1)})$  \\
   \hline
  $\underline{u}_{b+2}(t_{0(i-1)},t_{0i},t_{b(i-1)},t_{bi},t_{(b+2)i})$ & $\underline{u}_{b+2}(t_{0i},t_{0(i+1)},t_{bi},t_{b(i+1)},t_{(b+2)(i+1)})$  \\
  \hline
  \hline
  $\underline{y}_{bi}$ & $\underline{y}_{b(i+1)}$\\
  \hline
\end{tabular} 
\end{table*}
Full details for this process are explained in Appendix  \ref{proof-1CH1} while Table \ref{tab:II2-1} shows details for the transmission  in time. Both relays knowing $(\underline{v}_0,\underline{x}_b)$ decode $(\underline{u}_0,\underline{u}_b)$ in the same block. Then each destination, by using backward decoding, decodes all codebooks in the last block. The final region is a combination of all constraints from Marton coding and decoding, which reduce to the above region by using Fourier-Motzkin elimination. 
\end{proof} \vspace{1mm}

\begin{remark}
We have the following observations:
\begin{itemize}
\item The rates in Theorem \ref{thm:II2-1} coincide with the conventional rate based on partial DF \cite{Cover1979}, and moreover it is easy to verify that, by setting $(X_1,X_2,V_0)=\emptyset$, $U_3=U_1,U_4=U_2$ $Z_1=Y_1$ and $Z_2=Y_2$,  the rate region in Theorem \ref{thm:II2-1} is equivalent to Marton's region \cite{Marton1979},

\item The new region improves on the existent regions  for the general BRC in \cite{Behboodi2009} and for the BRC with common relay as depicted in Fig. \ref{figCH1:I-3}. By setting $X_1=X_2=V_0$ and $U_1=U_2=U_0$, the rate region in Theorem \ref{thm:II2-1} can be shown to be equivalent to the inner bound  in \cite{Kramer2005}. Whereas the next corollary shows that the novel rate region  is strictly large than that  in \cite{Kramer2005}.
\end{itemize}
\end{remark}
The following corollary provides a sharper inner bound on the capacity region of the BRC with common relay (BRC-CR). By dividing the help of relay into two components $V_0$ and $X_1$, the relay is also able to help private information of the first destination. This is in contrast to the encoding technique used in \cite{Kramer2005}, where the relay only  helps common information. As a consequence of this,  when $Y_2=\emptyset$ and the first destination is a physically degraded version of the relay the region in \cite{Kramer2005} cannot achieve the capacity of this channel. This is not the case of the next rate region. Furthermore, it will be shown later that a special case of this corollary reaches the capacity of the degraded Gaussian BRC-CR and semi-degraded BRC-CR. \vspace{1mm}

\begin{corollary}[BRC with common relay]
An inner bound on the capacity region of the BRC-CR $\mathcal{R}_{\textrm{BRC-CR}}\subseteq$ $ \mathcal{C}_{\textrm{BRC-CR}}$ is given by 
\begin{align*}
\mathcal{R}_{\textrm{BRC-CR}} = co &\displaystyle{\bigcup\limits_{P_{V_0U_0U_1U_3U_4X_1X}\in\mathcal{Q}}}  \Big\{\\
(R_0\geq 0, R_1\geq 0,&R_2 \geq 0):  \\ 
R_0+R_1 &\leq \min\{I_1+I_{1p},I_3+I_{3p}\}\\
&+I(U_3;Y_1\vert U_1,U_0,X_1,V_0), \\
R_0+R_2 &\leq I(U_0,V_0,U_4;Y_2)-I(U_0;X_1\vert V_0),  \\
R_0+R_1+R_2 &\leq \min\{I_2,I_3\}+I_{3p}\\
&+I(U_3;Y_1\vert U_1,U_0,X_1,V_0)\\
+I(U_4;&Y_2\vert U_0,V_0)-I(U_0;X_1\vert V_0)-I_M, \\
R_0+R_1+R_2 &\leq \min\{I_1,I_3\}+I_{1p}\\
&+I(U_3;Y_1\vert U_1,U_0,X_1,V_0)\\
+I(U_4;&Y_2\vert U_0,V_0)-I(U_0;X_1\vert V_0)-I_M, \\
2R_0+R_1+R_2  &\leq I(U_3;Y_1\vert U_1,U_0,X_1,V_0)\\
&+I(U_4;Y_2\vert U_0,V_0)+I_2\\
&+\min\{I_1+I_{1p},I_3+I_{3p}\}\\
&-I(U_0;X_1\vert V_0)-I_M \Big\}
\end{align*}
where the quantities are defined by
\begin{align*}
I_1&=I(U_0,V_0;Y_1), \\
I_2&=I(U_0,V_0;Y_2),\\
 I_3&=I(U_0;Z_1|X_1,V_0), \\
I_{1p}&=I(U_1X_1;Y_1|U_0,V_0), \\
I_{3p}&=I(U_1;Z_1|U_0,V_0,X_1), \\
I_M &=I(U_3;U_4\vert X_1,U_1,U_0,V_0),
\end{align*}
$co\{\cdot\}$ denotes the convex hull and  $\mathcal{Q}$ is the set of all joint PDs $P_{V_0U_0U_1U_3U_4X_1X}$ satisfying
\begin{equation*}
\begin{array} {c}
(V_0,U_0,U_1,U_3,U_4) \mkv (X_1,X) \mkv (Y_1,Z_1,Y_2).
\end{array} 
\end{equation*}
\label{thm:II2-2}
\end{corollary}

%%*********************************************************************************************************
%%II.3 CF-DF
%%*********************************************************************************************************
%\vspace{-8mm}
\subsection{Achievable region based on CF-DF strategy}

Consider now a broadcast relay channel where the source-to-relay channel  is stronger that the relay-to-destination channel for the first user  and weaker for the second one. Hence cooperation is better be based on DF scheme for user one and CF scheme for user two. Actually, the source must broadcast the information to the destinations based on a broadcast code combined with CF and DF schemes. This scenario may arise when the encoder does not know (e.g. due to user mobility and fading) whether the source-to-relay channel is much stronger or not than the  relay-to-destination channel. The next theorem presents the general achievable rate region for the case where the first relay employs DF scheme while the second relay uses CF scheme to help common and private information \cite{Behboodi2010A}.  \vspace{1mm}

\begin{theorem}[CF-DF region] An inner bound on the capacity region of the BRC $\mathcal{R}_{\textrm{DF-CF}}\subseteq \mathcal{C}_{\textrm{BRC}} $  with heterogeneous cooperative strategies is given by
\begin{align*}
\mathcal{R}_{\textrm{CF-DF}}&= co \displaystyle{\bigcup\limits_{P\in\mathcal{Q}}}  \Big\{(R_0\geq 0, R_1\geq 0,R_2 \geq 0):  \\ 
R_0+R_1 & \leq I_1,  \\
R_0+R_2 & \leq I_2-I(U_2;X_1\vert U_0,V_0),  \\
R_0+R_1+R_2 &\leq I_1+J_2-I(U_1,X_1;U_2\vert U_0,V_0), \\
R_0+R_1+R_2 &\leq J_1+I_2-I(U_1,X_1;U_2\vert U_0,V_0), \\   
2R_0+R_1+R_2 & \leq I_1+I_2-I(U_1,X_1;U_2\vert U_0,V_0) \Big\},\label{eqCH1:II3-1}
\end{align*}
where the quantities $(I_i,J_i,\Delta_0)$ with $i=\{1,2\}$ are given by 
\begin{align*}
I_1&= \min\big\{I(U_0,U_1;Z_1\vert X_1,V_0),I(U_1,U_0,X_1,V_0;Y_1)\big\},\\
I_2&= I(U_2,U_0,V_0;\hat{Z}_2,Y_2\vert X_2),   \\
J_1&= \min\big\{I(U_1;Z_1\vert X_1,U_0,V_0),I(U_1,X_1;Y_1\vert U_0,V_0)\big\},\\
J_2&= I(U_2;\hat{Z}_2,Y_2\vert X_2,U_0,V_0),   \\  
\end{align*}
$co\{\cdot\}$ denotes the convex hull and  the set of all admissible PDs $\mathcal{Q}$ is defined as 

\begin{align*}
%\vspace{4mm}
\mathcal{Q}= \Big\{&P_{V_0U_0U_1U_2X_1X_2XY_1Y_2Z_1Z_2\hat{Z}_2}= P_{V_0}P_{X_2}P_{X_1|V_0} \\
&P_{U_0|V_0} P_{U_2U_1|X_1U_0} P_{X|U_2U_1}P_{Y_1Y_2Z_1Z_2|XX_1X_2}\times\\
&P_{\hat{Z}_2|X_2Z_2},\,\,\, \\ 
&\textrm{satisfying }\,\,\,I(X_2;Y_2) \geq I(Z_2;\hat{Z}_2\vert X_2Y_2)\,\,\, \textrm{and }\\
(V_0,& U_0,U_1,U_2) \mkv (X_1,X_2,X) \mkv (Y_1,Z_1,Y_2,Z_2) \Big \}. \nonumber
\end{align*}  %\vspace{-2mm}
\label{thm:II3-1}
\end{theorem}

\begin{remark}
It should emphasized that it is possible to exchange the coding strategy between first and second relay and thus a bigger region is obtained by taking the convex hull of the union of both regions. 
\end{remark}

The proof of this theorem is relegated to Appendix \ref{proof-2}. Instead, here we discuss the relevant steps of it. In order to send common information while exploiting the help of DF relay at destination 1, we use regular encoding with block-Markov coding. The description $V_0$ is the part of $X_1$ to help the transmission of $U_0$, and the second relay helps destination 2 based on CF scheme (i.e. relay and source inputs are independently chosen). Regular encoding is used to superimpose the code of the current block over that of the previous block. The relay using DF scheme transmits the message from the previous block and hence the destination can exploit it for decoding as usually. But the relay using CF scheme seems to impose the decoding of two superimposed codes at the destination. By noting that the  codeword center carries the dummy message in the first block, the destination decodes the cloud  knowing the center, and then in the next block it continues by removing the center code. 

Nevertheless, this procedure leads to performance loss because one part of the transmitted code is indeed thrown away. Therefore, at this point the reader may think that superposition coding needed for DF should not work with CF scheme. Helpfully, this is not the case. By using backward decoding, the code can be exploited with CF scheme  as well and without loss of performance. The destination decoding CF scheme takes $V_0$ not as  the relay code but as part of the source code, over which $U_0$ is superimposed. Then, the last block $U_0$ carries the dummy message superimposed on $V_0$, which is the message from the last block. For instance, $(U_0,V_0)$ can be jointly decoded by exploiting both codes and without performance loss with respect to usual CF scheme.

Finally, we consider the compound relay channel, where the channel in operation is chosen from the set of relay channels. For simplicity, suppose that the set includes only two channels such that DF  compared to CF strategy yields a better rate for the first channel and a worse rate for the second one. The overall goal is to transmit at the best possible rate with arbitrary small error probability for both channels. Then using regular encoding, it can be seen that the best cooperative strategy can be selected for each channel because the first relay employs DF while the second one uses CF scheme. The next corollary  directly results from this observation.  \vspace{1mm}

\begin{corollary}[common-information]
A lower bound on the capacity of the compound  relay channel (or common-message BRC) is given by   all rates $R_0$ satysfing
\begin{equation*}
\begin{array} {rl}
R_0 \leq \max\limits_{P_{X_1X_2X}\in\mathcal{Q}}&\min\big\{I(X;Z_1\vert X_1),\\
&I(X,X_1;Y_1),I(X;\hat{Z}_2,Y_2\vert X_2)\big\}.
\end{array}    %\vspace{-2mm}
\label{eqCH1:II3-3}
\end{equation*}
\label{thm:II3-2}
\end{corollary}
\begin{corollary}[private information]   An inner bound on the capacity region of the BRC with heterogeneous cooperative strategies is given by the convex hull of the set of rates $(R_1,R_2)$ satisfying 
\begin{align*}
R_1 & \leq \min\big\{I(U_1;Z_1\vert X_1),I(U_1,X_1;Y_1)\big\}, \\
R_2 & \leq I(U_2;\hat{Z}_2,Y_2\vert X_2)-I(U_2;X_1), \\
R_1+R_2 & \leq \min\big\{I(U_1;Z_1\vert X_1),I(U_1,X_1;Y_1)\big\} \\
&+ I(U_2;\hat{Z}_2,Y_2\vert X_2) -I(U_1,X_1;U_2),
%\label{eqCH1:II3-4}
\end{align*}
for all joint PDs $P_{U_1U_2 X_1X_2XY_1Y_2Z_1Z_2\hat{Z}_2} \in \mathcal{Q}$. 
\label{thm:II3-3}
\end{corollary}\vspace{2mm}
Corollary \ref{thm:II3-2} follows from Theorem \ref{thm:II3-1} by choosing $U_1=U_2=U_0=X$, $V_0=X_1$. Whereas Corollary \ref{thm:II3-3} follows by setting $U_0=V_0=\emptyset$.

\begin{remark}
The region in Theorem \ref{thm:II3-1} is equivalent to Marton's region \cite{Marton1979} with $(X_1,X_2,V_0)=\emptyset$, $Z_1=Y_1$ and $Z_2=Y_2$. Observe that the rate corresponding to DF scheme which appears in Theorem \ref{thm:II3-1} coincides with the usual DF rate, whereas the CF rate appears with a little difference. In fact, $X$ is being decomposed into $(U,X_1)$, replacing it in the rate term corresponding to CF scheme. 
\end{remark}
%%*********************************************************************************************************
%%II.4 CF-CF
%%*********************************************************************************************************
\subsection{Achievable region based on CF-CF strategy}

Consider now another scenario where both  relay-to-destination channels are stronger than the others and hence the efficient coding strategy turns to be CF scheme for both users. The inner bound based on this strategy is stated in the following theorem \cite{Behboodi2010C} and its proof is presented in Appendix \ref{proof-3}. 

\begin{theorem}[CF-CF region] An inner bound on the capacity region of the BRC $\mathcal{R}_{\textrm{CF-CF}}\subseteq \mathcal{C}_{\textrm{BRC}}$ is given by
\begin{align*}
\mathcal{R}_{\textrm{CF-CF}} = co \displaystyle{\bigcup\limits_{P\in\mathcal{Q}}}  \Big\{(R_0&\geq 0, R_1\geq 0,R_2 \geq 0):  \\ 
R_0+R_1 & \leq I(U_0,U_1;Y_1,\hat{Z}_1\vert X_1),  \\
R_0+R_2 & \leq I(U_0,U_2;Y_2,\hat{Z}_2\vert X_2),  \\
R_0+R_1+R_2 &\leq I_0+I(U_1;Y_1,\hat{Z}_1\vert X_1,U_0)\\
+I(U_2;Y_2&,\hat{Z}_2\vert X_2,U_0)-I(U_1;U_2\vert U_0), 
\end{align*}

\begin{align*}
2R_0+R_1+R_2 & \leq I(U_0,U_1;Y_1,\hat{Z}_1\vert X_1)\\
+I(U_0,U_2&;Y_2,\hat{Z}_2\vert X_2)-I(U_1;U_2\vert U_0)\Big\},
\label{eqCH1:II4-1}
\end{align*}
where the quantity $I_0$ is defined by 
\begin{equation*}
\begin{array} {l}
I_0=\min\big\{I(U_0;Y_1,\hat{Z}_1\vert X_1),I(U_0;Y_2,\hat{Z}_2\vert X_2)\big\},\\
\end{array}   
\end{equation*}
$co\{\cdot\}$ denotes the convex hull and  the set of all admissible PDs $\mathcal{Q}$ is defined as 
\begin{align*}
\mathcal{Q}=\big \{&P_{U_0U_1U_2X_1X_2XY_1Y_2Z_1Z_2\hat{Z}_1\hat{Z}_2}= P_{X_2}P_{X_1}P_{U_0} \times\\
& P_{U_2U_1|U_0}P_{X|U_2U_1}P_{Y_1Y_2Z_1Z_2|XX_1X_2}\times\\
&P_{\hat{Z}_1|X_1Z_1}P_{\hat{Z}_2|X_2Z_2},  \\
&  I(X_1;Y_1) \geq I(Z_1;\hat{Z}_1\vert X_1,Y_1),\\
& I(X_2;Y_2) \geq I(Z_2;\hat{Z}_2\vert X_2,Y_2),\\
& (U_0,U_1,U_2) \mkv (X_1,X_2,X) \mkv (Y_1,Z_1,Y_2,Z_2) \big \}. 
%\label{eqCH1:II4-2}
\end{align*} 
\label{thm:II4-1}
\end{theorem}
Notice that by setting $(X_1,X_2)=\emptyset$, $Z_1=Y_1$ and $Z_2=Y_2$ this region is equivalent to Marton's region \cite{Marton1979}.
\begin{remark}
A general achievable rate region follows by applying time-sharing on the regions stated in Theorems \ref{thm:II2-1}, \ref{thm:II3-1} and \ref{thm:II4-1}. 
\end{remark}
%%*********************************************************************************************************
%%III Upper Bounds and Capacity Results
%%*********************************************************************************************************
\section{Outer Bounds and Capacity Results}
In this section, we first provide an outer-bound on the capacity region of the general BRC . Then some capacity results for the cases of semi-degraded BRC with common relay (BRC-CR)  and degraded Gaussian BRC-CR are stated. 
%%%%%%%%%%%%%%%%%%%%%%%%%%
\subsection{Outer bounds on the capacity region of general BRC}
%%%%%%%%%%%%%%%%%%%%%%%%%%
The next theorems provide general outer-bounds on the capacity regions of the BRC described in Fig. \ref{figCH1:I-4} and the BRC-CR where $X_1=X_2$ and $Z_1=Z_2$, respectively. The proof is presented in Appendix \ref{proof-8}.\vspace{2mm}

\begin{theorem}[outer-bound BRC]
The capacity region $\mathcal{C}_{\textrm{BRC}}$ of the BRC   is included in the set $\mathcal{C}_{\textrm{BRC}}^{\textrm{out}}$ of all rates  $(R_0,R_1,R_2)$ satisfying 
\begin{align*} 
&\mathcal{C}_{\textrm{BRC}}^{\textrm{out}} = co \displaystyle{\bigcup\limits_{P_{VV_1 U_1 U_2 X_1 X_2 X}\in\mathcal{Q}}} \Big\{(R_0\geq  0 ,R_1\geq 0,R_2\geq 0): \\
&R_0 \leq \min\big\{I(V;Y_2),I(V;Y_1)\big\},\\
&R_0+R_1 \leq \min\big\{I(V;Y_1),I(V;Y_2)\big\} + I(U_1;Y_1\vert V),\\
&R_0+R_2 \leq \min\big\{I(V;Y_1),I(V;Y_2)\big\} + I(U_2;Y_2\vert V),\\
&R_0+R_1 \leq \min\big\{I(V,V_1;Y_1,Z_1\vert X_1),I(V,V_1;Y_2,Z_2)\big\}\\
&\hspace{15mm}+ I(U_1;Y_1,Z_1\vert V,V_1,X_1),\\
&R_0+R_2 \leq \min\big\{I(V,V_1;Y_1,Z_1\vert X_1),I(V,V_1;Y_2,Z_2)\big\}\\
&\hspace{15mm}+I(U_2;Y_2,Z_2\vert V,V_1,X_1),
\end{align*}

\begin{align*} 
&R_0+R_1+R_2 \leq  I(V;Y_{1})+I(U_{2};Y_{2}\vert V)\\
&\hspace{15mm}+ I(U_1;Y_{1}|U_{2},V), \\
&R_0+R_1+R_2 \leq  I(V;Y_{2})+I(U_{1};Y_{1}\vert V)\\
&\hspace{15mm}+I(U_{2};Y_{2}|U_{1},V), \\
&R_0+R_1+R_2 \leq  I(V,V_{1};Y_{1},Z_{1}|X_{1})\\
&\hspace{15mm}+I(U_{2};Y_{2},Z_{2}\vert V,V_{1},X_{1})\\
&\hspace{15mm}+I(U_1;Y_{1},Z_{1}|X_{1},U_{2},V,V_{1}),  \\
&R_0+R_1+R_2 \leq  I(V,V_{1};Y_{2},Z_{2})\\
&\hspace{15mm}+I(U_{1};Y_{1},Z_{1}\vert V,V_{1},X_{1})\\
&\hspace{15mm}+I(U_2;Y_{2},Z_{2}|X_{1},U_{1},V,V_{1}) \Big\},
\end{align*}
where $co\{\cdot\}$ denotes the convex hull and $\mathcal{Q}$ is the set of all joint PDs $P_{VV_1 U_1 U_2 X_1 X_2 X}$ satisfying
$X_1 \mkv V_1 \mkv$ \\ $ (V,U_1,U_2,X)$ and $(V,U_1,U_2) \mkv (X,X_1,X_2) \mkv (Y_1,Y_2,Z_1,Z_2)$.
%where $co\{\cdot\}$ denotes the convex hull and $\mathcal{Q}$ is the set of all joint PDs $P_{VV_1U_1U_2X_1X}$ satisfying $X_1 \mkv V_1 \mkv (V,U_1,U_2,X)$. 
The cardinality of the auxiliary RVs are subjected to satisfy: 
\begin{align*}
\|\mathcal{V}\| &\leq \|\mathcal{X}\|\|\mathcal{X}_1\|\|\mathcal{X}_2\|\|\mathcal{Z}_1\|\|\mathcal{Z}_2\|+25 \ , \\ 
\|\mathcal{V}_1\| &\leq \|\mathcal{X}\|\|\mathcal{X}_1\|\|\mathcal{X}_2\|\|\mathcal{Z}_1\|\|\mathcal{Z}_2\|+17 \ , \\
\|\mathcal{U}_1\|,\|\mathcal{U}_2\| &\leq  \|\mathcal{X}\|\|\mathcal{X}_1\|\|\mathcal{X}_2\|\|\mathcal{Z}_1\|\|\mathcal{Z}_2\|+8. 
\end{align*}
\label{thm:II5-6}
\end{theorem}

\begin{remark}
We observe from the proof that $V_1$ is formed of causal and non-causal parts of the relay outputs. Hence $V_1$ can be intuitively seen as the help of the relays for $V$. It can also be inferred from the form of this rate region  that $V$ and $(U_1,U_2)$ represent common and private information, respectively.\\
\end{remark}

\begin{remark} We have the following observations:

\begin{itemize}
\item The outer-bound is valid for the general BRC. However, in the case of the SRC the outputs $(Z_b,Y_b)$ depend only on $(X,X_b)$ for $b=\{1,2\}$. By using these relations, the terms $I(U_b;Y_{b},Z_{b}|X_{b},T)$ and $I(U_b;Y_{b}|T)$ can be further bounded by $I(X;Y_{b},Z_{b}|X_{b},T)$ and $I(X,X_b;Y_{b}|T)$, respectively, for any variables $T\in\{V,V_1,U_1,U_2\}$. This simplifies the previous region. 

\item Moreover we can see that the rate region in Theorem \ref{thm:II5-6} is not totally symmetric. Thus, another upper bound can be derived by exchanging indices 1 and 2, i.e., by introducing $V_2$ and $X_2$ instead of $V_1$ and $X_1$. The final bound will be the intersection of these two regions.

\item If the relays are not present, i.e., $Z_1=Z_2=X_1=X_2=V_1=\emptyset$, it is not difficult to show that the previous bound reduces to the outer-bound for general broadcast channels, referred to as $UVW$-outer-bound \cite{Nair2007}. Furthermore, it was  recently shown that such bound is at least as good as all currently developed outer-bounds for the capacity region of broadcast channels \cite{Nair2011}. 
\end{itemize}
\end{remark}

The next theorem presents an outer-bound on the capacity region of the BRC with common relay. In this case, due to the fact that $Z_1=Z_2$ and $X_1=X_2$, we can  choose $V_1=V_2$ because of the definition of $V_b$ (cf. Appendix \ref{proof-8}). Therefore, based on the aforementioned symmetric property, the outer-bound in Theorem \ref{thm:II5-6}  yields the next result. \vspace{2mm}

\begin{theorem}[outer-bound BRC-CR]
The capacity region  $\mathcal{C}_{\textrm{BRC-CR}}$  of the BRC-CR is included in the set $\mathcal{C}_{\textrm{BRC-CR}}^{\textrm{out}}$ of all rate pairs  $(R_0,R_1,R_2)$ satisfying 
\begin{align*} 
&\mathcal{C}_{\textrm{BRC-CR}}^{\textrm{out}} = \displaystyle{co\bigcup\limits_{P_{VV_1U_1U_2X_1X}\in\mathcal{Q}}} \Big\{(R_0\geq 0 ,R_1\geq 0,R_2\geq 0):\\
&R_0 \leq \min\big\{I(V;Y_2),I(V;Y_1)\big\},\\
&R_0+R_1 \leq \min\big\{I(V;Y_1),I(V;Y_2)\big\} + I(U_1;Y_1\vert V),\\
&R_0+R_2 \leq \min\big\{I(V;Y_1),I(V;Y_2)\big\} + I(U_2;Y_2\vert V),\\
&R_0+R_1 \leq I(U_1;Y_1,Z_1\vert V,V_1,X_1)\\
&\hspace{8mm}+  \min\big\{I(V,V_1;Y_1,Z_1\vert X_1),I(V,V_1;Y_2,Z_1\vert X_1)\big\},\\
&R_0+R_2 \leq I(U_2;Y_2,Z_1\vert V,V_1,X_1)\\
&\hspace{8mm}+ \min\big\{I(V,V_1;Y_1,Z_1\vert X_1),I(V,V_1;Y_2,Z_1\vert X_1)\big\},\\
&R_0+R_1+R_2 \leq  I(V;Y_{1})+I(U_{2};Y_{2}\vert V)\\
&\hspace{8mm}+ I(U_1;Y_{1}|U_{2},V), \\
&R_0+R_1+R_2 \leq  I(V;Y_{2})+I(U_{1};Y_{1}\vert V)\\
&\hspace{8mm}+I(U_2;Y_{2}|U_{1},V), \\
&R_0+R_1+R_2 \leq  I(V,V_{1};Y_{1},Z_{1}|X_{1})\\
&\hspace{8mm}+I(U_{2};Y_{2},Z_{1}\vert V,V_{1},X_{1})\\
&\hspace{8mm}+I(U_1;Y_{1},Z_{1}|X_{1},U_{2},V,V_{1}),  \\
&R_0+R_1+R_2 \leq  I(V,V_{1};Y_{2},Z_{1}|X_{1})\\
&\hspace{8mm}+I(U_{1};Y_{1},Z_{1}\vert V,V_{1},X_{1})\\
&\hspace{8mm}+I(U_2;Y_{2},Z_{1}|X_{1},U_{1},V,V_{1}) \Big\},
\end{align*}
where $co\{\cdot\}$ denotes the convex hull and $\mathcal{Q}$ is the set of all joint PDs $P_{VV_1U_1U_2X_1X}$ verifying $X_1 \mkv V_1 \mkv (V,U_1,U_2,X)$ and $(V,U_1,U_2) \mkv (X,X_1) \mkv (Y_1,Y_2,Z_1)$, 
%where $co\{\cdot\}$ denotes the convex hull and $\mathcal{Q}$ is the set of all joint PDs $P_{VV_1U_1U_2X_1X}$ verifying $X_1 \mkv V_1 \mkv (V,U_1,U_2,X)$ 
where the cardinality of auxiliary RVs is subjected to satisfy:
 \begin{align*} 
 \|\mathcal{V}\| &\leq \|\mathcal{X}\|\|\mathcal{X}_1\|\|\mathcal{Z}_1\|+19 \ , \\
 \|\mathcal{V}_1\| &\leq \|\mathcal{X}\|\|\mathcal{X}_1\|\|\mathcal{Z}_1\|+11 \,  \\
 \|\mathcal{U}_1\|,\|\mathcal{U}_2\| &\leq \|\mathcal{X}\|\|\mathcal{X}_1\|\|\mathcal{Z}_1\|+8.
\end{align*}
\label{thm:II5-5}
\end{theorem}
\begin{proof}
It is enough to replace $Z_2$ with $Z_1$ in Theorem \ref{thm:II5-6}. Then the proof follows by taking the union with the symmetric region and using the fact that $I(V,V_{1};Y_{2},Z_{1}|X_{1})$ is less than $I(V,V_{1};Y_{2},Z_{1})$ due to the existing Markov relationship between $V_1$ and $X_1$.
\end{proof}

Finally, the next theorem presents an upper bound on capacity of the common-message BRC. This  is useful to evaluate the capacity of the compound relay channel. \vspace{2mm}

\begin{theorem}[upper bound on common-information]
An upper bound on the capacity of the common-message BRC (or  compound relay channel) is given by
\begin{equation*}
\begin{array} {rl}
R_0 \leq \max\limits_{P_{X_1X_2X}\in\mathcal{Q} }&\min\big\{I(X;Z_1Y_1\vert X_1),I(X,X_1;Y_1),\\
&I(X;Z_2,Y_2\vert X_2),I(X,X_2;Y_2)\big\}.
\end{array}
\label{eqCH1:II3-5}
\end{equation*}
\label{thm:II3-4}
\end{theorem}
\begin{proof}
The proof follows from conventional arguments \cite{Cover1979}. The common information $W_0$ is assumed to be decoded at both destinations. Moreover, the upper bound is the combination of the cut-set bound on each relay channel. 
\end{proof}

%%%%%%%%%%%%%%%%%%%%%%%%%%
\subsection{The degraded and the semi-degraded BRC with common relay}
%%%%%%%%%%%%%%%%%%%%%%%%%%
We now present inner and outer-bounds, and capacity results for a special class of broadcast relay channels with common relay (BRC-CR). Let us first define these classes of channels. \vspace{1mm}

\begin{definition}[degraded BRC-CR] \label{def-degraded}
A BRC-CR
 %as shown in Fig. \ref{figCH1:II-3} ,
  where $Z_1=Z_2$ and $X_1=X_2$, is said to be \emph{degraded}, respect to \emph{semi-degraded}, if the stochastic mapping $\big\{\mathbb{P}_{Y_1Z_1Y_2|XX_1}:\mathcal{X} \ex \mathcal{X}_1   \longmapsto \mathcal{Y}_1  \ex \mathcal{Z}_1 \ex \mathcal{Y}_2 \big\}$ satisfies at least one of the following conditions:
\begin{description}
\item[(I)] $X \mkv (X_1,Z_1) \mkv (Y_1,Y_2)$  and \\$(X,X_1) \mkv Y_1 \mkv Y_2$,

\item [(II)] $X \mkv (X_1,Z_1) \mkv Y_2$  and $X \mkv (Y_1,X_1) \mkv Z_1$,
\end{description}
where (I) is referred to as degraded BRC-CR and (II) to as semi-degraded BRC-CR. 
\end{definition}

Notice that the degraded BRC-CR can be seen as the combination of a degraded relay channel with a degraded BC. On the other hand, the semi-degraded case can be seen as the combination of a degraded BC with a reversely degraded relay channel. The capacity region of the semi-degraded BRC-CR is stated. %\vspace{1mm}

\begin{theorem}[semi-degraded BRC-CR]
The capacity region of the semi-degraded BRC-CR is given by the following rate region
\begin{align*} 
\mathcal{C}_{\textrm{BRC-CR}} &= \displaystyle{\bigcup\limits_{P_{UX_1X}\in\mathcal{Q}}} \Big\{(R_1\geq 0 ,R_2\geq 0): \\
R_2  & \leq \min\{I(U,X_1;Y_2),I(U;Z_1\vert X_1)\},\\ 
R_1+R_2 & \leq  \min\{I(U,X_1;Y_2), I(U;Z_1\vert X_1)\} \\
& +I(X;Y_1\vert X_1,U) \Big\}, %\label{eqCH1:II-4} 
\end{align*}
where $\mathcal{Q}$ is the set of all joint PDs $P_{UX_1X}$ satisfying $U \mkv (X_1,X) \mkv (Y_1,Z_1,Y_2)$, where the alphabet of $U$ is subjected to satisfy $\|\mathcal{U}\|\leq \|\mathcal{X}\|\|\mathcal{X}_1\|+2$. 
\label{thm:II5-4}
\end{theorem}\vspace{1mm}

\begin{proof}
It easy to show that  the rate region stated in Theorem \ref{thm:II5-4} directly follows from  that of Theorem \ref{thm:II2-1} by setting  $X_1=X_2=V_0$, $Z_1=Z_2$, $U_0=U_2=U_4=U$, and $U_1=U_3=X$. Whereas the converse proof is presented in Appendix \ref{proof-6}. 
\end{proof}

The next theorems provide outer and inner bounds on the capacity region of the degraded BRC-CR. \vspace{2mm}

\begin{theorem}[outer-bound degraded BRC-CR]
The capacity region $\mathcal{C}_{\textrm{BRC-CR}}$ of the degraded BRC-CR is included in the set of pair rates $(R_0,R_1)$ satisfying
\begin{align*} 
\mathcal{C}_{\textrm{BRC-CR}}^{\textrm{out}}&= \displaystyle{\bigcup\limits_{P_{UX_1X}\in\mathcal{Q}}} \Big\{(R_0\geq 0 ,R_1\geq 0): \\
R_0  \leq &I(U;Y_2),\\ 
R_1  \leq &\min\big\{I(X;Z_1\vert X_1,U),I(X,X_1;Y_1\vert U)\big\},  \\
R_0+R_1 \leq &\min\big\{I(X;Z_1\vert X_1),I(X,X_1;Y_1)\big\}\Big\}, %\label{eqCH1:II-4} 
\end{align*}
where $\mathcal{Q}$ is the set of all joint PDs $P_{UX_1X}$ satisfying $U \mkv (X_1,X) \mkv (Y_1,Z_1,Y_2)$, and the alphabet of  $U$ is subjected to satisfy 
$\|\mathcal{U}\|\leq \|\mathcal{X}\| \|\mathcal{X}_1\|+2$. 
\label{thm:II5-1}
\end{theorem}
By applying the degraded condition, it is easy to see that the outer-bound of  Theorem \ref{thm:II5-1} is included in that of Theorem \ref{thm:II5-5}. The proof of Theorem \ref{thm:II5-1} is presented in Appendix \ref{proof-4}.\vspace{2mm}

\begin{theorem}[inner bound degraded BRC-CR]
An inner bound on the capacity region $\mathcal{R}_{\textrm{BRC-CR}}$ of the BRC-CR is given by the set of rates $(R_0,R_1)$ \label{theo-region} satisfying
\begin{align*} 
\mathcal{R}_{\textrm{BRC-CR}} &= co \displaystyle{\bigcup\limits_{P_{UVX_1X}\in\mathcal{Q}}} \Big\{(R_0\geq 0 ,R_1\geq 0): \\
R_0\leq &I(U,V;Y_2)-I(U;X_1\vert V),\\ 
R_0+R_1 \leq &\min\big\{I(X;Z_1\vert X_1,V),I(X,X_1;Y_1)\big\},  \\
R_0+R_1 \leq &\min\big\{I(X;Z_1\vert X_1,U,V),I(X,X_1;Y_1\vert U,V)\big\}\\
&+ I(U,V;Y_2)-I(U;X_1\vert V)   \Big\},
\label{eqCH1:4} 
\end{align*} 
where $co\{\cdot\}$ denotes the convex hull for all PDs in $\mathcal{Q}$ verifying   $P_{UVX_1X}= P_{X|UX_1}  P_{X_1U|V} P_{V}$
with $(U,V) \mkv (X_1,X) \mkv (Y_1,Z_1,Y_2)$.
\label{thm:II5-2}
\end{theorem}
\begin{proof}
The proof of this theorem easily follows by choosing $U_0=U_2=U_4=U$, $V_0=V$, $U_1=U_3=X$  in Corollary \ref{thm:II2-2}. 
\end{proof}\vspace{1mm}

\begin{remark}
We observe that  in general the bounds in Theorems \ref{thm:II5-1} and \ref{thm:II5-2} do not coincide.   The difficulty arises in sharing the help of the relay between common and private information. In the inner bound, $V$ is seen as the help of relay for $R_0$. Notice that the choice of $V=\emptyset$ would remove the help of relay for the common information and hence when $Y_1=Y_2$  the region will be clearly suboptimal. Whereas the choice of $V=X_1$ will lead to a similar problem when $Y_2=\emptyset$.  Indeed, the code for common information cannot be superimposed on the whole relay code because it limits the relay help for private information. An alternative approach would be to superimpose common information on an additional description $V$, which plays the role of the relay help for common information. But this would cause another problem since $U$ is not superimposed on $X_1$, which implies that these descriptions do not have full dependence anymore. As a consequence of this, the converse does not 
seem to work. In other words, Marton coding removes the problem of  correlation at the price of deviating from the outer-bound. This is the main reason why the bounds are not tight for the degraded BRC with common relay.
\end{remark}

\begin{figure*}[t]
\centering
\subfigure[Degraded Gaussian BRC with common relay.]{
\includegraphics [width=.45 \textwidth] {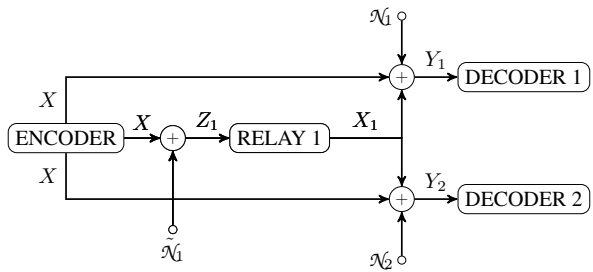}
\label{figCH1:DGBRC1}
}
\subfigure[Degraded Gaussian BRC with partial cooperation.]{
\includegraphics [width=.45 \textwidth] {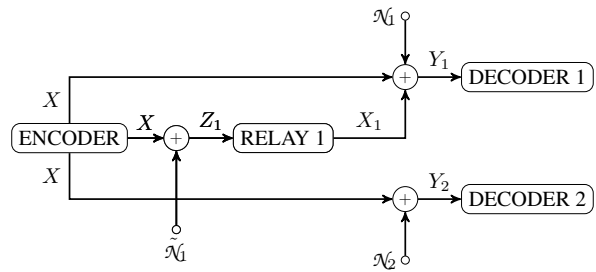}
\label{figCH1:DGBRC2}
}
\caption{Degraded Gaussian BRCs.}
\end{figure*}

%%%%%%%%%%%%%%%%%%%%%%%%%
\subsection{The degraded Gaussian BRC with common relay}
%%%%%%%%%%%%%%%%%%%%%%%%%

%%%%%%%%%%%%%%%%%%%%%%%%%
Interestingly, the inner and outer bounds in Theorems \ref{thm:II5-2} and \ref{thm:II5-1} coincide for the degraded Gaussian BRC with common relay, as shown in Fig. \ref{figCH1:DGBRC1}.  The degraded Gaussian BRC-CR is defined by the outputs: 
\begin{equation*}
\begin{array}{l}
Y_1=X+X_1+\mathpzc{N}_1, \\
Y_2=X+X_1+\mathpzc{N}_2, \\
Z_1=X+\tilde{\mathpzc{N}}_1,
\end{array}
\end{equation*}
where the source and the relay have power constraints $P,P_1$, and $\mathpzc{N}_1,\mathpzc{N}_2,\tilde{\mathpzc{N}}_1$ are independent Gaussian noises with variances $N_1,N_2,\tilde{N}_1$, respectively, such that the noises $\mathpzc{N}_1,\mathpzc{N}_2, \tilde{\mathpzc{N}}_1$ satisfy the necessary Markov conditions in definition \ref{def-degraded}. It is enough to assume physical degradedness of the receiver signals respect to the relay, and the stochastic degradedness of one receiver respect to the other one. Indeed, there exist $\mathpzc{N},\mathpzc{N}'$ such that:
\begin{equation*}
\begin{array}{l}
\mathpzc{N}_1=\tilde{\mathpzc{N}}_1+\mathpzc{N}, \\
\mathpzc{N}_2=\tilde{\mathpzc{N}}_1+\mathpzc{N}'
\end{array}
\end{equation*}
and also ${N}_1<{N}_2$. The following theorem holds as special case of Theorems \ref{thm:II5-1} and \ref{thm:II5-2}. \vspace{1mm}

\begin{theorem}[degraded Gaussian BRC-CR]
The capacity region of the degraded Gaussian BRC-CR is 
\begin{align*} 
\mathcal{C}_{\textrm{BRC-CR}}& = \displaystyle{\bigcup\limits_{0\leq\beta,\alpha\leq1}} \Big\{(R_0\geq 0 ,R_1\geq 0): \,\,\,  \\
R_0 & \leq \mathcal{C}\left(\frac{\alpha(P+P_1+2\sqrt{\overline\beta PP_1})}{\overline\alpha(P+P_1+2\sqrt{\overline\beta PP_1})+{N}_2}\right),\\
R_1 & \leq   \mathcal{C}\left(\frac{\overline\alpha(P+P_1+2\sqrt{\overline\beta PP_1})}{{N}_1}\right),\\
R_0+R_1 & \leq  \mathcal{C}\left(\frac{\beta P}{\tilde{N}_1}\right) \Big\}. %\label{eqCH1:II-4} 
\end{align*}
\label{thm:II5-3}
\end{theorem}

We shall not prove this theorem here since it was independently established  in \cite{Bhaskaran2008}. The original inner and outer-bounds initially provided had different forms, but their equivalence was established later using a tuning technique. In our case, these bounds can be simply derived from Theorems \ref{thm:II5-1} and \ref{thm:II5-2}. The outer-bound is the same as \cite{Bhaskaran2008} and the inner bound includes the result in \cite{Bhaskaran2008}. The  equivalence of these bounds can be then established. The inner bound in Theorem \ref{thm:II5-3} is obtained from Theorem \ref{thm:II5-1}  by choosing $U$ and $X_1$ conditionally independent given $V$. The source divides its power into $\theta P$ and $\overline{\theta} P$ for the first and the second user, respectively. The relay does the same with its power into $\theta_r P_1$ and $\overline{\theta_r} P_1$. Then $\gamma$ and $\rho$ represent the correlation coefficient between ($U$,$V$) and  ($X_1$,$X$), respectively. Parameters $\alpha$ and $\beta$ can be respectively interpreted as the power allocation at the source for both destinations and the correlation coefficient between source and relay signals. The inner bound is calculated by following \cite{Bhaskaran2008}. The outer-bound remains the same and it equals to the region in Theorem \ref{thm:II5-3}, but it is derived in a different way. 

%%%%%%%%%%%%%%%%%%%%%%%%%%
\subsection{Degraded Gaussian BRC with partial cooperation}
%%%%%%%%%%%%%%%%%%%%%%%%%%
We next present the capacity region of the Gaussian degraded BRC with partial cooperation, as depicted in Fig. \ref{figCH1:DGBRC2}. In this setting, there is no relay-destination cooperation for the second destination and the first destination is physically degraded respect to the relay signal.  Input and output relations are as follows:
\begin{equation*}
\begin{array}{l}
Y_1=X+X_1+\mathpzc{N}_1, \\
Y_2=X+\mathpzc{N}_2, \\
Z_1=X+\tilde{\mathpzc{N}}_1. 
\end{array}
\end{equation*}
The source and the relay have power constraints $P,P_1$, and $\mathpzc{N}_1,\mathpzc{N}_2,\tilde{\mathpzc{N}}_1$ are independent Gaussian noises with variances $N_1,N_2,\tilde{N}_1$. In addition to this,  there exists $\mathpzc{N}$ such that $\mathpzc{N}_1=\tilde{\mathpzc{N}}_1+\mathpzc{N}$, which means that $Y_1$ is physically degraded respect to $Z_1$  and we also assume ${N}_2<\tilde{N}_1$. The proof of the following theorem is presented in Appendix \ref{proof-5-1}.\vspace{1mm}

\begin{theorem}
\textit{(Gaussian degraded BRC with partial cooperation)} The capacity region of the Gaussian degraded BRC with partial cooperation is given by
\begin{align*} 
\mathcal{C}_{\textrm{BRC-PC}} &= \displaystyle{\bigcup\limits_{0\leq\beta,\alpha\leq1}} \Big\{(R_1\geq 0 ,R_2\geq 0): \\
R_1&\leq \underset{\beta\in[0, 1]}{\max} \min\Big\{ \mathcal{C}\left(\frac{\displaystyle \alpha\beta P}{\overline\alpha P+\widehat{N}_1}\right),
\end{align*}
\begin{align*} 
&\hspace{5mm}   \mathcal{C}\left(\displaystyle\frac{\displaystyle \alpha P+{P_1}+{2\sqrt{\overline\beta\alpha PP_1}}}{\overline\alpha P+{N}_1}\right)   \Big\},\\
R_2&\leq \mathcal{C}\left(\displaystyle\frac{\displaystyle \overline{\alpha}P}{N_2}\right)
\Big\}. %\label{eqCH1:II-4} 
\end{align*}
\label{thm:II5-3-1}
\end{theorem}

The proof of this theorem is indeed similar to Theorem \ref{thm:II5-4} for the capacity of the semi-degraded BRC. The source assigns power $\alpha P$ to carry the message to destination $Y_1$ and $\overline{\alpha} P$ to destination $Y_2$. Parameters $\alpha$ and $\beta$ are defined as well as in Theorem \ref{thm:II5-3}. Destination $Y_2$ is the best receiver so it can decode the message intended for destination $Y_1$, even after the help of the relay. It means that both the first relay and the destination appear to be  degraded respect to the second destination. So the second destination can correctly decode the interference of other users. However, we emphasize that $Z_1$ is not necessarily physically degraded respect to $Y_2$, which makes of Theorem \ref{thm:II5-3-1} a stronger result than that in Theorem \ref{thm:II5-4}. 

%%%%%%%%%%%%%%%%%%%%%%%%%
%\subsection{Capacity region of BRC with orthogonal components}
%%%%%%%%%%%%%%%%%%%%%%%%%%%

%%*********************************************************************************************************
%%III. Numerical Result and Guassian Examples
%%*********************************************************************************************************
\section{Gaussian Simultaneous and Broadcast Relay Channels} \label{BRC:IV}

\begin{figure}[ht!]
\centering
\includegraphics [width=.5 \textwidth] {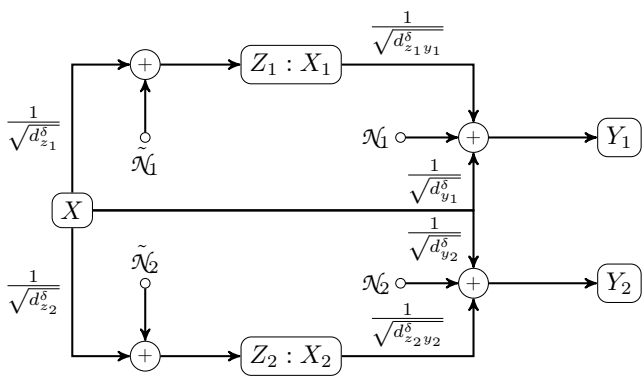}
\caption{Gaussian BRC }
\label{Fig-GaussianBRC}
\end{figure}

In this section, based on the rate regions presented in Section \ref{achievable-regions}, we compute achievable rate regions for the Gaussian BRC.  The Gaussian BRC is modeled as follows:
\begin{equation*}
\begin{array}{lcr}\displaystyle 
Y_{1i} =\frac{X_i}{\sqrt{d^\delta_{y_1}}}+\frac{X_{1i}}{\sqrt{d^\delta_{z_1y_1}}}+\mathpzc{N}_{1i},&  & \displaystyle Z_{1i} =\frac{X_i}{\sqrt{d^\delta_{z_1}}}+\tilde{\mathpzc{N}}_{1i},\\
\displaystyle Y_{2i} =\frac{X_i}{\sqrt{d^\delta_{y_2}}}+\frac{X_{2i}}{\sqrt{d^\delta_{z_2y_2}}}+\mathpzc{N}_{2i},&    & \displaystyle Z_{2i}  =\frac{X_i}{\sqrt{d^\delta_{z_2}}}+\tilde{\mathpzc{N}}_{2i}.
\end{array}
%\label{eqCH1:III-1}
\end{equation*}
The channel inputs $\{X_i\}$ and the relay inputs $\{X_{1i}\}$ and $\{X_{2i}\}$ must satisfy the power constraints
\begin{equation*}
\begin{array}{lcr}\displaystyle 
\sum_{i=1}^n X_i^2\leq n P ,& \textrm{and}  & \displaystyle \sum_{i=1}^n X_{ki}^2\leq n P_k,\,\,\, k=\{1,2\}. 
\end{array}
\end{equation*}
 The channel noises $\tilde{\mathpzc{N}}_{1i},\tilde{\mathpzc{N}}_{2i}$, ${\mathpzc{N}}_{1i},{\mathpzc{N}}_{2i}$ are zero-mean i.i.d. Gaussian RVs of variances $\tilde{{N}}_1,\tilde{{N}}_2,N_1,N_2$ and independent of the channel and the relay inputs. The distances {$(d_{y_1},d_{y_2})$} between the source and the destinations $1$ and $2$, respectively, are assumed to be fixed during the communication. Similarly, the distances between the relays and their destinations {$(d_{z_1y_1},d_{z_2y_2})$}. As shown in Fig. \ref{Fig-GaussianBRC}, notice that in this simultaneous Gaussian relay channel no interference is allowed from the relay $b$ to the destination $\overline{b}=\{1,2\}\setminus \{b\}$, for $b=\{1,2\}$. In the remainder of this section, we evaluate DF-DF, DF-CF and CF-CF regions, and outer-bounds. As for the classical BC, by using superposition coding, we decompose $X$ as the sum of two independent descriptions such that     
$\esp\left\{X_A^2\right\}=\alpha P$ and $\esp\left\{X_B^2\right\}=\overline{\alpha} P$, where $\overline{\alpha}=1-\alpha$.  The codewords $(X_A,X_B)$ contain informations for destinations $Y_1$ and $Y_2$, respectively.

%%*********************************************************************************************************
%%III.1 DF DF Case
%%*********************************************************************************************************
\subsection{DF-DF region for Gaussian BRC}

We aim to evaluate the rate region in Theorem \ref{thm:II2-1} for the presented Gaussian BRC.  To this end, we rely on well-known coding schemes for broadcast and relay channels.  A \emph{Dirty-Paper Coding} (DPC) scheme is needed for destination $Y_2$ to cancel the interference coming from the relay signal $X_1$. Similarly, a DPC scheme is needed for destination $Y_1$  to cancel the signal noise $X_B$ coming from the code of the other user. The auxiliary RVs $(U_1,U_2)$ are chosen as:
\begin{equation*}
\begin{array} {l}
U_1=X_A+\lambda\,\, X_B \textrm{ with }\,\,\displaystyle  X_A=\tilde{X}_A+\sqrt{\frac{\overline{\beta_1}\alpha P}{P_1}}X_1, \\
U_2=X_B+\gamma X_1\,\, \textrm{ with }\,\, \displaystyle X_B=\tilde{X}_B+\sqrt{\frac{\overline{\beta_2}\overline{\alpha}P}{P_1}}X_2,
\end{array} 
%\label{eqCH1:III1-13}
\end{equation*}
for some parameters $\beta_1,\beta_2,\alpha,\gamma,\lambda\in[0,1]$, where the encoder sends $X=X_A+X_B$. Now choose  in Theorem \ref{thm:II2-1} $V_0=U_0=\emptyset$, $U_1=U_3$ and $U_4=U_2$. It can be seen that this choice leads to $I_M=0$ and $I_i=J_i$ for $i=\{1,2\}$. Then for $R_0=0$ and based on the above RVs, the next rates are achievable:
\begin{align*}
R_1 \leq\min\big\{I(U_1;Z_1\vert X_1),&I(U_1,X_1;Y_1)\big\}\\
& -I(U_1;X_2,U_2|X_1),\\
R_2 \leq \min\big\{	I(U_2;Z_2\vert X_2),& I(U_2,X_2;Y_2)\big\}\\
&-I(X_1;U_2|X_2).
\end{align*}
%We try to evaluate these rates.

%%%%%%%%%
%%%DESTINATION 1
For destination 1, the achievable rate is the minimum of two mutual informations, where the first term is given by $R_{11}\leq I(U_1;Z_1\vert X_1)-I(U_1;X_2,U_2\vert X_1)$. The current problem becomes similar to the conventional DPC with $\tilde{X}_A$ as the main message, $X_B$ as the interference and $\tilde{N}_1$ as the noise.
Hence the corresponding rate writes as
 
\begin{align}
&R_{11}^{(\beta_1,\lambda)}=\nonumber\\
&\frac{1}{2}\log\left[\frac{\displaystyle \alpha\beta_1 P(\alpha\beta_1 P+\overline{\alpha}P+d^\delta_{z_1}\tilde{N}_1)}
{d^\delta_{z_1}\tilde{N}_1(\alpha\beta_1 P+\lambda^2\overline{\alpha}P)+(1-\lambda)^2\overline{\alpha}P\alpha{\beta_1}P}\right].
\label{eqCH1:III1-16}
\end{align}
The second term is $R_{12}=I(U_1,X_1;Y_1)-I(U_1;X_2,U_2\vert X_1)$, where the first mutual information can be decomposed into two terms $I(X_1;Y_1)$ and $I(U_1;Y_1\vert X_1)$. 
%--------------------
Notice that regardless of the former, the rest of the terms in the expression of rate $R_{12}$ are similar to $R_{11}$. 
%--------------------------
The main codeword is $\tilde{X}_A$, while $X_B$, ${\mathpzc{N}}_1$ are the random state and the noise. After adding the term $I(X_1;Y_1)$, we obtain

\begin{align}
&R_{12}^{(\beta_1,\lambda)}=\nonumber\\
&\frac{1}{2}\log\left[\frac{\displaystyle\alpha\beta_1 P d^\delta_{y_1} \left (\frac{P}{d^\delta_{y_1}}+\frac{P_1}{d^\delta_{z_1y_1}}+2\sqrt{\frac{\overline{\beta_1}\alpha PP_1}{d^\delta_{y_1}d^\delta_{z_1y_1}}}+N_1\right)}
{\displaystyle d^\delta_{y_1} {N}_1(\alpha\beta_1 P+\lambda^2\overline{\alpha}P)+(1-\lambda)^2 \overline{\alpha}P\alpha{\beta_1}P}\right].
%R_{12}=I(U_1X_1;Y_1)-I(U_1;U_2\vert X_1)  \\
\label{eqCH1:III1-17}
\end{align}
%%%%%%%%
\begin{figure*} [!t]

\begin{align}
R_{21}^{(\beta_1,\beta_2,\gamma)}=\frac{1}{2}\log\left[\frac{\displaystyle\overline{\alpha}\beta_2 P(\overline{\alpha}\beta_2 P+{\alpha}P+d^\delta_{z_2}\tilde{N}_2)}
{(d^\delta_{z_2}\tilde{N}_2+\alpha\beta_1 P)(\overline{\alpha}\beta_2 P+\gamma^2\overline{\beta_1}\alpha P)+(1-\gamma)^2\overline{\alpha}\beta_2 P\alpha\overline{\beta_1}P}\right],
\label{eqCH1:III1-15}
\end{align}
\hrule
\end{figure*}
\begin{figure*} [!t]

\begin{align}
R_{22}^{(\beta_1,\beta_2,\gamma)}=\frac{1}{2}\log\left[\frac{\displaystyle\overline{\alpha}\beta_2 Pd^\delta_{y_2} \left(\frac{P}{d^\delta_{y_2}}+\frac{P_2}{d^\delta_{z_2y_2}}+2\sqrt{\frac{\overline{\beta_2}\overline{\alpha} PP_2}{d^\delta_{y_2}d^\delta_{z_2y_2}}}+N_2\right)}
{(d^\delta_{y_2} {N}_2+\alpha\beta_1 P )( \overline{\alpha}\beta_2 P+\gamma^2\overline{\beta_1}{\alpha}P)+(1-\gamma)^2 \overline{\alpha}\beta_2P\alpha\overline{\beta_1}P }\right].
\label{eqCH1:III1-18}
\end{align}
\hrule
\end{figure*}

%%%%%%%%

Based on expressions \eqref{eqCH1:III1-16} and \eqref{eqCH1:III1-17}, the maximum achievable rate follows as
\begin{align*}
 R^*_1=\underset{0\leq\beta_1,\lambda\leq1}{\max} \min\left\{R_{11}^{(\beta_1,\lambda)},R_{12}^{(\beta_1,\lambda)}\right\}.
\end{align*}

For the destinations, the argument is similar to the one above with the difference that for the current DPC, where only $X_1$ can be canceled,  the rest of $X_A$ appears as noise for the destinations. So it becomes the conventional DPC with $\tilde{X}_B$ as the main message, $X_1$ as the interference, and $\tilde{\mathpzc{N}}_1$ and $\tilde{X}_A$ as the noises. The rates write as \eqref{eqCH1:III1-15} and \eqref{eqCH1:III1-18}.

And finally the maximum achievable rate follows as 
$$
R^*_2=\underset{0\leq\beta_2,\gamma\leq1}{\max} \min\left\{R_{21}^{(\beta_1,\beta_2,\gamma)},R_{22}^{(\beta_1,\beta_2,\gamma)}\right\}.
$$

%%*********************************************************************************************************
%%III.2 DF CF Case
%%*********************************************************************************************************
\subsection{DF-CF region for the Gaussian BRC}
% \input{III.2.DFCFex}
%\subsubsection{Inner Bounds on the private information}
As for the conventional BC, by using superposition coding, we decompose $X=X_A+X_B$ as a sum of two independent RVs such that $\esp\left\{X_A^2\right\}=\alpha P$ and $\esp\left\{X_B^2\right\}=\overline{\alpha} P$, where $\overline{\alpha}=1-\alpha$.  The codewords $(X_A,X_B)$ contain the information intended to receivers $Y_1$ and $Y_2$, respectively.  First, we identify two different cases for which DPC schemes are derived.  In the first case, the code is such that the CF destination is able to remove the interference caused by DF code. In the second case, the code is such that DF destination cancels the interference of CF code.

\subsubsection*{Case I}
A DPC scheme is applied to $X_B$ to cancel the interference  $X_A$ while the relay signal is similarly selected  to \cite{Cover1979}. Hence, the RVs $(U_1,U_2)$ are set to
\begin{align}
U_1 & =X_A=\tilde{X}_A+\sqrt{\frac{\overline{\beta}\alpha P}{P_1}}X_1,\\
U_2 & =X_B+\gamma X_A,
\label{eqCH1:III2-1}
\end{align}
where $\beta$ is the correlation coefficient between the relay and the source and, $\tilde{X}_A$ and $X_1$ are independent. Notice that  in this case, instead of only $Y_2$, we have also $\hat{Z}_2$ present which is chosen to as $\hat{Z}_2=Z_2+\hat{\mathpzc{N}}_2$. Thus, DPC should also be able to cancel the interference at both, received and compressed signals having different noise levels. Calculation should be done again with $(Y_2,\hat{Z}_2)$, which are the main message $X_B$ and the interference $X_A$. We can show that the optimum $\gamma$ has a similar form to the classical DPC with the noise term replaced by an equivalent noise which is like the harmonic mean of the noise in $(Y_2,\hat{Z}_2)$. The optimum $\gamma^*$ is given by
\begin{align}
\gamma^*&=\frac{\overline{\alpha}P}{\overline{\alpha}P+N_{t1}},\nonumber\\
N_{t1}&=\left[ (d^\delta_{z_2}(\tilde{N}_2+\widehat{N}_2))^{-1}+(d^\delta_{y_2}({N}_2))^{-1}\right]^{-1}.
\end{align}
As we can see the equivalent noise is twice of the harmonic mean of the other noise terms. 

From Corollary \ref{thm:II3-3},  we can see that the  optimal $\gamma^*$ and the current definitions yield the rates
\begin{align}
R_1^*&=\min\big\{I(U_1;Z_1\vert X_1),I(U_1,X_1;Y_1)\big\}\nonumber\\
&= \underset{0\leq\beta\leq1}{\max} \min\Big\{\mathcal{C}\left(\frac{\alpha\beta P}{\overline\alpha P+d^\delta_{z_1}\tilde{N}_1}\right),\nonumber\\
&\mathcal{C}\left(\frac{\displaystyle \alpha \frac{P}{d^\delta_{y_1}}+\frac{P_1}{d^\delta_{z_1y_1}}+2\sqrt{\frac{\overline\beta\alpha PP_1}{d^\delta_{y_1}d^\delta_{z_1y_1}}}}{\displaystyle \frac{\overline\alpha P}{d^\delta_{y_1}}+{N}_1}\right)\Big\}, \\
R_2^*&=I(U_2;Y_2,\hat{Z}_2\vert X_2)-I(U_1,X_1;U_2)
\end{align}
\begin{align}
&=\mathcal{C}\left(\frac{\overline{\alpha}P}{d^\delta_{y_2}N_2}+\frac{\overline{\alpha}P}{d^\delta_{z_2}(\widehat{N}_2+\tilde{N}_2)}\right).
\label{eqCH1:III2-2}
\end{align}
 Note that since $(X_A,X_B)$ are chosen independent, destination 1 sees $X_B$ as an additional channel noise. The compression noise is chosen as follows
\begin{equation}
\widehat{N}_2={\left(P\left(\frac{1}{d^\delta_{y_2}N_2}+\frac{1}{d^\delta_{z_2}\tilde{N}_2}\right)+1\right)}/{\frac{P_2}{d^\delta_{y_2}N_2}}.
\label{eqCH1:III2-3}
\end{equation}

\subsubsection*{Case 2} We use a DPC scheme for destination $Y_2$ to cancel the interference $X_1$, and next we use a DPC scheme for destination $Y_1$ to cancel $X_B$. For this case, the auxiliary RVs $(U_1,U_2)$ are chosen as
\begin{equation}
\begin{array} {l}
U_1=X_A+\lambda\,\, X_B \textrm{ with }\,\, X_A=\tilde{X}_A+\sqrt{\displaystyle \frac{\overline{\beta}\alpha P}{P_1}}X_1, \\
U_2=X_B+\gamma X_1.
\end{array} 
\label{eqCH1:III2-4}
\end{equation}
From Corollary \ref{thm:II3-3}, the corresponding rates with the current definitions are 
\begin{align}
R_1& =\min\big\{I(U_1;Z_1\vert X_1),I(U_1,X_1;Y_1)\big\}\nonumber\\
&\hspace{10mm}-I(U_1;U_2|X_1),\\
R_2 & = I(U_2;Y_2,\hat{Z}_2\vert X_2)-I(X_1;U_2).
\end{align}
The argument for destination 2  is similar than before but it differs in the DPC. Here only $X_1$ can be canceled and then $X_A$ remains as additional noise. The optimum $\gamma^*$ similar to \cite{Behboodi2009} is given by 
\begin{align}
\gamma^*&=\sqrt{\frac{\overline{\beta}\alpha P}{P_1}}\frac{\overline{\alpha}P}{\overline{\alpha}P+N_{t2}},\\
N_{t2}&=\big( (d^\delta_{z_2}(\tilde{N}_2+\widehat{N}_2)\nonumber\\
&\hspace{10mm}+{\beta}\alpha P)^{-1}+(d^\delta_{y_2}({N}_2)+{\beta}\alpha P)^{-1}\big)^{-1},
\end{align}
and
\begin{equation}
R_2^*=\mathcal{C}\left(\frac{\overline{\alpha}P}{d^\delta_{y_2}N_2+{\beta}\alpha P}+\frac{\overline{\alpha}P}{d^\delta_{z_2}(\widehat{N}_2+\tilde{N}_2)+{\beta}\alpha P}\right).
\label{eqCH1:III2-5}
\end{equation}
For destination 1, the achievable rate is the minimum of two terms, where the first one is given by 
\begin{align}
&R_{11}^{(\beta,\lambda)}= I(U_1;Z_1\vert X_1)-I(U_1;U_2\vert X_1)\nonumber\\
&=\frac{1}{2}\log\left(\frac{\alpha\beta P(\alpha\beta P+\overline{\alpha}P+d^\delta_{z_1}\tilde{N}_1)}
{d^\delta_{z_1}\tilde{N}_1(\alpha\beta P+\lambda^2\overline{\alpha}P)+(1-\lambda)^2\overline{\alpha}P\alpha{\beta}P}\right).
\label{eqCH1:III2-6}
\end{align}
The second term is $R_{12}=I(U_1X_1;Y_1)-I(U_1;U_2\vert X_1)$, where the first mutual information can be decomposed into two terms $I(X_1;Y_1)$ and $I(U_1;Y_1\vert X_1)$. 
%--------------------
Notice that regardless of the former, the rest of the terms in the expression of the rate $R_{12}$ are similar to $R_{11}$. 
%--------------------------
The main codeword is $\tilde{X}_A$, while $X_B$ and ${\mathpzc{N}}_1$ represent  the random state and the noise, respectively. After adding the term $I(X_1;Y_1)$, we obtain\\
\begin{align}
&R_{12}^{(\beta,\lambda)}=\nonumber\\
&\frac{1}{2}\log\left[\frac{\displaystyle \alpha\beta P d^\delta_{y_1}\left(\frac{P}{d^\delta_{y_1}}+\frac{P_1}{d^\delta_{z_1y_1}}+2\sqrt{\frac{\overline{\beta}\alpha PP_1}{d^\delta_{y_1}d^\delta_{z_1y_1}}}+N_1\right)}
{{N}_1 d^\delta_{y_1}(\alpha\beta P+\lambda^2\overline{\alpha}P)+(1-\lambda)^2\overline{\alpha}P\alpha{\beta}P}\right].
\label{eqCH1:III2-7}
\end{align}
Based on expressions \eqref{eqCH1:III2-7} and \eqref{eqCH1:III2-6}, the maximum achievable rate follows as
\begin{equation}
R^*_1=\underset{0\leq\beta,\lambda\leq1}{\max} \min\big\{R_{11}^{(\beta,\lambda)},R_{12}^{(\beta,\lambda)}\big\}.   \label{eqCH1:III2-7B}
\end{equation}
It should be noted that the constraint for $\widehat{N}_2$ is still the same as \eqref{eqCH1:III2-3}.
 % Corollary \ref{thm:2} and Theorem \ref{thm:4}:

%%*********************************************************************************************************
%%III.3 CF CF Case
%%*********************************************************************************************************
%\vspace{-3mm}
\subsection{CF-CF region for the Gaussian BRC}
We now investigate the Gaussian BRC for the CF-CF region, where the relays are collocated with the destinations. In this setting, the compression noises are chosen as follows:
\begin{align}
\hat{Z}_1&=Z_1+\hat{\mathpzc{N}}_1,\nonumber\\
\hat{Z}_2&=Z_2+\hat{\mathpzc{N}}_2,
\end{align}
where $\hat{\mathpzc{N}}_1,\hat{\mathpzc{N}}_2$ are zero-mean Gaussian noises of variances $\hat{N}_1,\hat{N}_2$. As for the conventional BC, by using superposition coding, we decompose $X=X_A+X_B$ as a sum of two independent RVs such that $\esp\left\{X_A^2\right\}=\alpha P$ and $\esp\left\{X_B^2\right\}=\overline{\alpha} P$, where $\overline{\alpha}=1-\alpha$.  The codewords $(X_A,X_B)$ contain the information intended to destinations $Y_1$ and $Y_2$. A DPC scheme is applied to $X_B$ to cancel interference  $X_A$ while the relay signal is similarly selected to \cite{Cover1979}. So the auxiliary RVs $(U_1,U_2)$ are set to
\begin{equation}
U_1=X_A, \,\,\,\, U_2=X_B+\gamma X_A.
\label{eqCH1:10}
\end{equation}
Notice that,  in this case, instead of only $Y_2$ we have also $\hat{Z}_2$ present in the rate. Thus, DPC should be also able to cancel the interference in both, received and compressed signals which have different noise levels. Calculation should be done again with $(Y_2,\hat{Z}_2)$ which are the main message $X_B$ and the interference $X_A$. It can be shown that the optimum $\gamma$ has a similar form to the classical DPC with the noise term replaced by an equivalent noise which is like the harmonic mean of the noises in $(Y_2,\hat{Z}_2)$. The optimum
\begin{align}
 \gamma^*&=\frac{\overline{\alpha}P}{\overline{\alpha}P+N_{t1}},\nonumber\\
 N_{t1}&=\left[ 1/(d^\delta_{z_2}(\tilde{N}_2+\widehat{N}_2))+1/(d^\delta_{y_2}{N}_2)\right]^{-1}.
 \end{align}
Observe that the equivalent noise is twice of the harmonic mean of the other noise terms. We use Theorem \ref{thm:II4-1} with  $U_0=\phi$ to find the following rates 
\begin{align}
R_1^*&=I(U_1;Y_1,\hat{Z}_1\vert X_1)\nonumber\\
&=\mathcal{C}\left(\frac{{\alpha}P}{d^\delta_{y_1}N_1+\overline{\alpha} P}+\frac{{\alpha}P}{d^\delta_{z_1}(\widehat{N}_1+\tilde{N}_1)+\overline{\alpha} P}\right),\nonumber\\ 
R_2^*&=I(U_2;Y_2,\hat{Z}_2\vert X_2)-I(U_1X_1;U_2)
\end{align}
\begin{align}
&=\mathcal{C}\left(\frac{\overline{\alpha}P}{d^\delta_{y_2}N_2}+\frac{\overline{\alpha}P}{d^\delta_{z_2}(\widehat{N}_2+\tilde{N}_2)}\right).
\label{eqCH1:11}
\end{align}
Note that since $(X_A,X_B)$ are chosen independent, destination 1 sees $X_B$ as additional channel noise. The compression noises are chosen as follows: 
\begin{align}
\hat{N}_1=\tilde{N}_1{\left[P\left(\frac{1}{d^\delta_{y_1}N_1}+\frac{1}{d^\delta_{z_1}\tilde{N}_1}\right)+1\right]}/{\frac{P_1}{d^\delta_{z_1y_1}N_1}},\nonumber\\
\hat{N}_2=\tilde{N}_2{\left[P\left(\frac{1}{d^\delta_{y_2}N_2}+\frac{1}{d^\delta_{z_2}\tilde{N}_2}\right)+1\right]}/{\frac{P_2}{d^\delta_{z_2y_2}N_2}}.
\label{eqCH1:12}
\end{align}

\subsubsection*{Common-rate} The goal is to send common-information at rate $R_0$. To this end, define $X=U_0$ and evaluate Theorem \ref{thm:II4-1} with $U_1=U_2=\phi$.  It is easy to verify that the following common-rate is achievable
\begin{align}
R_0\leq \min\Big\{&\mathcal{C}\left(\frac{P}{d^\delta_{y_1}N_1}+\frac{P}{d^\delta_{z_1}(\widehat{N}_1+\tilde{N}_1)}\right),\nonumber\\
& \mathcal{C}\left(\frac{P}{d^\delta_{y_2}N_2}+\frac{P}{d^\delta_{z_2}(\widehat{N}_2+\tilde{N}_2)}\right)\Big\}.
\label{eqCH1:19}
\end{align} 
The constraints for compression noises remain the same as before. 

%One can see the similarity between this case and the case of classical dirty paper coding. 
%%%%%%%%%%%%%%%%%%%%%%%%%%%%%%%%%%%%%%%%%%%%%%%%%%%%%%%%%%%%%%%%%%%55

\subsection{The source is oblivious to the cooperative strategy adopted by the relay}
In this setting, we deal with two different models referred to as the Compound relay channel (RC) and the Composite relay channel (RC). 

\subsubsection{Compound RC}  The goal is to send common-information at rate $R_0$ based on  the  DF-CF region. The definition of the channels remain the same. We set  $X=U+\sqrt{\displaystyle \frac{\overline{\beta}P}{P_1}}X_1$ and evaluate Corollary \ref{thm:II3-2}.  It is easy to verify that the achievable rate $R_{\textrm{DF}}$ for the destination $Y_1$ writes as
\begin{align}
R_{\textrm{DF}}\leq& \min\Big\{\mathcal{C}\left(\frac{\beta P}{d^\delta_{z_1}\tilde{N}_1}\right), \nonumber\\
&\mathcal{C}\left(\frac{\displaystyle \frac{P}{d^\delta_{y_1}}+\frac{P_1}{\displaystyle  d^\delta_{z_1y_1}}+2\sqrt{\displaystyle  \frac{ \displaystyle  \overline\beta PP_1}{d^\delta_{y_1}d^\delta_{z_1y_1}}}}{{N}_1}\right)\Big\}.
\label{eqCH1:III2-8B}
\end{align}  
For destination $Y_2$, the CF rate $I(X;Y_2,\hat{Z}_2|X_2)$ is as follows
\begin{equation}
R_{\textrm{CF}}\leq \mathcal{C}\left(\frac{P}{d^\delta_{y_2}N_2}+\frac{P}{d^\delta_{z_2}(\widehat{N}_2+\tilde{N}_2)}\right).
\label{eqCH1:III2-9}
\end{equation}
The upper bound from Theorem \ref{thm:II3-4} writes as the next rate  
\begin{align}
C\leq \underset{0\leq\beta_1,\beta_2\leq1}{\max} \min & \Big\{ \mathcal{C}\left(\displaystyle \beta_1 P\left[\frac{1}{\displaystyle  d^\delta_{z_1}\tilde{N}_1}+\frac{1}{\displaystyle  d^\delta_{y_1}{N}_1}\right]\right), \nonumber\\
&\mathcal{C}\left(\displaystyle \frac{\displaystyle \frac{P}{\displaystyle  d^\delta_{y_1}}+\displaystyle  \frac{P_1}{d^\delta_{z_1y_1}}+2\sqrt{\frac{\displaystyle  \overline\beta_1 PP_1}{d^\delta_{y_1}d^\delta_{z_1y_1}}}}{{N}_1}\right),\nonumber\\
& \mathcal{C}\left(\beta_2 P\left[\frac{1}{d^\delta_{z_2}\tilde{N}_2}+\frac{1}{d^\delta_{y_2}{N}_2}\right]\right),\nonumber\\
&\mathcal{C}\left(  \frac{\displaystyle \frac{P}{d^\delta_{y_2}}+\frac{P_2}{d^\delta_{z_2y_2}}+2\sqrt{\displaystyle  \frac{\displaystyle  \overline\beta_2 PP_2}{\displaystyle  d^\delta_{y_2}d^\delta_{z_2y_2}}}}{{N}_2}\right)
\Big\}. 
\end{align}
Observe that the rate \eqref{eqCH1:III2-9} is exactly the same as the Gaussian CF rate \cite{Kramer2005}. This means that DF based on regular encoding can be also  decoded with the CF strategy, as well as the case with collocated relay and receiver \cite{Katz2005}. By using the proposed coding, it is possible to send common information at the minimum rate between DF \eqref{eqCH1:III2-8B} and CF \eqref{eqCH1:III2-9} rates
$$
R_0=\min\{R_{\textrm{DF}},R_{\textrm{CF}}\}.
$$ 
For the case of private information, we have shown that any pair of rates $(R_{\textrm{DF}}\leq R^*_1,R_{\textrm{CF}}\leq R^*_2)$ given by  \eqref{eqCH1:III2-5} and \eqref{eqCH1:III2-7B} are admissible and thus $(R_{\textrm{DF}},R_{\textrm{CF}})$ can be simultaneously sent.  

 \begin{figure} [t]
\centering   % \vspace{-3mm} 
\includegraphics [width=0.5\textwidth] {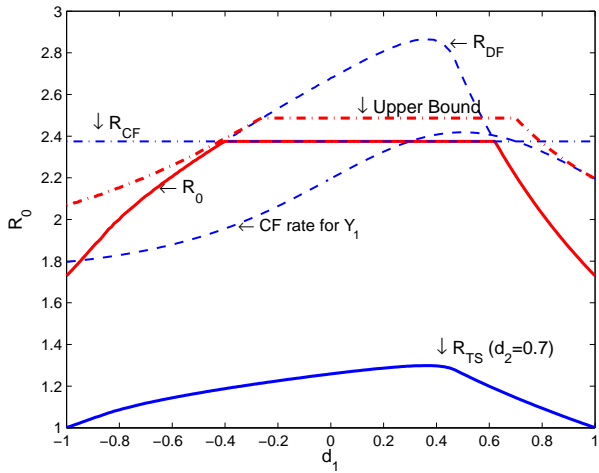}   
\caption{Common-rate of the Gaussian BRC with DF-CF strategies.}
\label{figCH1:4}
%\vspace{-4mm}
\end{figure}

Fig. \ref{figCH1:4} shows numerical evaluation of the common-rate $R_0$. All channel noises are set to the unit variance and $P=P_1=P_2=10$. The distance between $X$ and $(Y_1,Y_2)$ is one while $d_{z_1}=d_1$, $d_{z_1y_1}=1-d_1$, $d_{z_2}=d_2$, $d_{z_2y_2}=1-d_2$.  Relay 1 moves with $d_1\in\left[-1,1\right]$ and Fig. \ref{figCH1:4} presents rates as a function of $d_1$.  Whereas the position of relay 2 is assumed to be fixed to $d_2=0.7$ so $R_{\textrm{CF}}$ is a constant function of $d_1$, but $R_{\textrm{DF}}$ depends on $d_1$. For comparison, CF rate for destination $Y_1$ is also plotted which corresponds to the case where the first relay uses CF scheme.  This setting serves to compare the performances of coding respect to the relay position. We remark that one can achieve the minimum between CF and DF rates. These rates are also compared with a naive time-sharing strategy which consists on DF scheme $\tau\%$ of time and CF scheme  $(1-\tau)\%$ of time\footnote{Time-sharing in compound settings should 
not be confused with conventional time-sharing yielding a convex combination of rates.}. Time-sharing yields the following achievable rate 
$$
R_{\textrm{TS}}=\underset{0\leq\tau\leq1}{\max} \min\{\tau R_{\textrm{DF}},(1-\tau)R_{\textrm{CF}}\}.
$$ 
Notice that with the proposed coding scheme significant gains can be achieved when the relay is close to the source, i.e., DF scheme is more suitable, compared to the worst case.

\subsubsection{Composite RC} Consider now a composite model where the relay is collocated with the source with probability $p$ (refer to it as the first channel) and with the destination with probability $1-p$ (refer to it as the second channel). Therefore, DF scheme is the suitable strategy for the first channel while CF scheme performs better on the second one.  Define the expected rate as 
$$
R_{av}=R_0+pR_{1}+(1-p)R_{2},
$$ 
for any achievable triple of rates $(R_0,R_1,R_2)$.  Expected rate based on the proposed coding strategy is compared to conventional strategies. Alternative coding schemes for this scenario, where the encoder can simply invest on one coding scheme DF or CF, are possible. In fact, there are different ways to proceed:

\begin{itemize}
\item Send information via DF scheme at the best possible rate between both channels. Then the worst channel cannot decode and thus the expected rate becomes $p^{\max}_{\textrm{DF}} R^{\max}_{\textrm{DF}}$, where $R^{\max}_{\textrm{DF}}$ is the DF rate achieved on the best channel and $p^{\max}_{\textrm{DF}}$ is its probability.

\item Send information via the DF scheme at the rate of the worst (second) channel and hence both users can decode the information at rate $R^{\min}_{\textrm{DF}}$. Finally the next expected rate is achievable by investing on only one coding scheme   
$$
R_{av}^{\textrm{DF}}=\max\big\{p^{\max}_{\textrm{DF}} R^{\max}_{\textrm{DF}},R^{\min}_{\textrm{DF}}\big\}.
$$ 

\item By investing on CF scheme with the same arguments as before,  the expected rate writes as 
$$
R_{av}^{\textrm{CF}}=\max\big\{p^{\max}_{\textrm{CF}} R^{\max}_{\textrm{CF}},R^{\min}_{\textrm{CF}}\big\},
$$ 
with definitions of $(R^{\min}_{\textrm{CF}},R^{\max}_{\textrm{CF}},p^{\max}_{\textrm{CF}})$ similar to before. 
\end{itemize}

\begin{figure} [th!]
\centering  
\includegraphics [width=0.5\textwidth] {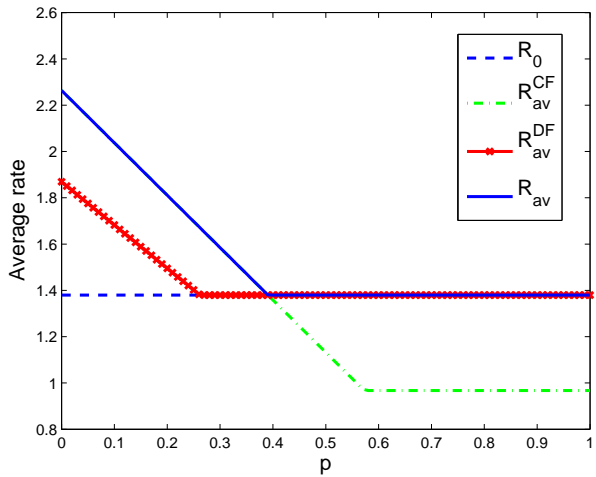}    
\caption{Expected rate for the composite Gaussian relay channel.}
\label{figCH1:3}
\end{figure}

Fig. \ref{figCH1:3} shows numerical evaluation of the average rate. All channel noises are set to have unit variance and $P=P_1=P_2=10$. The distance between $X$ and $(Y_1,Y_2)$ is $(3,1)$, while $d_{z_1}=1$, $d_{z_1y_1}=2$, $d_{z_2}=0.9$, $d_{z_2y_2}=0.1$. As one can see, the common-rate strategy provides a fixed rate all time which is always better than the worst case. However, at one corner full investment on one rate performs better because the high probability of one channel reduces the effect of the other. 
Based on the proposed coding scheme, i.e., using common and private messages, it is possible to cover all corner points performing better than both full investment strategies. It is worth to mention that the corner zone only requires private information of one channel. 

\subsection{The source is oblivious to the presence of relay}

We now focus on a scenario where the source  is unaware of the relay's presence. This arises, for example, when the informed relay decides by itself to help the destination whenever relaying is efficient (e.g. channel conditions are good enough). In this case, the BRC would have a single relay node. It is assumed here that there is no common information, then we set $X_2=\emptyset$ and $Z_2=Y_2$. The Gaussian BRC is defined as follows:
\begin{eqnarray}
Y_1&=&X+X_1+\mathpzc{N}_1,\nonumber  \\
Y_2&=&X+\mathpzc{N}_2, \nonumber \\   
Z_1&=&X+\widehat{\mathpzc{N}}_1.
\label{eqCH1:III1-18B}
\end{eqnarray}
As for the classical BC, by using superposition coding, we decompose $X$ as the sum of two independent descriptions such that $\esp\left\{X_A^2\right\}=\alpha P$ and $\esp\left\{X_B^2\right\}=\overline{\alpha} P$, where $\overline{\alpha}=1-\alpha$.  The codewords $(X_A,X_B)$ contain the information intended for destinations $Y_1$ and  $Y_2$, respectively. We use a DPC scheme applied to $X_B$ to cancel the interference $X_A$ while the relay signal is similarly chosen as in \cite{Cover1979}. Hence, the auxiliary RVs $(U_1,U_2)$ are set to
\begin{equation}
\begin{array} {l}
U_1=X_A=\tilde{X}_A+\sqrt{\displaystyle \frac{\overline{\beta}\alpha P}{P_1}}X_1, \\
U_2=X_B+\gamma X_A,
\end{array} 
\label{eqCH1:III1-20}
\end{equation}
where $\beta$ is the correlation coefficient between relay and source signals, and $\tilde{X}_A$ and $X_1$ are independent.

%From  the rate region obtained at the previous section, it is not difficult to obtain another scenario, which yields to the well-known scenario described in \cite{Kramer2005}. 

The distance between the relay and the source is denoted by $d_1$, between the relay and destination 1 by $1-d_1$ and between destination 2 and the source by $d_2$. The new Gaussian BRC writes as: 
$Z_1={X}/{d_1}+\widehat{\mathpzc{N}}_1$,  $Y_1=X+{X_1}/{(1-d_1)}+\mathpzc{N}_1$ and $Y_2={X}/{d_2}+\mathpzc{N}_2$. From the previous section, the achievable rates are  
\begin{align}
R_1^*&=\underset{\beta\in[0, 1]}{\max} \min\Big\{ \mathcal{C}\left(\frac{\displaystyle \alpha\beta P}{\overline\alpha P+d^2_1\widehat{N}_1}\right) ,   \nonumber\\
&\hspace{10mm} \mathcal{C}\left(\displaystyle\frac{\displaystyle \alpha P+\frac{P_1}{(1-d_1)^2}+\frac{2\sqrt{\overline\beta\alpha PP_1}}{\vert 1-d_1\vert}}{\overline\alpha P+{N}_1}\right)   \Big\}, \nonumber\\
R_2^*&=\mathcal{C}\left(\displaystyle\frac{\displaystyle \overline{\alpha}P}{d^2_2N_2}\right).
\label{eqCH1:III1-32}
\end{align}
Notice that since $(X_A,X_B)$ are  independent  then destination 1 sees $X_B$ as additional noise. 
%\begin{figure} [th]
%\centering    
%\includegraphics [width=0.7\textwidth] {DFDF.eps}
%\caption{Inner bound on the capacity of the Gaussian BRC.}
%\label{figCH1:III1-3}
%\end{figure}
The following outer-bound can be also derived for this channel:
%\begin{align*}
%R_1\leq& I(U,X_1;Y_1)\\
%R_1+R_2\leq& I(U;Z_1,Y_1\vert X_1)+I(X;Y_2\vert U,X_1)
%\end{align*}
%The calculation follows the same steps as the appendix \ref{proof-4} and we  do not present it here in detail. The final result is as follows: 
\begin{align}
R_1&\leq\underset{\beta\in[0, 1]}{\max} \min\Big\{ \mathcal{C}\left(\frac{\displaystyle \alpha\beta P}{\overline\alpha P+d^2_1\widehat{N}_1}+\frac{\displaystyle \alpha\beta P}{\overline\alpha P+{N}_1}\right), \nonumber  \\
&\hspace{10mm} \mathcal{C}\left(\displaystyle\frac{\displaystyle \alpha P+\frac{P_1}{(1-d_1)^2}+\frac{2\sqrt{\overline\beta\alpha PP_1}}{\vert 1-d_1\vert}}{\overline\alpha P+{N}_1}\right)   \Big\}, \nonumber\\
R_2&\leq \mathcal{C}\left(\displaystyle\frac{\displaystyle \overline{\alpha}P}{d^2_2N_2}\right).
\label{eqCH1:III1-32-1}
\end{align}
Note that if the relay channel is degraded, the bound in \eqref{eqCH1:III1-32-1} reduces to the rate  region in \eqref{eqCH1:III1-32} and thus we have the capacity of this channel according to Theorem \ref{thm:II5-3-1}. It can be seen that the broadcast strategy provides significant gains compare to the simple time-sharing scheme which consists in sharing over time the information for both destinations.   

%
%This is a slightly different result from the theorem \ref{thm:II5-3}. Here only the first relay channel is degraded and moreover the second destination 
% 

%Fig. \ref{figCH1:III1-3} shows a numerical evaluation of these rates. All channel noises are set to the unit variance and $P=P_1=10$. We assume that destination 2, which does not possess a relay, is the closest to the source $d_2=0.4$, while the distance between the relay and the source is set to $d_1=1.4$. 

%%*********************************************************************************************************
%%IV. Summary
%%*********************************************************************************************************
\section{Summary and Discussion}
In this paper we investigated cooperative strategies for simultaneous and broadcast relay channels. Several cooperative schemes have been proposed and the corresponding inner and outer-bounds on the capacity region were derived. The focus was on the simultaneous relay channel (SRC) with two relay channels, where the central idea is this problem can be turned into the broadcast relay channel (BRC). Then each branch of this new channel represents one of the possible relay channels. In this setting, the source wishes to send common information to guarantee a minimum amount of information regardless of the channel and additional private information to each of the destinations.

Depending on the nature of the channels involved, it is well-known that the best way to cover the information from the relays tothe  destinations is not the same. Based on the best known cooperative strategies, namely, \emph{Decode-and-Forward} (DF) and \emph{Compress-and-Forward} (CF), achievable rate regions for three different scenarios of interest have been derived. These are summarized as follows: (i) both relay nodes use DF scheme, (ii) one relay uses CF scheme while the other uses DF scheme, and (iii) both relay nodes use CF scheme. In particular, for region (ii) it is shown that \emph{superposition coding} can work with CF scheme without incurring performance losses.  These inner bounds are shown to be tight for some specific scenarios, yielding capacity results for the semi-degraded BRC with common relay (BRC-CR) and two classes of Gaussian degraded BRC-CRs. Whereas the bounds seem to be not tight for the general degraded BRC-CR. An outer-bound on the capacity  of the general BRC was also derived. 
One should emphasize  that when the relays are not present this bound reduces to the best known outer-bound for general broadcast channels (referred to as $UVW$-outer-bound).  Similarly, when only one relay channel is present at once this bound reduces to the cut-set bound for the general relay channel. 

Finally, application examples for Gaussian channels have been studied and achievable rates were computed for all inner bounds. Special attention was given to two models of practical importance for opportunistic and oblivious cooperation in wireless networks. The first model refers to the situation where the source must be oblivious to the cooperative strategy adopted by the relay (e.g. DF or CF scheme). The second one models the situation where the  source must be oblivious to the presence of a nearby relay which may help the communication between source and destination. Numerical results evaluate the gains that can be achieved with the proposed coding strategies compared to naive approaches.

As future work, it would be interesting to exploit these results in the context of composite relay networks with random parameters (e.g. fading, spatial position of nodes, etc.) where performance is measured in terms of capacity versus outage notions. Of particular interest is the investigation of novel rate regions based on (linear) structured coding, e.g., lattice codes \cite{Nazer2011}, which in some cases can improve on random coding.

%%*********************************************************************************************************
%%V. appendices
%%*********************************************************************************************************

\appendices
%\section{Proof of the Theorems}
%%*********************************************************************************************************
%%V.1 Proof DF DF
%%*********************************************************************************************************
\section{Sketch of Proof of Theorem \ref{thm:II2-1}} 
\label{proof-1CH1}
To prove the theorem, first split the private information $W_b$ into non-negative indices $(S_{0b},S_b,S_{b+2})$ with $b=\{1,2\}$. Then, merge the common information $W_0$ with a part of private information $(S_{01},S_{02})$ into a single message, as shown in Fig. \ref{figCH1:II-3-1}. 
Hence we obtain that $R_b= S_{b+2}+S_{b}+S_{0b}$. For notation simplicity, we denote $\underline{u}=u_1^n$ for every $u$. We now consider the main steps for codebook generation, encoding and decoding procedures.\vspace{1mm} \\
\textit{Code generation:}
\begin{enumerate}[(i)]
	\item Generate $2^{nT_0} $ i.i.d. sequences $\underline{v}_0$ each with PD   
	$$
	P_{V_0}(\underline{v}_0)=\prod_{j=1}^n p_{V_0}(v_{0j}),
	$$
	and index them as $\underline{v}_0(r_0)$ with $r_0= \left[1: 2^{nT_0} \right]$.
	\item 
	For each $\underline{v}_0(r_0)$, generate $ 2^{nT_0} $ i.i.d. sequences   $\underline{u}_0$ each with PD  
	$$
	P_{U_0| V_0}(\underline{u}_0\vert \underline{v}_0(r_0))=\prod_{j=1}^np_{U_0\vert V_0}(u_{0j}\vert v_{0j}(r_0)),
	$$
	and index them as $\underline{u}_0(r_0,t_0)$ with $t_0= \left[1:2^{nT_0}\right]$. 
	\item
	For $b\in\{1,2\}$ and each $\underline{v}_0(r_0)$, generate $2^{nT_b}$ i.i.d. sequences $\underline{x}_b$ each with PD 
	$$
	P_{X_b| V_0}(\underline{x}_b\vert \underline{v}_0(r_0))=\prod_{j=1}^np_{X_b\vert V_0}(x_{bj}\vert v_{0j}(r_0)),
	$$
	and index them as $\underline{x}_b(r_0,r_b)$ with $r_b= \left[1:2^{nT_b}\right]$.  
\item
Partition the set $\big\{1,\dots,2^{nT_0}\big\}$ into $2^{n(R_0+S_{01}+S_{02})}$ cells (similarly to \cite{Marton1979}) and label them as $S_{w_0,s_{01},s_{02}}$. In each cell there are $2^{n(T_0-R_0-S_{01}-S_{02})}$ elements. 	
\item
For each $b = \{1,2\}$ and every pair $\big(\underline{u}_0(r_{0},t_{0}),$ $\underline{x}_b(r_{0},r_{b}) \big)$ chosen in the bin $(w_{0},s_{01},s_{02})$, generate $2^{nT_b}$ i.i.d. sequences $\underline{u}_{b}$ each with PD
\begin{align*}
&P_{U_b|U_0X_bV_0}\big (\underline{u}_{b}\vert \underline{u}_0(r_{0},t_{0}),\underline{x}_b(r_{0},r_{b}), \underline{v}_0(r_{0}) \big) =\\
&\prod_{j=1}^np_{U_b\vert U_0X_bV_0}(u_{bj}\vert u_{0j}(r_{0},t_{0}),x_{bj}(r_{0},r_{b}),v_{0j}(r_{0})),
\end{align*}
and index them as $\underline{u}_b(r_{0},t_{0},r_{b},t_{b})$ with $t_b= \left[1:2^{nT_b}\right]$.  
\item
For $b=\{1,2\}$, partition the set $\big\{1,\ldots,2^{nT_b}\big\}$ into $2^{nS_b}$ cells and label them as $S_{s_b}$. In each cell there are $2^{n(T_b-S_b)}$ elements.  
\item
%$(\underline{u}_3,\underline{u}_4)$: 
For each $b=\{1,2\}$ and every pair of sequences $\big( \underline{u}_1(r_{0},t_{0},r_{1},t_{1}),$ $\underline{u}_2(r_{0},t_{0},r_{2},t_{2}) \big)$ chosen in the bin $(s_{1},s_{2})$, generate $2^{nT_{b+2}}$ i.i.d. sequences $\underline{u}_{b+2}$ each with PD
\begin{align*}
P_{U_{b+2}| U_b }(&\underline{u}_{b+2}\vert \underline{u}_b(r_{0},t_{0},r_{b},t_{b}))= \\
&\prod_{j=1}^np_{U_{b+2}\vert U_b}(u_{(b+2)j}\vert u_{bj}(r_{0},t_{0},r_{b},t_{b})). 
\end{align*}
Index them as $\underline{u}_{b+2}(r_{0},t_{0},r_{b},t_{b},t_{b+2})$ with $t_{b+2}\in \left[1,2^{nT_{b+2}}\right]$.
\item 
For $b=\{1,2\}$, partition the set $\big\{1,\ldots,2^{nT_{b+2}}\big\}$ into $2^{nS_{b+2}}$ cells and label them as $S_{s_{b+2}}$. In each cell there are $2^{n(T_{b+2}-S_{b+2})}$ elements. 	
\item 
Finally, use a deterministic function for generating $\underline{x}$ as $f\left(\underline{u}_3,\underline{u}_4\right)$ indexed by \\ $\underline{x}(r_{0},t_{0},r_{1},r_2,t_{1},t_2,t_{3},t_4)$.
\end{enumerate}%\vspace{2mm}
%%%%%%%%%%%%%%%%%%%%%%%%%%%%%%%%%%%%%%%%%%%%%%%%%%%%%%%%%%%%%%%%%%%%%%%%%Encoding
%%%%%%%%%%%%%%%%%%%%%%%%%%%%%%%%%%%%%%%%%%%%%%%%%%%%%%%%%%%%%%%%%%%%%%%%%%%%%%%%%%%%%%%%%%%%%%%%%%%%
\textit{Encoding Part:} Transmission is done over $B+1$ block where the encoding in block $i$ is as follows:
\begin{enumerate}[(i)]
\item
First, reorganize the current message $(w_{0i},w_{1i},w_{2i})$ into $(w_{0i},s_{01i},s_{02i},s_{1i},s_{2i},s_{3i},s_{4i})$.
\item
Then for each $b=\{1,2\}$, relay $b$ already knows about the index $(t_{0(i-1)},t_{b(i-1)})$, so it sends $\underline{x}_b\big(t_{0(i-1)},t_{b(i-1)}\big)$.
\item
For each $\underline{v}_0(t_{0(i-1)})$, the encoder searches for an index $t_{0i}$ at the cell $S_{w_{0i},s_{01i},s_{02i}}$ such that $\underline{u}_0\big(t_{0(i-1)},t_{0i}\big)$ is jointly typical with $\big(\underline{x}_{1}(t_{0(i-1)}, t_{1(i-1)}),\underline{x}_{2}(t_{0(i-1)}, t_{2(i-1)}),\underline{v}_0(t_{0(i-1)})\big)$. The success of this step requires that \cite{Marton1979}
\begin{equation}
T_0-R_0-S_{01}-S_{02}\geq I(U_0;X_1,X_2\vert V_0). 
\label{eqCH1:V1-1}
\end{equation}
\item \label{FirstMartonV1}
For each $b=\{1,2\}$ and every cell $S_{s_{bi}}$, define $\mathcal{L}_b$ as the set of all sequences  $\underline{u}_b\big(t_{0(i-1)},t_{0i},t_{b(i-1)},t_{bi}\big)$ for $t_{bi}\in{S}_{s_{bi}}$ which are jointly typical with 
\begin{align*}
\big( \underline{x}_{\overline{b}}(t_{0(i-1)}, &t_{\overline{b}(i-1)}),\underline{v}_0(t_{0(i-1)}),\\ &\underline{u}_0(t_{0(i-1)},t_{0i}),\underline{x}_b(t_{0(i-1)},t_{b(i-1)})\big)
\end{align*}
where $\overline{b}=\{1,2\}\setminus \{b\}$. In order to create $\mathcal{L}_b$, we look for the $\underline{u}_b$-index inside the cell $S_{s_{bi}}$ and find $\underline{u}_b$ such that it belongs to the set of $\epsilon$-typical $n$-sequences $\textsl{A}^n_\epsilon(V_0U_0X_1X_2U_b)$.
\item  \label{SecondMartonV1}
Look for a pair $(\underline{u}_1\in\mathcal{L}_1,\underline{u}_2\in\mathcal{L}_2)$ such that $\big(\underline{u}_1(t_{0(i-1)},t_{0i},t_{1(i-1)},t_{1i}),\underline{u}_2(t_{0(i-1)},t_{0i},t_{2(i-1)},t_{2i})\big) $ are jointly typical given the RVs $\big(\underline{v}_0(t_{0(i-1)}), \underline{x}_2(t_{0(i-1)},t_{2(i-1)}),$ $\underline{x}_1(t_{0(i-1)},t_{1(i-1)}),\underline{u}_0(t_{0(i-1)},t_{0i})\big)$. The success of coding steps (\ref{FirstMartonV1}) and (\ref{SecondMartonV1}) requires 
\begin{align}
T_b-S_b & \geq I(U_b;X_{\overline{b}}\vert X_b,U_0,V_0), \nonumber\\
T_1+T_2-S_1-S_2 & \geq I(U_1;X_2\vert X_1,U_0,V_0)\nonumber\\
& + I(U_2;X_1\vert X_2,U_0,V_0)\nonumber\\
&+I(U_2;U_1\vert X_1,X_2,U_0,V_0). \label{eqCH1:III-2}
\end{align}
Notice that the first inequality in the above expression, for $b= \{1,2\}$, guarantees the existence of non-empty sets $(\mathcal{L}_1,\mathcal{L}_2)$, and the last one is for the step (\ref{FirstMartonV1}).
\item
The encoder searches for indices $t_{3i}\in S_{s_{3i}}$ and $t_{4i}\in S_{s_{4i}}$ such that $\underline{u}_3\big(t_{0(i-1)},t_{0i},t_{1(i-1)},t_{1i},t_{3i}\big)$ and $\underline{u}_4\big(t_{0(i-1)},t_{0i},t_{2(i-1)},t_{2i},t_{4i}\big)$ are jointly typical given each typical pair of $\underline{u}_1(t_{0(i-1)},t_{0i},t_{1(i-1)},t_{1i})$ and $\underline{u}_2(t_{0(i-1)},t_{0i},t_{2(i-1)},t_{2i})$.
The success of this encoding step requires
\begin{equation}
\!\!\!\! \!\!\!\! T_3+T_4-S_{3}-S_{4}\geq I(U_3;U_4\vert U_1,U_2,X_1,X_2,U_0,V_0).  
\label{eqCH1:V1-3}
\end{equation}
\item
Once the encoder found $(t_{0i},t_{1i},t_{2i},t_{3i},t_{4i})$ (based on the code generation) corresponding to $(w_{0i},\\s_{01i},s_{02i},s_{1i},s_{2i},s_{3i},s_{4i})$, it transmits $\underline{x}(r_{0(i-1)},t_{0i},r_{1(i-1)},r_{2(i-1)},t_{1i},t_{2i},t_{3i},t_{4i})$. $t_{0i}$ carries the common message after bit recombination and Marton coding. The indices $t_{1i},t_{3i}$ and $t_{2i},t_{4i}$ are, respectively, private information for destinations $Y_1$ and $Y_2$. Whereas indices $t_{3i}$ and $t_{4i}$, corresponding to partial encoding, are directly transmitted  to the intended destinations. 
% which is randomly drawn from $P_{U_1U_2U_3U_4}(\underline{u}_1,\underline{u}_2,\underline{u}_3,\underline{u}_4)$.
\end{enumerate}%\vspace{2mm}
%%%%%%%%%%%%%%%%%%%%%%%%%%%%%%%%%%%%%%%%%%%%%%%%%%%%%%%%%%%%%%%%%%%%%%%%%%%%%%%%%%%%%%%%%%%%%%%%%%%%
%%%%%%%%%%%%%%Decoding
%%%%%%%%%%%%%%%%%%%%%%%%%%%%%%%%%%%%%%%%%%%%%%%%%%%%%%%%%%%%%%%%%%%%%%%%%%%%%%%%%%%%%%%%%%%%%%%%%%%%
\textit{Decoding Part:} 
In block $i$, in order to decode messages relays assume that all messages up to block $i-1$ have been correctly decoded and then decode the current messages in the same block. The destinations use backward decoding and assume that all messages until block $i+1$ have been  correctly decoded. 
\begin{enumerate}[(i)]
\item
First for $b=\{1,2\}$, the relay $b$ after receiving $\underline{z}_{bi}$ tries to decode $(t_{0i},t_{bi})$. The relay is aware of $(V_0,X_b)$ because it is supposed to know about $(t_{0(i-1)},t_{b(i-1)})$. The relay $b$ declares that the pair $(t_{0i},t_{bi})$ is sent if the following conditions are simultaneously satisfied:
\begin{enumerate} 
\item
$\underline{u}_0(t_{0(i-1)},t_{0i})$ is jointly typical with $\big(\underline{z}_{bi}$, $\underline{v}_0(t_{0(i-1)}),$ $\underline{x}_b (t_{0(i-1)},t_{b(i-1)}) \big)$. 
\item
$\underline{u}_b(t_{0(i-1)},t_{0i},t_{b(i-1)},t_{bi})$ is jointly typical with $\big(\underline{z}_{bi}$, $\underline{v}_0(t_{0(i-1)}),$ $\underline{x}_b(t_{0(i-1)},t_{b(i-1)}) \big)$. 
\end{enumerate} 
Notice that $\underline{u}_0$ has been generated independent of $\underline{x}_b$ and hence $\underline{x}_b$ does not appear 
in the given part of mutual information. This is an important issue that may increase the region. Constraints for reliable decoding are:
\begin{align}
T_b & < I(U_b;Z_b\vert U_0,V_0,X_b), \label{eqCH1:V1-4A} \\
\!\!\!\! T_b+T_0 & < I(U_b;Z_b\vert U_0,V_0,X_b)+I(U_0;Z_b,X_b\vert V_0). \!\!\!\! \label{eqCH1:V1-4B}
\end{align}
%The last inequality can be simplified using expression \eqref{eqCH1:III-1}.  After removing the common term of information, i.e., $I(U_0;X_b\vert V_0)$, we obtain the following inequality:
%\begin{align}  % 
%T_b+R_0+S_{01}+S_{02} & < I(U_0,U_b;Z_b\vert V_0,X_b) \nonumber \\
%& -I(U_0;X_{\bar{b}}\vert X_b,V_0), \label{eqCH1:III-5}
%\end{align}
\begin{remark}
The intuition behind expressions \eqref{eqCH1:V1-4A} and \eqref{eqCH1:V1-4B} is as follows. Since the relay knows $\underline{x}_{b(i-1)}$ we are indeed decreasing the cardinality of the set of possible $\underline{u}_0$,  which without additional knowledge is $2^{nT_0}$. The new set of possible $(\underline{u}_0$, $\mathcal{L}_{X_b})$ can be defined as all $\underline{u}_0$ jointly typical with $\underline{x}_{b(i-1)}$. It can be shown \cite{elgamal-vandermeulen-1981} that $\esp[\left\|\mathcal{L}_{X_b}\right\|]=2^{n [T_0-I(U_0;X_b\vert V_0) ] }$, which proves our claim on the reduction of cardinality.  
One can see that after simplification of expression \eqref{eqCH1:V1-4B} by using \eqref{eqCH1:V1-1}, $I(U_0;Z_b,X_b\vert V_0)$ is removed and the final bound reduces to $I(U_0,U_b;Z_b\vert V_0,X_b)$.
\end{remark}
\item 
For each $b\in\{1,2\}$ destination $b$, after receiving $\underline{y}_{b(i+1)}$, tries to decode  the relay-forwarded information $(t_{0i},t_{bi})$, knowing $(t_{0(i+1)},t_{b(i+1)})$. It also tries to decode the direct information $t_{(b+2)(i+1)}$. Backward decoding is used to  decode indices $(t_{0i},t_{bi})$. The decoder declares that $(t_{0i},t_{bi},t_{(b+2)(i+1)})$ is sent if the following constraints are simultaneously satisfied:
\begin{enumerate} 
\item 
$\big(\underline{v}_0(t_{0i}),\underline{u}_0(t_{0i},t_{0(i+1)}),\underline{y}_{b(i+1)}\big)$  are jointly typical,
\item \label{BackwardDecodingV1}
$\big(\underline{x}_b(t_{0(i)},t_{b(i)}), \underline{v}_0(t_{0i}), \underline{u}_0(t_{0i},t_{0(i+1)})  \big)$  and $\underline{y}_{b(i+1)}$ are jointly typical,
\item
$ \big( \underline{u}_b(t_{0i},t_{0(i+1)},t_{bi},t_{b(i+1)}),\underline{u}_{b+2}(t_{0i},t_{0(i+1)},t_{bi},$ $t_{b(i+1)}, t_{b(i+1)})  \big)$ and  $\big( \underline{y}_{b(i+1)}, \underline{v}_0(t_{0i})$,\\ $\underline{u}_0(t_{0i},t_{0(i+1)}),$ $\underline{x}_b\big(t_{0(i)},$ $ t_{b(i)}\big)\big) $ are jointly typical. 
\end{enumerate}
Notice that for decoding step (\ref{BackwardDecodingV1}), the destination knows $t_{0(i+1)}$ which has been chosen such that $(\underline{u}_0,\underline{x}_b)$ are jointly typical and this information contributes to decrease the cardinality of all possible $\underline{x}_b$. This is similarly to what happened with decoding at relay. Hence $U_0$ in step (\ref{BackwardDecodingV1}) does not  appear in the given part of mutual information. From this we have that the main constraints for successful decoding are as follows:
\begin{align}
T_{b+2} & < I(U_{b+2};Y_b\vert U_0,V_0,X_b,U_b),\label{eqCH1:V1-6A}\\
T_{b+2}&+T_b  < I(U_{b+2},U_b,X_b;Y_b\vert U_0,V_0), \label{eqCH1:V1-6B}\\
%\end{align}
%\begin{align}
T_{b+2}&+T_b+T_0   < I(V_0,U_0;Y_b)+I(X_b;Y_b,U_0\vert V_0) \nonumber\\
&+I(U_{b+2},U_b;Y_b\vert U_0,V_0,X_b). \label{eqCH1:V1-6C}
\end{align}
%Note that each term in \eqref{eqCH1:III-6C} corresponds to the contribution of the decoding constraints. 
Observe that $U_0$ increases the bound in \eqref{eqCH1:V1-6B}. Similarly,  by using \eqref{eqCH1:V1-1} and after removing the common term $I(U_0;X_b\vert V_0)$, one can simplify the bound in \eqref{eqCH1:V1-6C} to $I(U_{b+2},U_b,X_b,V_0,U_0;Y_b)$.
%, we obtain the following constraints with $\overline{b}=\{1,2\}\setminus\{b\}$:
%\begin{align}
%T_{b+2} & < I(U_{b+2};Y_b\vert U_0,V_0,X_b,U_b),\nonumber \\
%T_{b+2}+T_b & < I(U_{b+2},U_b,X_b;Y_b\vert U_0,V_0), \nonumber\\
%T_{b+2}+T_b&+R_0+S_{01}+S_{02}  < -I(U_0;X_{\overline{b}}\vert V_0,X_b)\nonumber \\
%& +I(U_{b+2},U_b,X_b,V_0,U_0;Y_b). \label{eqCH1:III-7}
%\end{align}
\item Theorem \ref{thm:II2-1} follows by applying Fourier-Motzkin elimination to  
%\eqref{eqCH1:III-7}, \eqref{eqCH1:III-5}, and 
expressions \eqref{eqCH1:V1-1}-\eqref{eqCH1:V1-6C} and using the non-negativity property of the rates, which concludes the proof.% similar to \cite{Nair2009}.
\end{enumerate}
%%*********************************************************************************************************
%%V.2 Proof DF CF
%%*********************************************************************************************************
\section{Sketch of Proof of Theorem \ref{thm:II3-1}} 
\label{proof-2}
Reorganize first private messages $w_i$, $i= \{1,2\}$ into $(s^\prime_i,{s}_i)$ with non-negative rates $(S^\prime_i,S_i)$ where $R_i=S'_i+S_i$. Merge $(s^\prime_1,s^\prime_2,w_0)$ to one message ${s}_0$ with rate $S_0=R_0+S^\prime_1+S^\prime_2$. For notation simplicity, we denote $\underline{u}=u_1^n$ for every $u$. We next consider the main steps for codebook generation, encoding and decoding procedures.\\ \vspace{2mm}
\textit{Code generation:}
\begin{enumerate}[(i)]
	\item Generate $2^{nS_0}$ i.i.d. sequences $\underline{v}_0$ with PD 
	$$
	P_{V_0}(\underline{v}_0)=\prod_{j=1}^np_{V_0}(v_{0j}),
	$$ 
	and index them as $\underline{v}_0(r_0)$ with $r_0=\left[1:2^{nS_0}\right]$.
	\item 
	For each $\underline{v}_0(r_0)$, generate $2^{nS_0}$ i.i.d. sequences $\underline{u}_0$ with PD 
	$$
	P_{U_0|V_0}(\underline{u}_0\vert \underline{v}_0(r_0))=  \prod_{j=1}^np_{U_0\vert V_0}(u_{0j}\vert v_{0j}(r_0)),
	$$ 
        and index them as $\underline{u}_0(r_0,s_0)$ with $s_0= \left[1:2^{nS_0}\right]$.
	\item
	For each $\underline{v}_0(r_0)$, generate $2^{nT_1}$ i.i.d. sequences $\underline{x}_1$ with PD 
	$$
	P_{X_1|V_0}(\underline{x}_1\vert \underline{v}_0(r_0))= \\ \prod_{j=1}^n p_{X_1\vert V_0}(x_{1j}\vert v_{0j}(r_0)),
	$$
	and index them as $\underline{x}_1(r_0,r_1)$ with $r_1= \left[1:2^{nT_1}\right]$.
\item 
	Generate  $2^{n{R}_{x_2}}$ i.i.d. sequences $\underline{x}_2$ with PD 
	$$
	P_{X_2}(\underline{x}_2)=\prod_{j=1}^np_{X_2}(x_{2j}),
	$$ 
	and index them as  $\underline{x}_2(r_2)$ with $r_2=\left[1:2^{n{R}_{x_2}}\right]$.
	\item
	For each $\underline{x}_2(r_2)$ generate $2^{n\hat{R}_2}$ i.i.d. sequences  $\underline{\hat{z}}_2$ with PD 
	$$
	P_{\hat{Z}_2|X_2}(\underline{\hat{z}}_2\vert \underline{x}_2(r_2))= \prod_{j=1}^np_{\hat{Z}_2\vert X_2}(\hat{z}_{2j}\vert x_{2j}(r_2)),
	$$ 
	and index them as $\underline{\hat{z}}_2(r_2,\hat{s})$, where $\hat{s}= \left[1:2^{n\hat{R}_2}\right]$.
\item   
	Partition the set $\big\{1,\dots,2^{n\hat{R}_2}\big\}$ into $2^{nR_2}$ cells and label them as $S_{r_2}$. In each cell there are $2^{n(\hat{R}_2-R_2)}$ elements. 
	\item
	For each pair $\big(\underline{u}_0(r_{0},s_{0}),\underline{x}_1(r_{0},r_{1})\big)$, generate $2^{nT_1}$ i.i.d. sequences $\underline{u}_{1}$ with PD
\begin{align*}
&P_{U_1|U_0X_1V_0}(\underline{u}_{1}\vert \underline{u}_0(r_{0},s_{0}),\underline{x}_1(r_{0},r_{1}),\underline{v}_0(r_{0}))=\\
&\prod_{j=1}^np_{U_1\vert U_0V_0X_1}(u_{1j}\vert u_{0j}(r_{0},s_{0}),x_{1j}(r_{0},r_{1}),v_{0j}(r_{0})),
\end{align*} 
and index them as $\underline{u}_1(r_{0},s_{0},r_{1},t_{1})$, where $t_1= \left[1:2^{nT_1}\right]$.
\item
	For each $\underline{u}_0(r_{0},s_{0})$, generate $2^{nT_2}$ i.i.d. sequences $\underline{u}_{2}$  with PD  
\begin{align*}
	P_{U_2| U_0V_0}(&\underline{u}_{2}\vert  \underline{u}_0(r_{0}, s_{0}),\underline{v}_0(r_{0}))\\
	&= \prod_{j=1}^np_{U_2\vert U_0V_0}(u_{2j}\vert u_{0j}(r_{0},s_{0}),v_{0j}(r_{0})), 
\end{align*} 
	and index them as $\underline{u}_2(r_{0},s_{0},t_{2})$, where $t_2= \left[1:2^{nT_2}\right]$.
	\item
For $b=\{1,2\}$, partition the set $\big\{1,\dots,2^{nT_b}\big\}$ into $2^{nS_b}$ subsets and label them as $S_{s_b}$. In each subset, there are $2^{n(T_b-S_b)}$ elements.
\item
Finally, use a deterministic function for generating $\underline{x}$ as $f\left(\underline{u}_1,\underline{u}_2\right)$ indexed by $\underline{x}(r_{0},s_{0},r_{1},t_{1},t_{2})$.
\end{enumerate}\vspace{2mm}
%%%%%%%%%%%%%%%%%%%%%%%%%%%%%%%%%%%%%%%%%%%%%%%Encoding Part
\textit{Encoding part:} In block $i$, the source wants to send messages $(w_{0i},w_{1i},w_{2i})$ by reorganizing them into $(s_{0i},s_{1i},s_{2i})$. Encoding steps are as follows:
\begin{enumerate}[(i)]
\item
DF relay  knows $(s_{0(i-1)},t_{1(i-1)})$ so it sends $\underline{x}_1\big(s_{0(i-1)},t_{1(i-1)}\big)$.
\item
CF relay knows from the previous block that $\hat{s}_{i-1} \in S_{r_{2i}}$ and it sends $\underline{x}_2(r_{2i})$.
\item \label{FirstMartonV2}
Then for each subset $S_{s_{2i}}$, create the set $\mathcal{L}$ consisting of those index $t_{2i}$ such that $t_{2i}\in S_{s_{2i}}$, and  $\underline{u}_2\big(s_{0(i-1)},s_{0i},t_{2i}\big)$ is jointly typical with $\underline{x}_{1}\big(s_{0(i-1)}, t_{1(i-1)}\big),\underline{v}_0\big(s_{0(i-1)}\big), \underline{u}_0\big(s_{0(i-1)},s_{0i}\big)$. 
\item  \label{SecondMartonV2}
Then look for $t_{1i}\in S_{s_{1i}}$ and $t_{2i}\in\mathcal{L}$ such that $\big(\underline{u}_1(s_{0(i-1)},s_{0i},t_{1(i-1)},t_{1i})$,$ \underline{u}_2(s_{0(i-1)},s_{0i},t_{2i})\big)$ are jointly typical given the codewords $\underline{v}_0(s_{0(i-1)}), \underline{x}_1(s_{0(i-1)},t_{1(i-1)}), $ and with $\underline{u}_0(s_{0(i-1)},s_{0i})$. The constraints for successful encoding of steps (\ref{FirstMartonV2}) and (\ref{SecondMartonV2}) are: 
\begin{align}
T_2-S_2 & \geq I(U_2;X_1\vert U_0,V_0), \label{eqCH1:V2-1A}\\
T_1+T_2-S_1-S_2 & \geq I(U_2;U_1,X_1\vert U_0,V_0). \label{eqCH1:V2-1B}
\end{align}
The first inequality guarantees the existence of non-empty sets $\mathcal{L}$. 
\item
From $(s_{0i},s_{1i},s_{2i})$, the source finds $(t_{1i},t_{2i})$ and sends $\underline{x}(s_{0(i-1)},s_{0i},t_{1(i-1)},t_{1i},t_{2i})$. 
\end{enumerate}%\vspace{2mm}
%%%%%%%%%%%%%%%%%%%%%%%%%%%%%%%%%%%%%%%%%%%%%%%Decoding Part
\textit{Decoding Part:} After the transmission of block $i+1$, DF relay starts to decode the messages of block $i+1$ with the assumption that all messages up to block $i$ have been correctly decoded. Destination 1 waits until the last block and uses backward decoding (similarly to \cite{Kramer2005}). The second destination first decodes $\hat{Z}_2$ and then uses it with $Y_2$ to decode the messages while the second relay tries to find $\hat{Z}_2$ in current block. 
\begin{enumerate}[(i)]
\item
DF relay  tries to decode $(s_{0(i+1)},t_{1(i+1)})$ and the conditions for reliable decoding are:
\begin{align}
T_1+S_0 & < I(U_0,U_1;Z_1\vert X_1V_0), \label{eqCH1:V2-2A}    \\
T_1 & < I(U_1;Z_1\vert U_0,V_0,X_1). 
\label{eqCH1:V2-2B}
\end{align}
\item
Destination $1$ tries to decode $(s_{0i},t_{1i})$ subject to
\begin{align}
T_1+S_0 & < I(X_1,V_0,U_0,U_1;Y_1),  \label{eqCH1:V2-3A} \\
T_1 & < I(U_1,X_1;Y_1\vert U_0,V_0). 
\label{eqCH1:V2-3B}
\end{align}
\item 
CF relay searches for $\hat{s}_{i}$ after receiving $\underline{z}_2(i)$ such that $\big(\underline{x}_{2}(r_{2i}),\underline{z}_{2}(i),\underline{\hat{z}}_{2}(\hat{s}_{i},r_{2i})\big)$ are jointly typical subject to
\begin{equation}
\hat{R}_2 \geq I(Z_2;\hat{Z}_2|X_2).
\label{eqCH1:V2-4}
\end{equation}
\item 
Destination $2$ searches for $r_{2(i+1)}$ such that $\big(\underline{y}_2(i+1),\underline{x}_{2}(r_{2(i+1)})\big)$ is  jointly typical. Then it finds  $\hat{s}_{i}$ such that $\hat{s}_{i}\in S_{r_{2(i+1)}}$ and $\big(\underline{\hat{z}}_{2}(\hat{s}_{i},r_{2i}),\underline{y}_2(i),\underline{x}_{2}(r_{2i})\big)$ is jointly typical. Conditions for reliable decoding are:
\begin{align}
R_{x_2} & \leq I(X_2;Y_2),\\
\hat{R}_2 & \leq R_{x_2}+I(\hat{Z}_2;Y_2|X_2).
\label{eqCH1:V2-5}
\end{align}
\item
Decoding of CF destination in block $i$ is done with the assumption of correct decoding of $(s_{0l},t_{2l})$ for $l\leq i-1$. The  pair $(s_{0i},t_{2i})$ is decoded as messages such that $(\underline{v_0}(s_{0(i-1)}),\underline{u_0}(s_{0(i-1)},s_{0i}),\underline{u_2}(s_{0(i-1)}\\,s_{0i},t_{2i}),\underline{y}_{2}(i),\underline{\hat{z}}_2(\hat{s}_i,,r_{2i}),\underline{x}_{2}(r_{2i}))$ and $(\underline{v}_0(s_{0i}),\underline{y}_{2}(i+1),\underline{\hat{z}}_2(\hat{s}_{i+1},r_{2(i+1)}),\underline{x}_{2}(r_{2(i+1)}))$ are all  jointly typical. This leads to the next constraints
\begin{align}
S_0+T_2 & \leq I(V_0,U_0,U_2;Y_2,\hat{Z}_2\vert X_2),\label{eqCH1:III-6A}\\
T_2 & \leq I(U_2;Y_2\hat{Z}_2\vert V_0,U_0,X_2). 
\label{eqCH1:V2-6B}
\end{align}
It is interesting to remark that regular encoding allows us to use the same code for DF and CF relays while keeping the same final CF rate.                                      
\end{enumerate}
After decoding of $(s_{0i},s_{1i},s_{2i})$ at destinations, the original messages $(w_{0i},w_{1i},w_{2i})$ can be extracted. Thus it can be shown  that the rate region in Theorem \ref{thm:II3-1} follows by applying Fourier-Motzkin elimination and form \eqref{eqCH1:V2-1A}-\eqref{eqCH1:V2-6B}, the equalities between the original and reorganized rates and the fact that all rates are positive. Similarly to \cite{Cover1979}, the necessary condition $I(X_2;Y_2)\geq I(Z_2;\hat{Z}_2\vert X_2,Y_2)$ follows from \eqref{eqCH1:V2-4} and \eqref{eqCH1:V2-5}.

%%*********************************************************************************************************
%%V.3 Proof CF CF
%%*********************************************************************************************************
\section{Sketch of Proof of Theorem \ref{thm:II4-1}} 
\label{proof-3}
Reorganize first private messages $w_i$, $i= \{1,2\}$ into $(s^\prime_i,{s}_i)$ with non-negative rates $(S^\prime_i,S_i)$ where $R_i=S^\prime_i+S_i$. Merge $(s^\prime_1,s^\prime_2,w_0)$ to one message ${s}_0$ with rate $S_0=R_0+S^\prime_1+S^\prime_2$. For notation simplicity, we denote $\underline{u}=u_1^n$ for every $u$. \vspace{2mm}\\
\textit{Code generation:}
\begin{enumerate}[(i)]
	\item Generate $2^{nS_0}$ i.i.d. sequences $\underline{u}_0$ with PD
	$$
	P_{U_0}(\underline{u}_0)=\prod_{j=1}^np_{U_0}(u_{0j}),
	$$ 
	and index them as $\underline{u}_0(s_0)$ with $s_0= \left[1:2^{nS_0}\right]$.
	\item 
	Generate  $2^{n{R}_{x_b}}$ i.i.d. sequences $\underline{x}_b$ with PD 
	$$
	P_{X_b}(\underline{x}_b)=\prod_{j=1}^np_{X_b}(x_{bj}),
	$$ 
	and index them as  $\underline{x}_b(r_b)$, where $r_b= \left[1:2^{n{R}_{x_b}}\right]$ for $b=\{1,2\}$.
	\item
	For each $\underline{x}_b(r_b)$ generate $2^{n\hat{R}_b}$ i.i.d. sequences  $\underline{\hat{z}}_b$ each with PD
	$$
	P_{\hat{Z}_b|X_b}(\underline{\hat{z}}_b\vert \underline{x}_b(r_b))= \\ \prod_{j=1}^np_{\hat{Z}_b\vert X_b}(\hat{z}_{bj}\vert x_{bj}(r_b)),
	$$ 
	and index them as $\underline{\hat{z}}_b(r_b,\hat{s}_b)$, where $\hat{s}_b= \left[1:2^{n\hat{R}_b}\right]$ for $b=\{1,2\}$.
\item   
	Partition the set $\big\{1,\ldots,2^{n\hat{R}_b}\big\}$ into $2^{nR_{x_b}}$ cells and label them as $S_{r_2}$. In each cell there are $2^{n(\hat{R}_b-R_{x_b})}$ elements. 
	\item
	For each pair $\underline{u}_0(s_{0})$, generate $2^{nT_b}$ i.i.d. sequences $\underline{u}_{b}$ with PD  
	$$
	P_{U_b|U_0}(\underline{u}_{b}\vert \underline{u}_0(s_{0}))=\prod_{j=1}^np_{U_b\vert U_0}(u_{bj}\vert u_{0j}(s_{0})),
	$$ 
	and index them as $\underline{u}_b(s_{0},t_{b})$, where $t_b=\left[1:2^{nT_b}\right]$ for $b=\{1,2\}$.
\item
For $b=\{1,2\}$, partition the set $\big\{1,\ldots,2^{nT_b}\big\}$ into $2^{nS_b}$ subsets and label them as $S_{s_b}$. In each subset there are $2^{n(T_b-S_b)}$ elements, for $b=\{1,2\}$.
\item
Finally, use a deterministic function for generating $\underline{x}$ as $f\left(\underline{u}_1,\underline{u}_2\right)$ indexed by $\underline{x}(s_{0},t_{1},t_{2})$.
\end{enumerate}\vspace{2mm}
%%%%%%%%%%%%%%%%%%%%%%%%%%%%%%%%%%%%%%%%%%%%%%%Encoding Part
\textit{Encoding part:} In block $i$, the source wants to send messages $(w_{0i},w_{1i},w_{2i})$ by reorganizing them into $(s_{0i},s_{1i},s_{2i})$. Encoding steps are as follows:
\begin{enumerate}[(i)]
\item
Relay $b$ knows from the previous block that $\hat{s}_{b(i-1)} \in S_{r_{bi}}$ and it sends $\underline{x}_b(r_{bi})$ for $b=\{1,2\}$.
\item  \label{SecondMartonV3}
Look for $t_{1i}\in S_{s_{1i}}$ and $t_{2i}\in S_{s_{2i}}$ such that $\big(\underline{u}_1(s_{0i},t_{1i})$,$ \underline{u}_2(s_{0i},t_{2i})\big)$ are jointly typical given the codeword $\underline{u}_0(s_{0i})$. The constraint to guarantee the success of this step is given by
\begin{align}
T_1+T_2-S_1-S_2 & \geq I(U_2;U_1\vert U_0). \label{eqCH1:V3-1B}
\end{align}
At the end, choose one pair $(t_{1(i-1)},t_{2(i-1)})$ satisfying these conditions.
\item
From $(s_{0i},s_{1i},s_{2i})$, the source finds $(t_{1i},t_{2i})$ and sends $\underline{x}(s_{0i},t_{1i},t_{2i})$. 
\end{enumerate}\vspace{2mm}
%%%%%%%%%%%%%%%%%%%%%%%%%%%%%%%%%%%%%%%%%%%%%%%Decoding Part
\textit{Decoding part:} In each block the relays start to find $\hat{s}_{bi}$ for that block. After the transmission of the block $i+1$, the destinations decode $\hat{s}_{bi}$ and then use it to find $\hat{Z}_b$ which along with $Y_b$ is used to decode the messages. 
\begin{enumerate}[(i)]
\item
Relay $b$ searches for $\hat{s}_{bi}$ after receiving $\underline{z}_b(i)$ such that $\big(\underline{x}_{b}(r_{bi}),\underline{z}_{b}(i),\underline{\hat{z}}_{b}(\hat{s}_{bi},r_{bi})\big)$ is jointly typical subject to
\begin{equation}
\hat{R}_b \geq I(Z_b;\hat{Z}_b|X_b).
\label{eqCH1:V3-4}
\end{equation}
\item 
Destination $b$ searches for $r_{b(i+1)}$ such that $\big(\underline{y}_b(i+1),\underline{x}_{b}(r_{b(i+1)})\big)$ is  jointly typical. Then in finds  $\hat{s}_{bi}$ such that $\hat{s}_{bi}\in S_{r_{b(i+1)}}$ and $\big(\underline{\hat{z}}_{b}(\hat{s}_{bi},r_{bi}),\underline{y}_b(i),\underline{x}_{b}(r_{bi})\big)$ are jointly typical. Conditions for reliable decoding are:
\begin{align}
R_{x_b} & \leq I(X_b;Y_b),\nonumber\\
\hat{R}_b &\leq R_{x_b}+I(\hat{Z}_b;Y_b|X_b).
\label{eqCH1:V3-5}
\end{align}
\item
Decoding in block $i$ is done such that $(\underline{u_0}(s_{0i}),\underline{u_b}(s_{0i},t_{bi}),\underline{y}_{b}(i),\underline{\hat{z}}_b(\hat{s}_{bi},r_{bi}),\underline{x}_{b}(r_{bi}))$ are all jointly typical. This leads to the next constraints
\begin{align}
S_0+T_b & \leq I(U_0,U_b;Y_b\hat{Z}_b\vert X_b),\label{eqCH1:V3-6A}\\
T_b & \leq I(U_b;Y_b,\hat{Z}_b\vert U_0,X_b). 
\label{eqCH1:V3-6B}
\end{align}
\end{enumerate}
After decoding indices  $(s_{0i},s_{1i},s_{2i})$ at the destinations, the original messages $(w_{0i},w_{1i},w_{2i})$ can be extracted. It is not difficult to show that the rate region in Theorem \ref{thm:II4-1} follows by applying Fourier-Motzkin elimination and form equations \eqref{eqCH1:V3-1B}-\eqref{eqCH1:V3-6B}, the equalities between original and reorganized rates, and the fact that all rates are positive. Similarly to \cite{Cover1979}, the necessary condition $I(X_b;Y_b)\geq I(Z_b;\hat{Z}_b\vert X_b,Y_b)$ follows from \eqref{eqCH1:V3-4} and \eqref{eqCH1:V3-5}, for $b=\{1,2\}$.

%%*********************************************************************************************************
%%V.8 Proof General Upper Bound
%%*********************************************************************************************************
\begin{figure*} [!b]
\hrule
\begin{equation}
\begin{array} {l}
\displaystyle\sum_{i=1}^nI(T_{1(i+1)}^n,T_{2(i+1)}^n,\dots,T_{N(i+1)}^n;S_{1i},S_{2i},\dots,S_{Mi}\vert W,S_{1}^{i-1},S_{2}^{i-1},\dots,S_{M}^{i-1})=\\ 
\,\,\,\,\,\,\,\,\,\,\,\,\,\,\,\,\,\,\,\,\,\,\,\,\,\,\,\,\,\,\,\,\,\,\,\,\,\,\,\displaystyle\sum_{i=1}^n I(S_{1}^{i-1},S_{2}^{i-1},\dots,S_{M}^{i-1};T_{1i},T_{2i},\dots,T_{Ni}\vert W,T_{1(i+1)}^n,T_{2(i+1)}^n,\dots,T_{N(i+1)}^n).
\end{array}
\label{CKidentity}
\end{equation}
\end{figure*}

\section{Sketch of Proof of Theorem \ref{thm:II5-6}} 
\label{proof-8}
We first state the Csiszar-Korner identity, formulated in a different way.
\begin{lemma}
For any RV $W$ and an ensemble of $n$ RVs $\mb{S}_{j}=(S_{j1},S_{j2},\dots,S_{jn})$ with $j\in \{1,2,\dots,M\}$ and $\mb{T}_{k}=(T_{k1},T_{k2},\dots,T_{kn})$ for $k\in \{1,2,\dots,N\}$, the equality \eqref{CKidentity} holds.
\label{lemma:V7-1}
\end{lemma}
The proof of this lemma easily follows as in \cite{Korner1977}. The next identity is also used during the proof:
\begin{equation}
I(A;B|D)-I(A;C|D)=I(A;B\vert C,D)-I(A;C\vert B,D).
\label{eqCH1:V7-3}
\end{equation}
%%%%%%%%%%%%%%%%%%
For any code $(n,\mc{W}_0,\mc{W}_1,\mc{W}_2,P_{e}^{(n)})$ with rates $(R_0,R_1,R_2)$, Fano's inequality yields
\begin{align*}
H(W_0|\mb{Y}_2) & \leq P_{e}^{(n)}nR_0+1\stackrel{\Delta}{=}n\epsilon_0, \\
H(W_1|\mb{Y}_1) &\leq H(W_0,W_1|\mb{Y}_1) \\
&\leq P_{e}^{(n)}n(R_0+R_1)+1\stackrel{\Delta}{=}n\epsilon_1, \\
H(W_2|\mb{Y}_2) &\leq H(W_0,W_2|\mb{Y}_2) \\
&\leq P_{e}^{(n)}n(R_0+R_2)+1\stackrel{\Delta}{=}n\epsilon_2.
\end{align*}
We start with the following inequality:
\begin{align}
n(R_0+&R_1+R_2)-n(\epsilon_0+\epsilon_1+\epsilon_2) \nonumber\\
& \leq I(W_0;\mb{Y}_1)+I(W_1;\mb{Y}_1)+I(W_2;\mb{Y}_2) \nonumber\\
& \leq I(W_0;\mb{Y}_1)+I(W_1;\mb{Y}_1,W_0,W_2)\nonumber\\
&\hspace{10mm}+I(W_2;\mb{Y}_2,W_0) \nonumber\\
& \leq I(W_0,W_1,W_2;\mb{Y}_1)-I(W_2;\mb{Y}_1|W_0)\nonumber\\
&\hspace{10mm}+I(W_2;\mb{Y}_2|W_0), \label{sumrate-1}
\end{align}
where we can bound the first term on the right hand side of \eqref{sumrate-1} as
\begin{align*}
I(W_0,W_1,W_2;\mb{Y}_1)&=\displaystyle\sum_{i=1}^nI(W_0,W_1,W_2;Y_{1i}\vert Y_{1}^{i-1})\\
& \leq I(W_0,W_1,W_2,Y_{1}^{i-1},Y_{2(i+1)}^{n};Y_{1i}) \\
% &=\displaystyle\sum_{i=1}^n\big[H(Y_{1i}\vert Y_{1}^{i-1})-H(Y_{1i}\vert Y_{1}^{i-1},W_0,W_1,W_2)\big]\\
% &\stackrel{(a)}{\leq} \displaystyle\sum_{i=1}^n\big[H(Y_{1i})-H(Y_{1i}\vert Y_{1}^{i-1},W_0,W_1,W_2,Y_{2(i+1)}^{n})\big]\\
& \stackrel{(a)}{=} \displaystyle\sum_{i=1}^nI(V_i,U_{1i},U_{2i};Y_{1i}),
\end{align*}
where (a) is based on the definitions of $V_i=(W_0,Y_{1}^{i-1},Y_{2(i+1)}^{n})$, $U_{1i}=(W_1,Y_{1}^{i-1},Y_{2(i+1)}^{n})$ and $U_{2i}=(W_2,Y_{1}^{i-1},Y_{2(i+1)}^{n})$. Now for the rest of terms in \eqref{sumrate-1}, we have:
\begin{align}
&I(W_2;\mb{Y}_2|W_0)-I(W_2;\mb{Y}_1|W_0)\nonumber\\
&=\displaystyle\sum_{i=1}^n\big[I(W_2;Y_{2i}\vert W_0,Y_{2(i+1)}^{n})-I(W_2;Y_{1i}\vert W_0,Y_{1}^{i-1})\big]\nonumber\\
&=\displaystyle\sum_{i=1}^n \big[I(W_2,Y_{1}^{i-1};Y_{2i}\vert W_0,Y_{2(i+1)}^{n})\nonumber\\
&\hspace{20mm}-I(Y_{1}^{i-1};Y_{2i}\vert W_2,W_0,Y_{2(i+1)}^{n})\nonumber
\end{align}
\begin{align}
&\hspace{20mm}-I(W_2,Y_{2(i+1)}^{n};Y_{1i}\vert W_0,Y_{1}^{i-1})\nonumber\\
&\hspace{20mm}+I(Y_{2(i+1)}^{n};Y_{1i}\vert W_2,W_0,Y_{1}^{i-1})\big] \nonumber\\
&\stackrel{(b)}{=} \displaystyle\sum_{i=1}^n\big[I(W_2,Y_{1}^{i-1};Y_{2i}\vert W_0,Y_{2(i+1)}^{n})\nonumber\\
&\hspace{20mm}-I(W_2,Y_{2(i+1)}^{n};Y_{1i}\vert W_0,Y_{1}^{i-1})\big]\nonumber\\
% &=\displaystyle\sum_{i=1}^n\big[I(W_2;Y_{2i}\vert W_0,Y_{1}^{i-1},Y_{2(i+1)}^{n})+I(Y_{1}^{i-1};Y_{2i}\vert W_0,Y_{2(i+1)}^{n})\\
% &-I(W_2;Y_{1i}\vert W_0,Y_{1}^{i-1},Y_{2(i+1)}^{n})-I(Y_{2(i+1)}^{n};Y_{1i}\vert W_0,Y_{1}^{i-1})\big] \\
&\stackrel{(c)}{=} \displaystyle\sum_{i=1}^n\big[I(W_2;Y_{2i}\vert W_0,Y_{1}^{i-1},Y_{2(i+1)}^{n})\nonumber\\
&\hspace{20mm}-I(W_2;Y_{1i}\vert W_0,Y_{1}^{i-1},Y_{2(i+1)}^{n})\big]\nonumber\\
&=I(U_{2i};Y_{2i}\vert V_i)-I(U_{2i};Y_{1i}\vert V_i),
\label{Proof:V7-1}
\end{align}
where (b) and (c) are due to Lemma \ref{lemma:V7-1} by choosing $M=N=1$ and $\mb{T}_1=\mb{Y}_1,\mb{S}_1=\mb{Y}_2$. Hence, the right hand side of \eqref{sumrate-1} writes as 
\begin{align}
&n(R_0+R_1+R_2)-n(\epsilon_0+\epsilon_1+\epsilon_2) \nonumber\\
&\leq \displaystyle\sum_{i=1}^n\big[I(V_{i},U_{1i},U_{2i};Y_{1i})+I(U_{2i};Y_{2i}\vert V_i)\nonumber\\
&\hspace{20mm}-I(U_{2i};Y_{1i}\vert V_i)\big]\nonumber\\ 
% &=\displaystyle\sum_{i=1}^n\big[I(V_i;Y_{1i})+I(U_{2i};Y_{2i}\vert V_i)+
% I(U_{1i};Y_{1i}|U_{2i},V_i)\big]\nonumber\\ 
&\stackrel{(d)}{=}\displaystyle\sum_{i=1}^n\big[I(V_i;Y_{1i})+I(U_{2i};Y_{2i}\vert V_i)+
I(U_{1i};Y_{1i}|U_{2i},V_i)\big],\label{8ineqCH1:1}
\end{align}
yielding the final inequality, where $(d)$ is due to standard manipulations.  We consider now the next inequality
\begin{align}
&n(R_0+R_1+R_2)-n(\epsilon_0+\epsilon_1+\epsilon_2)  \nonumber\\
&\leq I(W_0,W_1,W_2;\mb{Y}_1)-I(W_2;\mb{Y}_1|W_0)\nonumber\\
&\hspace{20mm}+I(W_2;\mb{Y}_2|W_0)\nonumber\\ 
& \leq I(W_0,W_1,W_2;\mb{Y}_1,\mb{Z}_1)-I(W_2;\mb{Y}_1,\mb{Z}_1|W_0)\nonumber\\
&+I(W_2;\mb{Y}_2,\mb{Z}_2|W_0). \label{sumrate-2}
\end{align}
Similarly as before, we obtain
\begin{align*}
I(W_0&,W_1,W_2;\mb{Y}_1,\mb{Z}_1)\\
&=\displaystyle\sum_{i=1}^nI(W_0,W_1,W_2;Y_{1i},Z_{1i}\vert Y_{1}^{i-1},Z_{1}^{i-1})\\
&\stackrel{(e)}{=}\displaystyle\sum_{i=1}^nI(W_0,W_1,W_2;Y_{1i},Z_{1i}\vert Y_{1}^{i-1},Z_{1}^{i-1},X_{1i})
\end{align*}
\begin{align*}
&\stackrel{(f)}{\leq} \displaystyle\sum_{i=1}^nI(W_0,W_1,W_2,Y_{1}^{i-1},Z_{1}^{i-1},Y_{2(i+1)}^{n}\\
&\hspace{20mm},Z_{2(i+1)}^{n};Y_{1i},Z_{1i}\vert X_{1i}) \\
&=\displaystyle\sum_{i=1}^n I(V_i,V_{1i},U_{1i},U_{2i};Y_{1i},Z_{1i}\vert X_{1i}),
\end{align*}
where (e) follows because $X_{1i}$ is a function of the past relay output, (f) is due to properties of mutual information and $V_{1i}$ is denoted by $(Z_{1}^{i-1},Z_{2(i+1)}^{n})$. In a similar way to \eqref{Proof:V7-1}, we can obtain
\begin{align*}
I&(W_2;\mb{Y}_2,\mb{Z}_2|W_0)-I(W_2;\mb{Y}_1,\mb{Z}_1|W_0)\\
% &=\displaystyle\sum_{i=1}^n\big[I(W_2;Y_{2i},Z_{2i}\vert W_0,Y_{2(i+1)}^{n},Z_{2(i+1)}^{n})-I(W_2;Y_{1i},Z_{1i}\vert W_0,Y_{1}^{i-1},Z_{1}^{i-1})\big]\\
&\stackrel{(g)}{\leq} \displaystyle\sum_{i=1}^n\big[I(W_2;Y_{2i},Z_{2i}\vert W_0,X_{1i},Y_{1}^{i-1},Z_{1}^{i-1}\\
&\hspace{20mm},Y_{2(i+1)}^{n},Z_{2(i+1)}^{n})\\
&-I(W_2;Y_{1i}\vert W_0,X_{1i},Y_{1}^{i-1},Z_{1}^{i-1},Y_{2(i+1)}^{n},Z_{2(i+1)}^{n})\big],
\end{align*}
where the step $(g)$ can be proven by using the same procedure as the steps in \eqref{Proof:V7-1}. Then 
\begin{align}
n&(R_0+R_1+R_2)-n\epsilon' \\
 &\leq \displaystyle\sum_{i=1}^n\big[I(V_i,V_{1i},U_{1i},U_{2i};Y_{1i},Z_{1i}|X_{1i})\nonumber\\
 & \hspace{10mm}+I(U_{2i};Y_{2i},Z_{2i}\vert V_i,V_{1i},X_{1i}) \nonumber\\
& \hspace{10mm}-I(U_{2i};Y_{1i},Z_{1i}\vert V_i,V_{1i},X_{1i})\big]\nonumber\\ 
&{=}\displaystyle\sum_{i=1}^n\big[I(V_i,V_{1i};Y_{1i},Z_{1i}|X_{1i})\nonumber\\
&\hspace{10mm}+I(U_{2i};Y_{2i},Z_{2i}\vert V_i,V_{1i},X_{1i})\nonumber\\
&\hspace{10mm}+ I(U_{1i};Y_{1i},Z_{1i}|X_{1i},U_{2i},V_i,V_{1i})\big],\label{8ineqCH1:2} 
\end{align}
by definning $(\epsilon'=\epsilon_0+\epsilon_1+\epsilon_2)$. Consider now the following inequality
\begin{align}
n(R_0&+R_1+R_2)-n(\epsilon_0+\epsilon_1+\epsilon_2)\nonumber\\
 &  \leq I(W_0;\mb{Y}_2)+I(W_1;\mb{Y}_1)+I(W_2;\mb{Y}_2) \nonumber\\
& \leq I(W_0,W_1,W_2;\mb{Y}_2)-I(W_1;\mb{Y}_2|W_0)\nonumber\\
&\hspace{10mm}+I(W_1;\mb{Y}_1|W_0). \label{sumrate-3}
\end{align}
Notice that this is the symmetrical version of \eqref{sumrate-1} and thus it can be bound in the same way. Now we simplify the right hand side of \eqref{sumrate-3} to
\begin{align}
n&(R_0+R_1+R_2)-n(\epsilon_0+\epsilon_1+\epsilon_2)\nonumber\\
 &\leq \displaystyle\sum_{i=1}^n\big[I(V_i,U_{1i},U_{2i};Y_{2i})+I(U_{1i};Y_{1i}\vert V_i)
\nonumber\\
&\hspace{10mm}-I(U_{1i};Y_{2i}\vert V_i)\big]\nonumber\\ 
&{=}\displaystyle\sum_{i=1}^n\big[I(V_i;Y_{2i})+I(U_{1i};Y_{1i}\vert V_i)+
I(U_{2i};Y_{2i}|U_{1i},V_i)\big].\label{8ineqCH1:3}
\end{align}
Another inequality which is symmetric to \eqref{sumrate-2} is the following and can be proved in a same way:
%%%%%
\begin{align}
n&(R_0+R_1+R_2)-n(\epsilon_0+\epsilon_1+\epsilon_2)\nonumber\\
 & \leq I(W_0,W_1,W_2;\mb{Y}_2)-I(W_1;\mb{Y}_2|W_0)\nonumber\\
 &\hspace{15mm}+I(W_1;\mb{Y}_1|W_0)\nonumber\\ &  \leq I(W_0,W_1,W_2;\mb{Y}_2,\mb{Z}_2)+I(W_1;\mb{Y}_1,\mb{Z}_1|W_0)\nonumber \\
 &\hspace{15mm}-I(W_1;\mb{Y}_2,\mb{Z}_2|W_0). \label{sumrate-4}
\end{align}
Now by following similar steps as before,  we can show
\begin{align*}
I(&W_0,W_1,W_2;\mb{Y}_2,\mb{Z}_2)\nonumber\\
&=\displaystyle\sum_{i=1}^nI(W_0,W_1,W_2;Y_{2i},Z_{2i}\vert Y_{2(i+1)}^{n},Z_{2(i+1)}^{n})\\
&\stackrel{(h)}{=}\displaystyle\sum_{i=1}^n \big[I(V_i,V_{1i};Y_{2i},Z_{2i})\nonumber\\
&\hspace{15mm}+I(U_{1i},U_{2i};Y_{2i},Z_{2i}\vert V_i,V_{1i},X_{1i})\big],
\end{align*}
where $(h)$ is because $X_{1i}$ is a function of the past relay output $V_{1i}$. Along the same lines, we can show
\begin{align*}
&I(W_1;\mb{Y}_1,\mb{Z}_1|W_0)-I(W_1;\mb{Y}_2,\mb{Z}_2|W_0)\\
&=\displaystyle\sum_{i=1}^n \big[I(W_2;Y_{1i},Z_{1i}\vert W_0,Y_{1}^{i-1},Z_{1}^{i-1})\\
&\hspace{10mm} -I(W_1;Y_{2i},Z_{2i}\vert W_0,Y_{2(i+1)}^{n},Z_{2(i+1)}^{n})\big]\\
& {\leq} \displaystyle\sum_{i=1}^n\big[I(W_1;Y_{1i}\vert W_0,X_{1i},Y_{1}^{i-1},Z_{1}^{i-1},Y_{2(i+1)}^{n},Z_{2(i+1)}^{n})\\
&-I(W_1;Y_{2i},Z_{2i}\vert W_0,X_{1i},Y_{1}^{i-1},Z_{1}^{i-1},Y_{2(i+1)}^{n},Z_{2(i+1)}^{n})\big].
\end{align*} 
Finally, we obtain
\begin{align}
&n(R_0+R_1+R_2)-n(\epsilon_0+\epsilon_1+\epsilon_2) \nonumber\\
&\leq \displaystyle\sum_{i=1}^n\big[I(V_i,V_{1i};Y_{2i},Z_{2i})\nonumber\\
&\hspace{30mm}+I(U_{1i},U_{2i};Y_{2i},Z_{2i}|V_i,V_{1i},X_{1i})\nonumber\\ 
&\hspace{30mm}+I(U_{1i};Y_{1i},Z_{1i}\vert V_i,V_{1i},X_{1i})\nonumber\\
&\hspace{30mm}-I(U_{1i};Y_{2i},Z_{2i}\vert V_i,V_{1i},X_{1i})\big]\nonumber\\ 
&{=}\displaystyle\sum_{i=1}^n\big[I(V_i,V_{1i};Y_{2i},Z_{2i})\nonumber\\
&\hspace{30mm}+I(U_{2i};Y_{2i},Z_{2i}\vert V_i,V_{1i},U_{1i},X_{1i})\nonumber\\
&\hspace{30mm}+ I(U_{1i};Y_{1i},Z_{1i}|X_{1i},V_i,V_{1i})\big].\label{8ineqCH1:4} 
\end{align}
The inequalities \eqref{8ineqCH1:1}, \eqref{8ineqCH1:2}, \eqref{8ineqCH1:3} and \eqref{8ineqCH1:4} are related to the sum of $R_0,R_1,R_2$. For the rest of the proof we focus on the following inequalities:
\begin{align*}
nR_0  & \leq I(W_0;\mb{Y}_2)+n\epsilon_0,\\
n(R_0+R_1) &  \leq I(W_0;\mb{Y}_2)+I(W_1;\mb{Y}_1\vert W_0)+n(\epsilon_0+\epsilon_1), \\
n(R_0+R_2) &  \leq I(W_0;\mb{Y}_1)+I(W_2;\mb{Y}_2\vert W_0)+n(\epsilon_0+\epsilon_2). 
\end{align*}
%%%%%%%%%%%%%%%%%%%%%
%%%%%%%%%%%%%%%%%%%%%
Starting from the last inequality, we have
\begin{align}
&n(R_0+R_1)-n(\epsilon_0+\epsilon_1) \leq  I(W_0;\mb{Y}_2)+I(W_1;\mb{Y}_1\vert W_0)\nonumber\\
&=\displaystyle\sum_{i=1}^n\big[ I(W_0;Y_{2i}\vert Y_{2(i+1)}^{n})+I(W_1;Y_{1i}\vert Y_{1}^{i-1},W_0)\big]\nonumber\\
&=\displaystyle\sum_{i=1}^n\big[ I(W_0,Y_{1}^{i-1};Y_{2i}\vert Y_{2(i+1)}^{n})\nonumber\\
&-I(Y_{1}^{i-1};Y_{2i}\vert W_0,Y_{2(i+1)}^{n})
+I(W_1;Y_{1i}\vert Y_{1}^{i-1},W_0)\big]\nonumber\\
&\stackrel{(a')}{=} \displaystyle\sum_{i=1}^n \big[I(W_0,Y_{1}^{i-1};Y_{2i}\vert Y_{2(i+1)}^{n})\nonumber\\
&\hspace{20mm}-I(Y_{2(i+1)}^{n};Y_{1i}\vert W_0,Y_{1}^{i-1})\nonumber\\
&\hspace{20mm}+I(W_1;Y_{1i}\vert Y_{1}^{i-1},W_0)\big]\nonumber\\
&\stackrel{(b')}{=} \displaystyle\sum_{i=1}^n \big[I(W_0,Y_{1}^{i-1};Y_{2i}\vert Y_{2(i+1)}^{n})\nonumber\\
&\hspace{20mm}+I(W_1;Y_{1i}\vert Y_{2(i+1)}^{n},Y_{1}^{i-1},W_0)\nonumber\\
&\hspace{20mm}-I(Y_{2(i+1)}^{n};Y_{1i}\vert W_1,W_0,Y_{1}^{i-1})\big]\nonumber\\
&\leq \displaystyle\sum_{i=1}^n \big[ I(W_0,Y_{2(i+1)}^{n},Y_{1}^{i-1};Y_{2i})\nonumber\\
&\hspace{20mm}+I(W_1;Y_{1i}\vert Y_{1}^{i-1},Y_{2(i+1)}^{n},W_0)\big]\nonumber\\
&\leq \displaystyle\sum_{i=1}^n\big[I(V_i;Y_{2i})+I(U_{1i};Y_{1i}\vert V_i)\big],
\label{8ineqCH1:5} 
\end{align}
where $(a')$ comes from Lemma \ref{lemma:V7-1} by choosing $M=N=1$, $S_1=Y_1, T_1=Y_2, W=W_0$, and $(b')$ comes from \eqref{eqCH1:V7-3}. With a similar procedure, it can be seen that
\begin{align}
n(R_0+R_2-\epsilon_0+\epsilon_2) &\leq  I(W_0;\mb{Y}_1)+I(W_2;\mb{Y}_2\vert W_0)\nonumber\\
&\leq \displaystyle\sum_{i=1}^n\big[I(V_i;Y_{1i})+I(U_{2i};Y_{2i}\vert V_i)\big].
\label{8ineqCH1:6} 
\end{align}
%%%%%%%%%%%%%%%%%
Now we move to the next inequality which is proved similar to \eqref{8ineqCH1:5}
\begin{align*}
&n(R_0+R_1)-n(\epsilon_0+\epsilon_1)\\
& \leq I(W_0;\mb{Y}_2)+I(W_1;\mb{Y}_1\vert W_0) \\
&\leq I(W_0;\mb{Y}_2,\mb{Z}_2)+I(W_1;\mb{Y}_1,\mb{Z}_1\vert W_0)\\
&=\displaystyle\sum_{i=1}^n\big[I(W_0;Y_{2i},Z_{2i}\vert Y_{2(i+1)}^{n},Z_{2(i+1)}^{n})\\
&\hspace{1mm}+I(W_1;Y_{1i},Z_{1i}\vert Y_{1}^{i-1},Z_{1}^{i-1},W_0)\big]\\
&{\leq} \displaystyle\sum_{i=1}^n\big[I(W_0,Y_{1}^{i-1},Z_{1}^{i-1},Z_{2(i+1)}^n,Y_{2(i+1)}^{n};Z_{2i},Y_{2i})
\end{align*}
\begin{align*}
&\hspace{1mm} + I(W_1;Y_{1i},Z_{1i}\vert Y_{1}^{i-1},Z_{1}^{i-1},Y_{2(i+1)}^{n},Z_{2(i+1)}^n,W_0)\big]\\
&\stackrel{c'}{=} \displaystyle\sum_{i=1}^n\big[I(W_0,Y_{1}^{i-1},Z_{1}^{i-1},Z_{2(i+1)}^n,Y_{2(i+1)}^{n};Z_{2i},Y_{2i})\\
&\hspace{1mm} + I(W_1;Y_{1i},Z_{1i}\vert Y_{1}^{i-1},Z_{1}^{i-1},Y_{2(i+1)}^{n},Z_{2(i+1)}^n,W_0,X_{1i})\big],
\end{align*}
where $(c')$ 
% comes from the Lemma \ref{lemma:V7-1} by choosing $M=N=2$, $T_1=Y_2, S_1=Y_1, T_2=Z_2, S_2=Z_1, W=W_0$, $(d')$ comes from \eqref{eqCH1:V7-3}, $(e')$ 
is due to the fact that $X_{1i}$ is a function of $Z_{1}^{i-1}$. By using the previous definitions, we obtain
\begin{equation}
\begin{array} {l}
n(R_0+R_1)-n(\epsilon_0+\epsilon_1) \\ =\displaystyle\sum_{i=1}^n\big[I(V_i,V_{1i};Z_{2i},Y_{2i}) 
+ I(U_{1i};Y_{1i},Z_{1i}\vert V_i,V_{1i},X_{1i})\big].
\end{array}
\label{8ineqCH1:7}
\end{equation}
%%%%%%%%%%%%%%%%%%%%
%%%%%%%%%%%%%%%%%%%%
And finally the proof of the final sum rate is as follows
%%%%%%%%%%%
\begin{align*}
&n(R_0+R_2)-n(\epsilon_0+\epsilon_2) \leq I(W_0;\mb{Y}_1)+I(W_2;\mb{Y}_2\vert W_0) \\
&\leq I(W_0;\mb{Y}_1,\mb{Z}_1)+I(W_2;\mb{Y}_2,\mb{Z}_2\vert W_0)\\
&=\displaystyle\sum_{i=1}^n\big[I(W_0;Y_{1i},Z_{1i}\vert Y_{1}^{i-1},Z_{1}^{i-1})\\
&+I(W_2;Y_{2i},Z_{2i}\vert Y_{2(i+1)}^{n},Z_{2(i+1)}^n,W_0)\big]\\
% &=\displaystyle\sum_{i=1}^n\big[I(W_0,Y_{2(i+1)}^{n},Z_{2(i+1)}^n;Z_{1i},Y_{1i}\vert Y_{1}^{i-1},Z_{1}^{i-1}) \\
% &\hspace{10mm}-I(Y_{2(i+1)}^{n},Z_{2(i+1)}^n;Y_{1i},Z_{1i}\vert W_0,Y_{1}^{i-1},Z_{1}^{i-1})\\
% &\hspace{10mm}+ I(W_2;Y_{2i},Z_{2i}\vert Y_{2(i+1)}^{n},Z_{2(i+1)}^n,W_0)\big]\\
% &\stackrel{(f')}{=} \displaystyle\sum_{i=1}^n\big[I(W_0,Y_{2(i+1)}^{n},Z_{2(i+1)}^n;Z_{1i},Y_{1i}\vert Y_{1}^{i-1},Z_{1}^{i-1}) \\
% &\hspace{10mm}-I(Y_{1}^{i-1},Z_{1}^{i-1};Y_{2i},Z_{2i}\vert W_0,Y_{2(i+1)}^{n},Z_{2(i+1)}^n)\\
% &\hspace{10mm}+ I(W_2;Y_{2i},Z_{2i}\vert Y_{2(i+1)}^{n},Z_{2(i+1)}^n,W_0)\big]\\
% &\stackrel{(g')}{=} \displaystyle\sum_{i=1}^n\big[I(W_0,Y_{2(i+1)}^{n},Z_{2(i+1)}^n;Z_{1i},Y_{1i}\vert Y_{1}^{i-1},Z_{1}^{i-1}) \\
% &\hspace{10mm}+ I(W_2;Y_{2i},Z_{2i}\vert Y_{1}^{i-1},Z_{1}^{i-1},Y_{2(i+1)}^{n},Z_{2(i+1)}^n,W_0) \\
% &\hspace{10mm}-I(Y_{1}^{i-1},Z_{1}^{i-1};Y_{2i},Z_{2i}\vert W_2,W_0,Y_{2(i+1)}^{n},Z_{2(i+1)}^n)\big]\\
&\leq \displaystyle\sum_{i=1}^n\big[I(W_0,Y_{2(i+1)}^{n},Z_{2(i+1)}^n;Z_{1i},Y_{1i}\vert Y_{1}^{i-1},Z_{1}^{i-1}) \\
&\hspace{1mm}+ I(W_2;Y_{2i},Z_{2i}\vert Y_{1}^{i-1},Z_{1}^{i-1},Y_{2(i+1)}^{n},Z_{2(i+1)}^n,W_0)\big] \\
&\stackrel{(d')}{=} \displaystyle\sum_{i=1}^n\big[I(W_0,Y_{2(i+1)}^{n},Z_{2(i+1)}^n;Z_{1i},Y_{1i}\vert Y_{1}^{i-1},Z_{1}^{i-1},X_{1i})\\
&\hspace{1mm}+ I(W_2;Y_{2i},Z_{2i}\vert Y_{1}^{i-1},Z_{1}^{i-1},Y_{2(i+1)}^{n},Z_{2(i+1)}^n,W_0,X_{1i})\big] \\
&{\leq} \displaystyle\sum_{i=1}^n\big[I(W_0,Y_{1}^{i-1},Z_{1}^{i-1},Y_{2(i+1)}^{n},Z_{2(i+1)}^n;Z_{1i},Y_{1i}\vert X_{1i})\\
&\hspace{1mm}+ I(W_2;Y_{2i},Z_{2i}\vert Y_{1}^{i-1},Z_{1}^{i-1},Y_{2(i+1)}^{n},Z_{2(i+1)}^n,W_0,X_{1i})\big].
%\label{eqCH1:V8-4}
\end{align*}
Again using previous definitions we obtain
\begin{align} 
n(R_0+R_2)&-n(\epsilon_0+\epsilon_2)\nonumber\\ 
&\leq\displaystyle\sum_{i=1}^nI(V_i,V_{1i};Z_{1i},Y_{1i}\vert X_{1i}) \nonumber\\
&+ I(U_{2i};Y_{2i},Z_{2i}\vert V_i,V_{1i},X_{1i}),
\label{8ineqCH1:8} 
\end{align} 
% where $(f')$ comes from the Lemma \ref{lemma:V7-1} with the choice $M=N=2$, $S_1=Y_1, T_1=Y_2, S_2=Z_1, T_2=Z_2, W=W_0$, $(g')$ comes from \eqref{eqCH1:V7-3},
 where $(d')$ is due to the fact that $X_{1i}$ is a function of $Z_{1}^{i-1}$. Finally, we prove the reminding first inequalities
\begin{align}
n(R_0+R_1)&-n(\epsilon_0+\epsilon_1) \leq I(W_0,W_1;\mb{Y}_1)\nonumber\\ 
&=\displaystyle\sum_{i=1}^nI(W_0,W_1;Y_{1i}\vert Y_{1}^{i-1})\nonumber\\ 
% & \leq \displaystyle\sum_{i=1}^nI(Y_1^{i-1},W_{0},W_1;Y_{1i}) \nonumber\\
&\leq \sum_{i=1}^nI(Y_{2(i+1)}^{n},Y_1^{i-1},W_0,W_1;Y_{1i}) \nonumber\\
&= \sum_{i=1}^nI(V_{i},U_{1i};Y_{1i}),
\label{8ineqCH1:9}
\end{align}
and similarly we derive:
\begin{align}
n(R_0+R_2)-n(\epsilon_0+\epsilon_2) \leq \sum_{i=1}^nI(V_{i},U_{2i};Y_{2i}). 
\label{8ineqCH1:10}
\end{align}
The next step is to prove another bound on the sum rate $R_0+R_1$:
\begin{align}
&n(R_0+R_1)-n(\epsilon_0+\epsilon_1) \leq I(W_0,W_1;\mb{Y}_1,\mb{Z}_1)\nonumber\\ 
& =\displaystyle\sum_{i=1}^nI(W_0,W_1;Y_{1i},Z_{1i}\vert Y_{1}^{i-1},Z_{1}^{i-1})\nonumber\\ 
& = \displaystyle\sum_{i=1}^nI(W_{0},W_1;Y_{1i},Z_{1i}\vert Y_1^{i-1},Z_{1}^{i-1},X_{1i}) \nonumber\\
&\leq \sum_{i=1}^nI(Y_{2(i+1)}^{n},Z_{2(i+1)}^{n},Y_1^{i-1},Z_1^{i-1},W_0,W_1;\nonumber\\
&\hspace{10mm} Y_{1i},Z_{1i}\vert X_{1i})\nonumber\\
&=\sum_{i=1}^nI(V_{i},V_{1i},U_{1i};Y_{1i},Z_{1i}\vert X_{1i}). 
\label{8ineqCH1:11}
\end{align}
Similarly, for the sum rate $R_0+R_2$:
\begin{align}
&n(R_0+R_2)-n(\epsilon_0+\epsilon_2)\nonumber\\
& \leq I(W_0,W_2;\mb{Y}_2,\mb{Z}_2)\nonumber\\
&=\displaystyle\sum_{i=1}^nI(W_0,W_2;Y_{2i},Z_{2i}\vert Y_{2(i+1)}^{n},Z_{2(i+1)}^{n})\nonumber\\ 
% & \leq \displaystyle\sum_{i=1}^nI(Y_{2(i+1)}^{n},Z_{2(i+1)}^{n},W_{0},W_2;Y_{2i},Z_{2i}) \nonumber\\
&\leq \sum_{i=1}^nI(Y_{2(i+1)}^{n},Z_{2(i+1)}^{n},Y_1^{i-1},Z_1^{i-1},W_0,W_2;Y_{2i},Z_{2i}) \nonumber\\
&= \sum_{i=1}^n\big[ I(V_{i},V_{1i};Y_{2i},Z_{2i})+I(U_{2i};Y_{2i},Z_{2i}\vert V_{i},V_{1i})\big]\nonumber\\
&\stackrel{(e')}{=} \sum_{i=1}^n\big[ I(V_{i},V_{1i};Y_{2i},Z_{2i})+I(U_{2i};Y_{2i},Z_{2i}\vert V_{i},V_{1i},X_{1i})\big],
\label{8ineqCH1:12}
\end{align}
where $(e')$ is due to the fact that $X_{1i}$ is function of $Z_1^{i-1}$ and so function of $V_{1i}$. And at last we bound the rate $R_0$,
\begin{align}
nR_0&-n\epsilon_0\leq I(W_0;\mb{Y}_1) \nonumber\\ 
& =\displaystyle\sum_{i=1}^nI(W_0;Y_{1i}\vert Y_{1}^{i-1})\nonumber\\ 
% & \leq \displaystyle\sum_{i=1}^nI(Y_1^{i-1},W_{0};Y_{1i}) \nonumber\\
&\leq \sum_{i=1}^nI(Y_{2(i+1)}^{n},Y_1^{i-1},W_0;Y_{1i})=\sum_{i=1}^nI(V_{i};Y_{1i}). 
\label{8ineqCH1:13}
\end{align}
Similarly for destination $Y_2$,
\begin{align}
nR_0-n\epsilon_0&\leq I(W_0;\mb{Y}_2)\leq \sum_{i=1}^nI(V_{i};Y_{2i}). 
\label{8ineqCH1:14}
\end{align}
The rest of the proof is as usual with resort to an independent time-sharing RV $Q$ applying it to \eqref{8ineqCH1:1}-\eqref{8ineqCH1:14} which yields the final region and concludes the proof. 
%%*********************************************************************************************************
%%V.6 Proof Capacity Semi Degraded
%%*********************************************************************************************************
\section{Sketch of Proof of Theorem \ref{thm:II5-4}} 
\label{proof-6}
We emphasize that the upper bound can be seen to be a special case of the outer-bound presented in Theorem \ref{thm:II5-5} for the semi-degraded BRC. However, for sake of clarity, we independently prove the converse in the Theorem \ref{thm:II5-4}. We start with the fact that the user $1$ must decode full information. For any code $(n,\mc{W}_1,\mc{W}_2,P_{e}^{(n)})$ (i.e. $(R_1,R_2)$), from Fano's inequality we obtain:
\begin{align*}
H(W_2|\mb{Y}_2) & \leq P_{e}^{(n)}nR_2+1\stackrel{\Delta}{=}n\epsilon_0, \\
H(W_1|\mb{Y}_1) & \leq P_{e}^{(n)}nR_1+1\stackrel{\Delta}{=}n\epsilon_1,  
\end{align*}
and          
\begin{align*}
nR_2  & \leq I(W_2;\mb{Y}_2)+n\epsilon_0,\\
n(R_1+R_2)-n\epsilon_0-n\epsilon_1&  \leq I(W_2;\mb{Y}_2)+I(W_1;\mb{Y}_1)\\ 
& \leq I(W_2;\mb{Y}_2)+I(W_1;\mb{Y}_1,W_2) \\
& \leq I(W_2;\mb{Y}_2)+I(W_1;\mb{Y}_1\vert W_2).
\end{align*}\vspace{2mm}
Before starting the proof, we state the following lemma.
\begin{lemma}
The following relation holds for the BRC-CR under the condition $X \mkv (Y_1,X_1) \mkv Z_1$, 
\begin{equation*}
H(Y_{1i}\vert Y_{1}^{i-1},W_2)=H(Y_{1i}\vert Y_{1}^{i-1},Z_{1}^{i-1},X_{1}^i,W_2).
\end{equation*}
\label{lemma:V6-1}
\end{lemma}

\begin{proof}
\begin{align*}
H&(Y_{1i}\vert Y_{1}^{i-1},W_2) =H(Y_{1i}\vert Y_{11},Y_{12},\dots,Y_{1(i-1)},W_2)\\
&\stackrel{(a)}{=}H(Y_{1i}\vert Y_{11},X_{11},Y_{12},\dots,Y_{1(i-1)},W_2)\\
&\stackrel{(b)}{=}H(Y_{1i}\vert Y_{11},X_{11},Z_{11},Y_{12},\dots,Y_{1(i-1)},W_2)\\
&\stackrel{(c)}{=} H(Y_{1i}\vert Y_{11},X_{11},Z_{11},X_{12},Y_{12},\dots,Y_{1(i-1)},W_2)\\
\vdots\\
&=H(Y_{1i}\vert Y_{11},X_{11},Z_{11},Y_{12},X_{12},Z_{12}\dots,Y_{1(i-1)}\\
&\hspace{15mm},X_{1(i-1)},Z_{1(i-1)},X_{1i},W_2)\\
&=H(Y_{1i}\vert Y_{1}^{i-1},Z_{1}^{i-1},X_{1}^i,W_2),
\end{align*} 
where $(a)$ follows since $X_{1i}=f_{1,i}({Z}_1^{i-1})$, for $i=1$, $X_{11}$ is chosen as constant because the argument of the function is empty, so it can be added for free, $(b)$ is due to the Markov chain assumption of the lemma where given $X_{11},Y_{11}$, $Z_{11}$ can be added for free. Since $X_{12}=f_{1,2}({Z}_{11})$ and it can be added for free, this justifies step $(c)$. With the same argument, we can continue to add first $Z_{1(j-1)}$ given $Y_{1(j-1)},X_{1(j-1)}$ and then $X_{1j}$ given $Z_{1(j-1)}$ until $j=i$ and this will conclude the proof of the lemma.
\end{proof}%\vspace{2mm}

By setting $U_i=(Y_{2}^{i-1},Z_{1}^{i-1},X_{1}^{i-1},W_2)$, it can be shown that
\begin{align*}
&I(W_1;\mb{Y}_1\vert W_2)=
\displaystyle\sum_{i=1}^nI(W_1;Y_{1i}\vert Y_{1}^{i-1},W_2)\\
&=\displaystyle\sum_{i=1}^n \left[H(Y_{1i}\vert Y_{1}^{i-1},W_2)-H(Y_{1i}\vert Y_{1}^{i-1},W_2,W_1)\right] \\
&\stackrel{(a)}{\leq}\displaystyle\sum_{i=1}^n \big[H(Y_{1i}\vert Y_{1}^{i-1},Z_{1}^{i-1},X_{1}^{i},W_2)\\
&\hspace{15mm}-H(Y_{1i}\vert X_{i},X_{1i},Y_{1}^{i-1},W_2,W_1)\big] \\
&\displaystyle \stackrel{(b)}{=}\sum_{i=1}^n \big[H(Y_{1i}\vert Y_{1}^{i-1},Y_{2}^{i-1},Z_{1}^{i-1},X_{1}^{i},W_2)\\
&\hspace{15mm}-H(Y_{1i}\vert X_{i},X_{1i},Y_{1}^{i-1},W_2,W_1)\big]\\
% &\displaystyle \stackrel{(c)}{=}\sum_{i=1}^n \left[H(Y_{1i}\vert Y_{1}^{i-1},Y_{2}^{i-1},Z_{1}^{i-1},X_{1}^{i},W_2)-H(Y_{1i}\vert X_{i},X_{1i})\right]\\
&\stackrel{(c)}{\leq} \displaystyle\sum_{i=1}^n \big[H(Y_{1i}\vert Y_{2}^{i-1},Z_{1}^{i-1},X_{1}^{i-1},W_2,X_{1i})\\
&\hspace{15mm}-H(Y_{1i}\vert X_{i},X_{1i},Y_{2}^{i-1},Z_{1}^{i-1},X_{1}^{i-1},W_2)\big] \\
&=\displaystyle\sum_{i=1}^nI(X_{i};Y_{1i}\vert Y_{2}^{i-1},Z_{1}^{i-1},X_{1}^{i-1},W_2,X_{1i})\\
&= \sum_{i=1}^nI(X_{i},X_{1i};Y_{1i}\vert U_i,X_{1i}), 
%\label{eqCH1:V6-1} 
\end{align*}
where $(a)$ results from Lemma \ref{lemma:V6-1}, $(b)$ results from the Markov chain $Y_{2i}\mkv(Z_{1i},X_{1i})\mkv X_{i}$, and $(c)$ is because $Y_{1i}$ depends only on $(X_{i},X_{1i})$. 

For the next bound, we have
\begin{align*}
&I(W_2;\mb{Y}_2)\leq I(W_2;\mb{Y}_2,\mb{Z}_1)\\
&=\displaystyle\sum_{i=1}^n I(W_2;Y_{1i},Z_{1i}\vert Y_{1}^{i-1},Z_{1}^{i-1})\\
&=\displaystyle\sum_{i=1}^n \left[H(W_2\vert Y_{1}^{i-1},Z_{1}^{i-1})-H(W_2\vert Y_{1}^{i},Z_{1}^{i})\right] \\
&\stackrel{(d)}{\leq}\displaystyle\sum_{i=1}^n \left[H(W_2 \vert Z_{1}^{i-1},X_{1}^{i})-H(W_2\vert X_{1}^{i},Z_{1}^{i})\right] \\
&=\displaystyle\sum_{i=1}^n \big[H(Z_{1i}\vert Z_{1}^{i-1},X_{1}^{i-1},X_{1i})\\
&\hspace{15mm}-H(Z_{1i}\vert X_{1i},X_{1}^{i-1},Z_{1}^{i-1},W_2)\big]\\
&\stackrel{(e)}{=}\displaystyle\sum_{i=1}^n \big[H(Z_{1i}\vert Z_{1}^{i-1},X_{1}^{i-1},X_{1i})\\
&\hspace{15mm}-H(Z_{1i}\vert X_{1i},Z_{1}^{i-1},X_{1}^{i-1},Y_2^{i-1},W_2)\big] 
\end{align*}
\begin{align*}
&{\leq}\displaystyle \sum_{i=1}^n \left[H(Z_{1i}\vert X_{1i})-H(Z_{1i}\vert X_{1i},Z_{1}^{i-1},X_{1}^{i-1},Y_2^{i-1},W_2)\right] \\
&=\displaystyle\sum_{i=1}^nI(Z_{1}^{i-1},X_{1}^{i-1},Y_2^{i-1},W_2;Z_{1i}\vert X_{1i})\\
& = \sum_{i=1}^nI(U_{i};Z_{1i}\vert X_{1i}),
\end{align*}
where $(d)$ follows since $X_{1i}$ is available given $Z_1^{i-1}$, but $Z_1^{i-1}$ also includes $Z_1^{j}$ for all the $j\leq i-1$, therefore given $Z_1^{i-1}$, $X_{11},X_{12},\dots,X_{1(i-1)}$ and thus $X_{1}^{i}$ are also available, and step $(e)$ follows since with $Z_1^{i-1},X_1^{i-1}$ and using the Markov chain  between $(Z_{1},X_1)$ and $Y_2$,  the output $Y_2^{i-1}$ is also available given $Z_1^{i-1}$.  For the last inequality, we have
\begin{align*}
I(W_2;\mb{Y}_2)& =\displaystyle\sum_{i=1}^nI(W_2;Y_{2i}\vert Y_{2}^{i-1}) \\
&\leq \displaystyle\sum_{i=1}^nI(Y_2^{i-1},W_{0};Y_{2i}) \\
& \leq \sum_{i=1}^nI(Z_{1}^{i-1},X_{1}^{i-1},Y_2^{i-1},W_2;Y_{2i})\\
& = \sum_{i=1}^nI(U_{i};Y_{2i}).
\end{align*}
Finally, the bound can be proved using an independent time-sharing RV $Q$.
%%*********************************************************************************************************
%%V.4 Proof Upper Bound
%%*********************************************************************************************************
\section{Sketch of Proof of Theorem \ref{thm:II5-1}} 
\label{proof-4}
We now prove the outer-bound in Theorem \ref{thm:II5-1}. First, notice that the second bound is the capacity of a degraded relay channel, shown in \cite{Cover1979}. Regarding the fact that destination $1$ is decoding all the information, the bound can be reached by using the same method. Therefore the focus is on the other bounds. For any code $(n,\mc{W}_0,\mc{W}_1,P_{e}^{(n)})$  with rates $(R_0,R_1)$, we want to show that if the error probability goes to zero then the rates satisfy the conditions in Theorem \ref{thm:II5-1}. From Fano's inequality we have that  
\begin{align*}
H(W_0|\mb{Y}_2) & \leq P_{e}^{(n)}nR_0+1\stackrel{\Delta}{=}n\epsilon_0, \\
H(W_1|\mb{Y}_1) & \leq H(W_0,W_1|\mb{Y}_1) \\
& \leq P_{e}^{(n)}n(R_0+R_1)+1\stackrel{\Delta}{=}n\epsilon_1,  
\end{align*}
and           
\begin{align*}
nR_0  & \leq I(W_0;\mb{Y}_2)+n\epsilon_0,\\
n(R_0+R_1) &-n\epsilon_0-n\epsilon_1 \leq I(W_0;\mb{Y}_2)+I(W_1;\mb{Y}_1) \\
&\leq I(W_0;\mb{Y}_2)+I(W_1;\mb{Y}_1,W_0), \\
& \leq I(W_0;\mb{Y}_2)+I(W_1;\mb{Y}_1\vert W_0).
\end{align*}
By setting $U_i=(Y_2^{i-1},W_0)$, it can be shown that
\begin{align*}
&I(W_1;\mb{Y}_1\vert W_0) =\displaystyle\sum_{i=1}^n \big[I(W_1;Y_{1i}\vert Y_{1}^{i-1},W_0)\big]\\
& =\displaystyle\sum_{i=1}^n \big[H(Y_{1i}\vert Y_{1}^{i-1},W_0)-H(Y_{1i}\vert Y_{1}^{i-1},W_0,W_1)\big]\\
& \stackrel{(a)}{\leq} \displaystyle\sum_{i=1}^n \big[H(Y_{1i}\vert Y_{2}^{i-1},W_0)\\
&\hspace{15mm}-H(Y_{1i}\vert X_{i},X_{1i},Y_{1}^{i-1},W_0,W_1)\big] \\
& \displaystyle \stackrel{(b)}{=}\sum_{i=1}^n \big[H(Y_{1i}\vert Y_2^{i-1},W_0)-H(Y_{1i}\vert X_{i},X_{1i})\big]  \\
%\displaystyle\sum_{i=1}^nH(Y_{1i}\vert Y_2^{i-1},W_0)-H(Y_{1i}\vert X_{i},X_{1i},Y_2^{i-1},W_0) = \\
& \stackrel{(c)}{\leq} \displaystyle\sum_{i=1}^n \big[I(X_{i},X_{1i};Y_{1i}\vert Y_2^{i-1},W_0)\\
&= \sum_{i=1}^nI(X_{i},X_{1i};Y_{1i}\vert U_i)\big],  \label{eqCH1:IV-3-2} 
\end{align*}
where $(a)$ results from the degradedness between $Y_{1}$ and $Y_2$, where $(b)$ and $(c)$ require the Markov chain between $Y_{1i}$ and $(X_{i},X_{1i})$. Similarly, we have that
\begin{align*}
&I(W_1;\mb{Y}_1\vert W_0) \leq I(W_1;\mb{Y}_1,\mb{Z}_1\vert W_0)\\
&=\displaystyle\sum_{i=1}^n \big[ I(W_1;Y_{1i},Z_{1i}\vert Y_{1}^{i-1},Z_{1}^{i-1},W_0)\big]\\
&=\displaystyle\sum_{i=1}^n \big[ H(W_1\vert Y_{1}^{i-1},Z_{1}^{i-1},W_0)-H(W_1\vert Y_{1}^{i},Z_{1}^{i},W_0)\big]\\
& \stackrel{(d)}{\leq}\displaystyle\sum_{i=1}^n \big[ H(W_1 \vert Z_{1}^{i-1},X_{1i},W_0)-H(W_1\vert X_{1i},Z_{1}^{i},W_0)\big]\\
&=\displaystyle\sum_{i=1}^n \big[ H(Z_{1i}\vert Z_{1}^{i-1},X_{1i},W_0)\\
&\hspace{15mm}-H(Z_{1i}\vert X_{1i},Z_{1}^{i-1},W_0,W_1)\big]\\
&\leq \displaystyle\sum_{i=1}^n \big[ H(Z_{1i}\vert Z_{1}^{i-1},X_{1i},W_0)\\
&\hspace{15mm}-H(Z_{1i}\vert X_{i},X_{1i},Z_{1}^{i-1},W_0,W_1)\big]\\
%\displaystyle\sum_{i=1}^nH(Z_{1i}\vert Z_{1}^{i-1},Y_2^{i-1},X_{1i},W_0)-H(Z_{1i}\vert X_{i},X_{1i}) \leq\\
&\stackrel{(e)}{\leq} \displaystyle \sum_{i=1}^n \big[ H(Z_{1i}\vert Y_2^{i-1},X_{1i},W_0)-H(Z_{1i}\vert X_{i},X_{1i}) \big]\\
&\stackrel{(f)}{=} \displaystyle\sum_{i=1}^n \big[H(Z_{1i}\vert Y_2^{i-1},X_{1i},W_0)\\
&\hspace{15mm}-H(Z_{1i}\vert X_{i},X_{1i},Y_2^{i-1},W_0)\big]\\
&=\displaystyle\sum_{i=1}^n I(X_{i};Z_{1i}\vert X_{1i},Y_2^{i-1},W_0)\\
&= \sum_{i=1}^nI(X_{i};Z_{1i}\vert X_{1i},U_i),
\end{align*}
where steps $(d)$ and $(e)$ result since $X_{1i}$ can be obtained via $Z_1^{i-1}$, so given $Z_1^{i-1}$ one can have $X_1^{i-1}$, and then with $Z_1^{i-1},X_1^{i-1}$ and using the Markov chain between $(Z_{1},X_1)$ and $(Y_1,Y_2)$, one can say that $(Y_1^{i-1},Y_2^{i-1})$ is also available given $Z_1^{i-1}$, and  steps $(e)$ and $(f)$ follow from the Markov chain between $Z_{1i}$ and $(X_{i},X_{1i})$. For the first inequality, we have
\begin{align*}
I(W_0;\mb{Y}_2) &=\displaystyle{\sum_{i=1}^n I(W_0;Y_{2i}\vert Y_{2}^{i-1}) } \leq  \sum_{i=1}^n I(U_{i};Y_{2i}). 
\end{align*}
Finally, the bound can be proved using an independent time-sharing RV $Q$. 

%%*********************************************************************************************************
%%V.7 Proof General Upper Bound for BRC-CR
%%*********************************************************************************************************
\section{Sketch of Proof of Theorem \ref{thm:II5-3-1}} 
\label{proof-5-1}
The direct part can be easily proved by using expression \eqref{eqCH1:III1-32} by removing $d_1$ and $d_2$ from the definition of the channel. Regarding the converse proof, we start with the following lemma.
\begin{lemma}
Any pair of rates $(R_1,R_2)$ in the capacity region $\mathcal{C}_{\textrm{BRC-PC}}$ of the degraded Gaussian BRC-PC satisfy the following inequalities:
\begin{align*}
nR_1\leq& \sum_{i=1}^n I(U_i,X_{1i};Y_{1i})+n\epsilon_1,\\
n(R_1+R_2)\leq& \sum_{i=1}^n I(U_i;Z_{1i}\vert X_{1i})\\
&\hspace{20mm}+I(X_i;Y_{2i}\vert U_i,X_{1i})+n\epsilon_2.
\end{align*}
\end{lemma}
\begin{proof}
This lemma can be obtained by taking $U_i=(W_1,Y_1^{i-1},Z_1^{i-1},Y_{2(i+1)}^{n})$ and similar steps as in Appendix \ref{proof-8}. For this reason, we will not repeat the proof here. Note that only the degradedness between the relay and the first destination is necessary for the proof. 
\end{proof}

Now for the Gaussian degraded BRC-PC defined as before, we calculate the preceding bounds. The calculation follows the same steps as in Appendix \ref{proof-4}. We start by bounding $h({Z}_{1i}|U_i,X_{1i})$ where it can be seen that
\begin{align*}
h(\tilde{\mathpzc{N}}_{1i})&=h({Z}_{1i}|U_i,X_{i},X_{1i})\leq h({Z}_{1i}|U_i,X_{1i})\\
&\leq h({Z}_{1i})= h(X_{i}+\tilde{\mathpzc{N}}_{1i}).
\end{align*}
Using this fact it can be said that
\begin{align*}
\displaystyle{\frac{n}{2}\log\left [2\pi e\tilde{N}_1 \right]}&=\sum_{i=1}^n h(\tilde{\mathpzc{N}}_{1i})\\
&\leq \sum_{i=1}^n h({Z}_{1i}|U_i,X_{1i})\\
&\leq \sum_{i=1}^n h(X_{i}+\tilde{\mathpzc{N}}_{1i})\\
&=\displaystyle{\frac{n}{2}\log\left [2\pi e(\tilde{N}_1+P) \right]}.
\end{align*}
The previous condition implies that there is $\alpha\in[0,1]$ such that 
$$
\sum_{i=1}^n h({Z}_{1i}|U_i,X_{1i})=\displaystyle{\frac{n}{2}\log\left [2\pi e(\tilde{N}_1+\overline{\alpha} P) \right]}.
$$ 
Note that the previous condition means that
\begin{align*}
\frac{1}{n}\sum_{i=1}^n \esp\esp^2(X_i|U_i,X_{1i})=\alpha P.
\end{align*}
Now take the following inequalities
\begin{align*}
0 &\leq \frac{1}{n}\sum_{i=1}^n \esp\esp^2(X_i|X_{1i})\\
&\leq \frac{1}{n}\sum_{i=1}^n \esp\esp^2(X_i|U_i,X_{1i})=\alpha P.
\end{align*}
This is the result of $\esp\esp^2(X|Y)\leq\esp\esp^2(X|Y,Z)$ which can be proved using Jensen's inequality. Similarly, the previous condition implies that there exists $\beta\in[0,1]$ such that 
\begin{align*}
\frac{1}{n}\sum_{i=1}^n \esp\esp^2(X_i|X_{1i})=\overline{\beta}\alpha P.
\end{align*}
From this equality, we get the following inequalities by following the same technique as \cite{Cover1979}
$$
\sum_{i=1}^n h({Z}_{1i}|X_{1i})\leq\displaystyle{\frac{n}{2}\log\left [2\pi e(\tilde{N}_1+\overline{\alpha}P+ \alpha\beta P) \right]}.
$$ 
Also exploiting the fact that $h(Y_{1i})$ can be bounded by
$$
\sum_{i=1}^n h({Y}_{1i})\leq\displaystyle{\frac{n}{2}\log\left [2\pi e({N}_1+P+P_1+2\sqrt{\alpha\overline{\beta}PP_1)} \right]}.
$$ 
From the degradedness of $Y_1$ respect to $Z_1$ and $Y_2$, and using  entropy power inequality, we obtain
\begin{align*}
\sum_{i=1}^n h({Y}_{1i}|U_i,X_{1i})\geq&\displaystyle{\frac{n}{2}\log\left [2\pi e({N}_1+\overline{\alpha} P) \right]},\\
\sum_{i=1}^n h({Y}_{2i}|U_i,X_{1i})\leq&\displaystyle{\frac{n}{2}\log\left [2\pi e(N_2+\overline{\alpha} P) \right]},
\end{align*}
which prove the upper bound and conclude the proof.

%%%%%%%%%%%%%%%%%%%%%%%%%%%%%%%%%%%%%%%%%%%%%%%%%%%%%%%%%%%%%%%%%%%%%%%%%%%%%%%%%%%%%%%%%%%%%%%%%%%%%%%%%%%%%%%%%%%%%%%

\section*{Acknowledgment}
The authors are grateful to Prof.\ Gerhard Kramer for many helpful discussions, and the Associate Editor Prof. Elza Erkip and the anonymous reviewers for their helpful comments and suggestions on earlier drafts of this paper.

\bibliographystyle{IEEEtran}
\bibliography{biblio}

\newpage

\begin{IEEEbiographynophoto}{Arash Behboodi} (S'08) reveived the B.Sc. and M.Sc. degrees in electrical engineering- communication systems from Sharif University of Technology, Tehran, Iran in 2005 and 2007 and Ph.D. degree from \'{E}cole Sup\'{e}rieure d'\'{e}lectricit\'{e} (Sup\'{e}lec), Gif-sur-Yvette, France in 2012. He is currently a posdoctoral researcher in SUPELEC. His research interests are information theory and stochastic geometry with application in complex communication networks.
\end{IEEEbiographynophoto}
\vspace{-180mm}
\begin{IEEEbiographynophoto}{Pablo Piantanida}
Pablo Piantanida has received the B.Sc. and M.Sc degrees in Electrical Engineering from the University of Buenos Aires (Argentina), in 2003, and the Ph.D. from the Paris-Sud University (France) in 2007. In 2006, he has been with the Department of Communications and Radio-Frequency Engineering at Vienna University of Technology (Austria). Since October 2007 he has joined in 2007 the Department of Telecommunications, SUPELEC, as an Assistant Professor in network information theory. His research interests include multi-terminal information theory, Shannon theory, cooperative communications, physical-layer security and coding theory for wireless applications. 
\end{IEEEbiographynophoto}

\end{document}